%% file: diss_lib-sub.tex
\documentclass[aps,amsmath,amssymb,11pt]{revtex4}
\usepackage{tikz,graphicx}
\usepackage{hyperref}
\usepackage{slashed}
\usepackage{stmaryrd}
\usepackage{bbold}
\usepackage{eepic}
\usepackage{pst-all}
\usepackage{pstricks-add}
\usepackage{lscape}
\usepackage{rotating}
\usepackage{multirow}
\newcommand{\RGB}[2]{\newrgbcolor{mycol}{#1}{\mycol#2}}
\begin{document}
\pagenumbering{roman}
\pagestyle{plain}
\thispagestyle{empty}
\begin{center}
\hrulefill\\\vspace{.3cm}
\textbf{\Large \vspace{0.7cm}Pionless Effective Field Theory in Few-Nucleon Systems}\\
\vspace{.01cm}
{\huge\textbf{$\slashed{\pi}$}}\\
\vspace{.3cm}
\hrulefill\\
\vspace{3.5cm}
by Johannes Kirscher\\
\vspace{0.2cm}
Diplom Physiker, March 2006, Friedrich-Alexander-Universit\"at Erlangen\\
\vspace{4cm}
\setlength{\baselineskip}{12pt}
A Dissertation submitted to\\
The Faculty of\\
The Columbian College of Arts and Sciences\\
of The George Washington University\\
in partial fulfillment of the requirements for the degree of\\
\textit{Doctor of Philosophy}\\
\vspace{1.55cm}
January 31, 2011\\
\vspace{1.55cm}
Dissertation directed by\\
Harald W. Grie\ss hammer\\
(Assistant Professor of Physics)\\
\end{center}
\setlength{\baselineskip}{20pt}
\newpage
\vspace{2cm}
The Columbian College of Arts and Sciences of The George Washington University certifies that Johannes Kirscher has
passed the Final Examination for the degree of Doctor of Philosophy as of October 26, 2010. This is the final and
approved form of the dissertation.\vspace{1cm}\\
\begin{center}
\textbf{\Large Pionless Effective Field Theory in Few-Nucleon Systems}\\
Johannes Kirscher\\
\end{center}
\vspace{2cm}
Dissertation Research Committee:
\begin{itemize}
\item[] Harald W. Grie\ss hammer, Assistant Professor of Physics, Dissertation Director
\item[] Gerald Feldman, Professor of Physics, Committee Member
\item[] Andrei Alexandru, Assistant Professor of Physics, Committee Member
\end{itemize}
\newpage
\newpage
\thispagestyle{plain}
\begin{acknowledgements}
\vspace{1cm}
Meinem Doktorvater Professor Harald Grie\ss hammer danke ich vor allem f\"ur seine Beharrlichkeit bei der Erziehung eines sturen und oft zu eitlen Sch\"ulers.
\begin{center}
\textit{``Lassen sie doch das Gestelze Herr Hesse!''}\\
\end{center}
Ohne die durch anhaltende Unterst\"utzung Professor Hartmut Hofmanns erreichten Erkentisse das RGM betreffend, w\"aren Ergebnisse derartiger Qualit\"at nicht in einer
solch kurzen Zeit zu erreichen gewesen.
Ai stimatissimi professori, signora Giuseppina Orlandini e signor Winfried Leidemann vorrei esprimere i miei sinceri
ringraziamenti per la critica costruttiva sullo sviluppo della base di RGM manifestata durante due soggiorni indimenticabili
nella bellissima citt\`{a} di Trento.
I do not want to miss the lesson Professor Michael Birse taught me about being thorough during a
discussion about cutoff dependence of deuteron and triton binding energies.
Dr. Matthias Schindler hat nicht nur durch Akribie beim Verbessern mehrerer Probeversionen dieser Arbeit ihre Qualit\"at entscheidend zum Guten ver\"andert,
sondern hat w\"ahrend vieler Unterhaltungen bei Kunstlicht und beim Spaziergang zur k\"orperlichen Ert\"uchtigung so manches ehrliche L\"acheln in mir geweckt.\\
I am in debt to my fellow students, Craig Pelissier in particular, for memorable, multi-cultural gatherings educating the stereotype of a foreign student.
W\"ahrend meiner Aufenthalte in Erlangen, wurden die morgentlichen Kaffeerunden, nach 75 min\"utiger Radanreise, mit Dr. Felix Karbstein zu Sinnbildern
der gepflegten Konversation.\\
Meinen Freunden Thorsten Bachmann, Simon Heugel und Joachim Schmidt danke ich sich als so weit entfernte Projektionsfl\"achen f\"ur zunehmend
nihilistische Gedanken geopfert zu haben.\\
Nichts macht mich gl\"ucklicher als das Wissen um eine Familie, die mir bedingungsloser R\"uckhalt war und sein wird, obgleich sie nur die negativen Folgen dieser Arbeit
auf mein Gem\"ut zu sp\"uren bekam. Vielen Dank Johanna und Markus daf\"ur das Erreichte besser einordnen zu k\"onnen.
Meinem Bruder Uwe, meinem Vater Manfred und meiner lieben Mutter Renate widme ich diese Arbeit.
\end{acknowledgements}
\newpage
\thispagestyle{plain}
\section*{Abstract}
\begin{center}
{\large\textbf{Pionless Effective Field Theory in Few-Nucleon Systems}}\\
\end{center}
\vspace{.3cm}
A systematic description of low-energy observables in light nuclei is presented. The effective field theory formalism without pions is extended to:
\begin{itemize}
\itemindent=2cm
\item predictions with next-to-leading-order accuracy for\item[] the 4-helium binding energy $B(\alpha)$, the triton charge radius,\item[] and the 3-helium-neutron scattering length;
\item phase shifts for neutron-deuteron scattering and $\alpha$-neutron\item[] low-energy scattering at leading order;
\item the ground states of the 5-helium (with and without Coulomb interaction)\item[] and 6-helium isotopes up to next-to-leading order;
\end{itemize}
The convergence from leading- to next-to-leading order of the theory is demonstrated for correlations between:
\begin{itemize}
\itemindent=3cm
\item the triton binding energy $B(t)$ and the triton charge radius;
\item $B(t)$ and the 4-helium binding energy $B(\alpha)$;
\end{itemize}
Furthermore, a correlation between $B(t)$ and the scattering length in the singlet S-wave channel of neutron-helium-3 scattering is discovered, and a
model-independent estimate for the trinucleon binding energy splitting is provided.
The results provide evidence for the usefulness of the applied power-counting scheme, treating next-to-leading-order interactions nonperturbatively
and four-nucleon interactions as, at least, one order higher. The 5- and 6-helium ground states are analyzed with a power-counting scheme which
includes the momentum-dependent next-to-leading order vertices perturbatively.
All calculations include a full treatment of the Coulomb interaction.
The assessment of numerical uncertainties associated with the solution of the few-body equation of
motion through the Resonating Group Method parallels the report of the results for light nuclei in order to establish this method as practical
for the analysis of systems with up to six particles interacting via short-range interactions.
\newpage
\thispagestyle{plain}
\setlength{\baselineskip}{16pt}
\tableofcontents
\setlength{\baselineskip}{20pt}
\newpage
\thispagestyle{empty}
\begin{center}{\small
\begin{minipage}[c]{.75\textwidth}
\vspace{3.5cm}
\begin{center}
\textit{[...]}
\end{center}
\par
\textsl{There is, or was, a poet. His name was Lallafa, and he wrote what are widely regarded throughout the Galaxy as being the finest poems in existence, the ``Songs of the Long Land''.}
\par
\textsl{They are/were unspeakably wonderful. That is to say, you couldn't speak very much of them at once without being so overcome with emotion, truth and a sense of wholeness and oneness
of things that you wouldn't pretty soon need a brisk walk round the block, possibly pausing at a bar on the way back for a quick glass of perspective and soda. They were that good.}
\par
\textsl{Lallafa had lived in the forests of the Long Lands of Effa. He lived there, and he wrote his poems there. He wrote them on pages made of dried habra leaves, without the benefit
of education or correcting fluid. He wrote about the light in the forest and what he thought about that. He wrote about the darkness in the forest, and what he thought about that. He
wrote about the girl who had left him and precisely what he thought about that.}
\par
\textsl{Long after his death his poems were found and wondered over. News of them spread like morning sunlight. For centuries they illuminated and watered the lives of many people
whose lives might otherwise have been darker and drier.}
\par
\textsl{Then, shortly after the invention of time travel, some major correcting fluid manufacturers wondered whether his poems might have been better still if he had had access
to some high-quality correcting fluid, and whether he might be persuaded to say a few words to that effect.}
\par
\begin{center}
\textit{[...]}
\end{center}
\vspace{.2cm}
\begin{flushright}
Douglas Adams \textit{``Life, the Universe and Everything''}
\end{flushright}
\end{minipage}}
\end{center}
\newpage
\pagenumbering{arabic}
\section{Introduction}\label{sec_intro}
Since nucleons are composed of quarks and gluons, the nuclear interaction can be understood as an effective theory of the strong interaction at a low-energy scale.
Effective field theories (EFTs), in general, are appropriate representations of particular theories in regions of the parameter space of our physical
world where$^\text{\cite{georgi-eft}}$ the latter are assumed to be valid but cannot be solved practically. In this sense,
Newtonian gravitational dynamics between two point masses relates General Relativity to a region of small momenta relative to the interacting masses.
Another example is the classical electrodynamic interaction energy between a point charge and a charge distribution - localized within a radius $R$ - which can be approximated by
the first terms of a multipole expansion for distances $\gg R$. A quantum EFT, describing properties of electrons ($e^-$) and photons at energies below $1$~MeV,
is Quantum Electrodynamics which does not take into account the weak interaction and its accompanying effects of other Leptons.
The last example is an EFT with a differing particle content, namely $e^-$ and photon gauge fields without $\mu$- and $\tau$-leptons and the respective
neutrinos. The Lagrange density is subject to the same space-time, \textit{i.e.}, Lorentz-, symmetry but a $U(1)$ instead of a $SU(2)\times U(1)$ gauge symmetry compared to
the underlying weak Lagrangean.
\par
In this work, a ``pionless'' EFT (EFT($\slashed{\pi}$)$^\text{\cite{eft-mam-rev}}$), with Chiral Perturbation Theory
($\chi$PT, see $\text{\cite{chipt-scherer}}$ for a pedagogical review and references) as the fundamental interaction,
is applied to few-nucleon systems. This EFT is devised for momenta of the interacting nucleons below the pion production threshold and consequently
considers solely neutrons and protons as degrees of freedom. Similar to the effect of $W-$ and $Z-$bosons in $e^-$ scattering, pion contributions to nucleon
amplitudes at low energies can be taken into account by proper renormalization.
\par
The goal of this work is to assess the region of the physical parameter space where EFT($\slashed{\pi}$) is applicable. This region is known
to include two- and three-nucleon bound- and scattering data with and without electromagnetic external probes, but its usefulness for the description of amplitudes
involving more than three nucleons in the asymptotic states is not known.
Knowledge about the range of applicability of EFT($\slashed{\pi}$) will further the understanding of few-nucleon systems. First, by providing a
minimal set of parameters for a required accuracy in observables of interest. Therefore, numerical calculations are not unnecessarily complicated
without gain in accuracy.
\newpage
Second, peculiar features of a system, \textit{e.g.}, the unbound five-nucleon state and the halo structure of 6-helium, are related to either $A\leq 4$ parameters
(throughout the text, $A$ denotes the number of particles in the asymptotic states and $B(A)$ is a binding energy of an $A$-body system which is not necessarily
related to a physical state)
or to unique properties of the respective five- and six-body systems. Analogously, the triton bound state is a unique property of
the three-nucleon system. Its binding energy, for instance, is in general not predicted correctly independently of the accuracy of the description of
the two-nucleon interaction. In effect, a system of two atoms can behave like a neutron and a proton while at the same time the similarity breaks down
for observables in the respective three-body systems. The three-body properties are unique to the atomic and the nucleon system.
Yet there are properties in the larger systems which follow from parameters of the interaction of the smaller subsystems - little Matryoschka knows
her mother to be bigger but cannot tell whether she is smiling.
We aim to identify universal properties  common to systems which share certain sets of $A\leq 4$ parameters. While universal
features of three- and four-body systems have been studied theoretically$^\text{\cite{bra-ham}}$ and experimentally$^\text{\cite{cold-bosons}}$,
universality in $A\geq 5$ systems is new territory.
The focus is on bound systems where no \textit{a priori} arguments about applicability or breakdown of the theory are known at the moment.
Specifically, no practical quantitative procedure is known to properly weigh the high-energy components present in the momentum distribution of a bound state,
and by that conclude from the distribution of nucleon momenta inside the nucleus directly on the breakdown of an EFT.
The gravitational analog to the nuclei considered here would be a ``solar'' system of three objects with masses
and momenta such that bulk properties - average extension and total angular momentum, for instance - are to some level described by the Newtonian theory,
but the influence of a fourth object could potentially lead to an increase of the momenta to an extent that space-time curvature can no longer be disregarded,
even in the leading approximation. While the Newtonian theory is
apt for \textit{the} solar system with its $\geq 8$ planets, one may na\"ively conclude from the relatively deeply bound four-nucleon state that a description
of nuclei composed of more than four nucleons in terms of those constituents - namely neutrons and protons - alone, \textit{i.e.}, with EFT($\slashed{\pi}$), is inappropriate.
Inappropriate in the sense
that components of a bound state with nucleon momenta larger than the mass of the lightest field of the underlying theory that was not included in
EFT($\slashed{\pi}$) - the mass of the pion $m_\pi$ - are significant.
Significant meaning that the order-by-order removal of parametrically small
dependencies on renormalization parameters - a feature of all EFTs - cannot be performed.
We will provide evidence that this breakdown of the EFT does not occur.
\par
Related to this puzzle of developing an interaction theory for few-nucleon systems (here, few means $A\leq 6$)
systematically from $\chi$PT and through that from Quantum Chromo Dynamics, is the difficulty of solving the few-body equations of motion. An approach refined over decades to yield very accurate predictions of both bound- and scattering nuclear observables has been to solve the $A-$body Schr\"odinger equation,
{\small\begin{eqnarray}\label{eq_a-body-sgl}
\hat{H}\,\vert\phi^{(n)}\rangle&=&
\left(\hat{T}_\text{cm}+\sum_{i<j}^A\frac{(\vec{p}_i-\vec{p}_j)^2}{2Am_\text{\tiny N}}+
\sum_{i<j}^AV_\text{NN}(\vec{r}_i,\vec{r}_j)+\sum_{i<j<k}^AV_{3\text{NI}}(\vec{r}_i,\vec{r}_j,\vec{r}_k)+\ldots\right)\vert\phi^{(n)}\rangle\nonumber\\
&=&E^{(i)}\,\vert\phi^{(n)}\rangle\;\;\;,
\end{eqnarray}}
where the center-of-mass motion can be treated independently and the nucleons have the same mass $m_\text{\tiny N}$.
The operator $\hat{T}_\text{cm}$ corresponds to the kinetic energy of the center of mass, and $\vec{r}_j$ ($\vec{p}_j$) is the coordinate (momentum) of one
of the nucleons in the $n$-th $A$-particle eigenstate $\vert\phi^{(n)}\rangle$.
Recent models$^\text{\cite{av18,bonn,ill,uix}}$
for the potentials $\hat{V}_{\text{NN},3\text{NI}}$- dots represent possible $4-$ and up to $A-$body interactions - are based on boson- and meson-exchange
phenomenology and symmetry
considerations. Here, the nonperturbative parts of the EFT($\slashed{\pi}$) interaction are related to a potential, too, while perturbative parts are treated as such.
The solution to eq.~(\ref{eq_a-body-sgl}) is obtained numerically, and while one can think of this as a technicality, aloof from the theory behind the potential,
the method chosen clarifies how uncertainties introduced by the numerical technique can be related to the residual renormalization scheme dependence.
\par
Specifically, our method of choice to solve the Schr\"odinger equation is a variational technique which allows an economic calculation of $A\leq 6$ bound and
scattering observables. The wave function ansatz follows the idea$^\text{\cite{rgm-a-wheeler,rgm-b-wheeler}}$ of the resonating group structure which combines
the total wave function of a particular state of a nucleus from components that divide the constituent neutrons and protons into separate groups. A first approximation
to the 4-helium ground state would accordingly be a neutron interacting with a 3-helium core, plus a proton attached to a triton, plus two deuterons bound together.
Constraining the variational basis with a grouping like this is not done here for $A\leq 4$ bound states, for which all possible groupings are taken into
account - \textit{i.e.}, di-neutron plus di-proton and two singlet deuterons in addition to the above mentioned - and a certain grouping merely defines the spin-
and orbital angular momentum structure.  The 6-helium calculations, in contrast, will show how a single configuration, namely a 4-helium core with two orbiting
neutrons, is sufficient for an appropriate description of the ground state. In contrast to the most recent incarnation of the method$^\text{\cite{rgm-nav-qual}}$ which
expands the bound fragment states - the triton, 3-helium, and deuteron in the example above - into a shell model basis, the method used here employs a
nonorthogonal Gau{\ss}ian basis allowing for an association of single basis vectors with high- or low-energy modes. Therefore, at least a heuristic relation of the basis
truncation to a regularization of the interaction can be drawn. Loop integrals are regularized two times: with a finite square-integrable basis
and the additional vertex regulator. Only with an infinite number of width parameters a complete basis would expand the momentum-modes responsible for
divergences in an amplitude.
\par
We will utilize the freedom in choosing a regulator for EFT($\slashed{\pi}$) to make the numerical calculations highly economical. This efficiency will allow us to
analyze correlations amongst few-nucleon observables by a variation of a three-body force parameter even for four-nucleon scattering observables at next-to-leading
order. Correlated observables are a natural consequence of an EFT, which claims to predict low-energy data within defined error margins once its
coupling constants are fitted to other low-energy data.
Independent of the specific regularization, one set of input data is thus correlated to another, \textit{e.g.}, with an interaction fitted to the triton binding energy $B(t)$,
predictions for the neutron-deuteron scattering length $^2a_\text{\tiny nd}$ are found in good agreement with data.
The approach will be used here for two reasons:
\begin{itemize}
 \item To reveal potential inconsistencies in the power counting as follows. Assuming a sound power counting, a change in the renormalization scheme of the
EFT - induced here by a variation in the regulator, the input data, and the three-body force - should only result in variations of predicted observables consistent
with the error estimate of the order at which the EFT calculation is conducted. Exceeding this uncertainty margin indicates an enhancement of an interaction
which was falsely demoted to a higher order (see the promotion of a three-nucleon force to leading order when considering the correlation between $B(t)$ and
$^2a_\text{\tiny nd}$, see \textit{e.g.}~\cite{tni-gang-1}).
\item With a power counting established at a specific order, discrepancies between (inconsistent data)/(theoretical predictions) based on different
experiments/(interaction models or numerical techniques) could be resolved. For example, we will calculate the scattering length for elastic 3-helium-neutron
scattering $a_0(^3\text{{\small He-n}})$ correlated to $B(t)$. If $B(t)$ is used as fitting input, $a_0(^3\text{{\small He-n}})$ is predicted with an error
assessed by a change in the renormalization scheme. Any measurement which predicts $a_0(^3\text{{\small He-n}})$
outside of this EFT uncertainty margin must be reevaluated. For this method, the knowledge of the ``true amount'' of the uncertainty is crucial. In the discussion of
neutron-deuteron scattering we will discuss how various probes of unobservable short-distance structure assess uncertainty quite differently. Limiting
the probe to a change in a regulator parameter, only, will be found to severely overestimate the accuracy of EFT($\slashed{\pi}$) at leading order.
\end{itemize}
The application of EFT($\slashed{\pi}$) to systems of more than four particles is of particular interest because of its implications for other, non-nuclear systems.
Peculiar properties associated with certain features of the subsystems of nuclei would translate to systems with the same number but different species of
particles. Specifically, the investigation here aims to resolve whether the peculiar unbound five-nucleon system and the halo structure of the 6-helium bound
state are universal properties of systems with unnaturally large two-body scattering lengths and a three-body bound state. Such a finding could be related to
atomic systems, where experimentally$^\text{\cite{cold-fermions}}$ similar conditions regarding $a_{s,t}$ and $B(t)$ can be realized.
\par
In this work, we study nuclear systems only, while the results can be related to other systems which interact via forces of the same universality class.
I start with a general overview of the EFT formalism.
\section{Nuclear Effective Field Theory}\label{sec_eft}
\textbf{\large Generalities.}\hspace{.5cm}
An effective field theory exploits the existence of different scales in a system. While an underlying theory is valid at least up to an energy scale $M$ with a Lagrange density
$\mathcal{L}\left(\Psi\right)$ constrained by spacetime and internal symmetries, the EFT is useful only for momenta $Q\ll M$. In general, either the ignorance about
the underlying theory, or the inability/impracticality to solve it in the energy interval of interest, motivate the model-independent EFT approach. The partition function of the effective theory,
\begin{equation}\label{eq_partition-function}
Z=\int\mathcal{D}\psi e^{i\int d^4x\mathcal{L}^{(\delta_m)}\left(\psi\right)}\;\;\;,
\end{equation}
from which the $S$-matrix is derived, depends on a different set of degrees of freedom (DoF) $\psi$, namely either only low-momentum modes of $\Psi$ or composed states
of the fundamental fields, for example, nucleons instead of quarks or $\alpha$ particles instead of pairs of neutrons and protons.
The symmetry group is the same or an approximation of the original, constraining the operators $\mathcal{Q}$ with mass dimension $\delta$ in the effective Lagrangean,
\begin{equation}\label{eq_general-eft-lagrangean}
\mathcal{L}^{(\delta_m)}\left(\psi\right)=\sum_{\delta<\delta_m}\sum_{i(\delta)}^{N(\delta)}g_i\left(\Lambda\right)\mathcal{Q}_i\left(\psi\right)\;\;\;.
\end{equation}
If expressed in the same DoF, the $S$-matrix of the underlying theory is recovered for $\delta_m\to\infty$ and momenta of order $Q$. The low-energy coupling
constants (LEC) $g_i$ depend on a renormalization parameter $\Lambda$
which specifies a regularization scheme and renormalization conditions and thus the balance of loops and vertices for a specific contribution to the $S$-matrix.
Assuming a natural size for the LECs (see eq.~(\ref{eq_naturalness})),
a truncation of the sum in
eq.~(\ref{eq_general-eft-lagrangean}) at $N_\text{max}$ will yield a Lagrange density which contributes all terms, up to a certain power, of an expansion of
the $S$-matrix in the small parameter
$Q/M$. In addition to this ordering scheme of the interaction terms in $\mathcal{L}$ - the vertex expansion -
the EFT framework requires an ordering of iterations of those operators - the loop expansion.

Two examples of phenomena determining the scale $M$ are:
\begin{itemize}
\item the mass of a particle associated with a DoF of the EFT whose mass is too large relative to its momentum to be produced or to decay; this is relevant in
deriving the kinetic-energy term for nucleons below;
\item a scalar field operator composed of DoF from the underlying theory with a non-zero vacuum expectation value; relevant here:
$\Psi\in\lbrace\text{neutron(n), proton(p),pion}(\pi),\Delta-\text{resonance}\rbrace$.
\end{itemize}
Having set $M$, the theory is expected to be useful for observables which are dominated by external momenta $Q\ll M$, and to break down if $Q$ is of the same magnitude or larger
than $M$.
The contributions from the most general Lagrangean compatible with the symmetries of the system, conjectured to yield the most general $S$-matrix with the
same symmetries for $Q\ll M$, are subjected to an ordering scheme in powers of $Q$. This scheme is to be verified \textit{a posteriori} by the analysis of the dependence of low-energy
observables on the renormalization scheme parameterized by $\Lambda$. If this dependence exceeds the contributions expected from operators with $N>N_\text{max}$,
the power counting was based on false assumptions, and one has to consider, \textit{e.g.}:
\begin{itemize}
\item LEC(s) of unnatural size;
\item Typical momenta in a system which are too large to put a certain observable within the range of applicability of the EFT.
\end{itemize}
\par\vskip10pt
\textbf{\large Degrees of Freedom.}\hspace{.5cm}
The objects of interest are nuclear systems with $A\leq 6$.
For scattering observables the interest is in center of mass energies $E_\text{cm}<10~$MeV. The na\"ive estimates for $Q$ in the
bound systems considered in this work together with relevant data are shown in table~\ref{tab_q-estimate}.
\begin{table*}
\renewcommand{\arraystretch}{1.5}
  \caption{\label{tab_q-estimate}{\small Data for light nuclei. To estimate the low-momentum scale $m_\text{\tiny N}=938.858~$MeV is used.}}
\footnotesize
\begin{tabular}{ll|ccccc}
\hline
      &&deuteron(d)&triton($^3$H,t)&3-helium($^3$He)&4-helium($\alpha$)&6-helium($^6$He)\\
\hline\hline
$B$&[MeV]&$2.2245727(20)^\text{\cite{exp-B_d}}$&
$8.481855(13)^\text{\cite{exp-B_t}}$&$7.718109(10)^\text{\cite{exp-B_t}}$&$28.295875(26)^\text{\cite{exp-B_t}}$&$29.2712(10)^\text{\cite{exp-B_t}}$\\
$\langle r^2\rangle_{\text{\tiny ch}}^{1/2}$&[fm]&$2.0952(60)^\text{\cite{exp-d_rch}}$&
$1.63(3)^\text{\cite{exp-t_rch}}$&$1.976(15)^\text{\cite{exp-4he_rch}}$&$1.671(14)^\text{\cite{exp-4he_rch}}$&$2.054(14)^\text{\cite{exp-6he_rch}}$\\
$Q\sim\sqrt{\frac{2m_\text{\tiny N}}{A}B}$&[MeV]&45.70&72.86&69.50&115.3&117.2\\
\hline
    \end{tabular}
\end{table*}
Nuclear processes at momenta of order $Q$ do not probe details of the internal structure of the nucleons, \textit{i.e.}, quarks and gluons. Transitions to excited states
such as the $\Delta$ or the decay into lighter particles, \textit{e.g.}, $\pi$- or $\rho$-mesons, with high momenta, are strongly suppressed. With $Q<m_\pi$ (mass of the pion),
even virtual pion exchanges can be approximated by contact interactions.
As a consequence, the theory is formulated with non-relativistic fermions, namely neutrons and protons, as the
sole DoF. The number of nucleons $A$ in the incoming state is conserved and only affected by interactions represented by operators with $\leq 2A$ nucleon fields.
The scales of the ``pionless'' effective field theory (EFT($\slashed{\pi}$)$^\text{\cite{rev-pionless-platter}}$) are:
\begin{itemize}
\item high-momentum-/breakdown-scale: $M\sim m_\pi$
\item low-momentum scale: $Q\sim\sqrt{m_\text{\tiny N}B(d)}$
\end{itemize}
The low-momentum scale is estimated here by the assumed typical momentum of a nucleon in the deuteron with binding energy $B(d)$. For scattering calculations,
the effective theory is useful for external momenta up to this order. In bound systems, the usefulness depends on the momentum distribution of individual nucleons.
In general, the bound state is a superposition of states with relative nucleon momenta ranging from zero to infinity. Only if the momentum distribution function
supresses states corresponding to momenta larger than $Q$ sufficiently, a useful description of the bound system can be expected. For the deuteron, the
formula in the last line of table~\ref{tab_q-estimate} estimates typical momenta of the nucleons in it well, if the high-momentum states
are strongly suppressed. For larger systems the angular momentum structure is more complicated. It is, for instance, possible to find two of four particles in a relative
D-wave state while the other two are in a relative S-wave.
Hence, the na\"ive estimate for a typical momentum given in table~\ref{tab_q-estimate} is no \textit{a priori} argument for the breakdown
of EFT($\slashed{\pi}$) in the respective system.
For a defined external momentum, however, the expansion of observables provided by this theory is appropriate for $Q\ll\mathcal{O}(m_\pi)$, becomes less useful for
$Q\lesssim\mathcal{O}(m_\pi)$, and is not applicable for $Q\gtrsim\mathcal{O}(m_\pi)$.
\par\vskip10pt
\textbf{\large Symmetries.\hspace{.5cm}}
Having defined the DoF, the interaction terms in the Lagrange density are constrained by \textit{spacetime} and \textit{internal} symmetries. Nucleons with small momenta
relative to their rest mass can be approximated by Pauli instead of Dirac spinors, and the Lagrangean should only be invariant under Galilei transformations and not under
elements of the proper Lorentz group. Parity remains as a symmetry of this theory for the strong interaction. The transition from a relativistic formulation of the kinetic energy
density to its low-energy version is summarized following$^\text{\cite{eft-repara,eft-hqlm}}$, to make the connection between the relativistic and non-relativistic theory.
\par
\begin{equation*}
\mathcal{L}_\text{kin}=
n^\dag(i\slashed{\partial}-m_\text{\tiny N})n\;\;\;\stackrel{\ref{enum_L-kin-1}.}{\rightarrow}\;\;\;\mathcal{L}_\text{kin}=\sum_vn_v^\dag(i\slashed{\partial}-m_\text{\tiny N})n_v\;\;\;
\stackrel{\ref{enum_L-kin-2}.}{\longrightarrow}\;\;\;\mathcal{L}_\text{kin}=\sum_vN^\dag_v\slashed{v}v_\nu\partial^\nu N_v
\end{equation*}
\begin{equation}\label{eq_L-kin}
\stackrel{\ref{enum_L-kin-3}.}{\rightarrow}\;\;\;
\mathcal{L}_\text{kin}=\sum_vN^\dag_v\left(i v_\nu\partial^\nu-\frac{\partial_\nu\partial^\nu}{2m_\text{\tiny N}}\right)N_v\;\;\;
\stackrel{\ref{enum_L-kin-4}.}{\rightarrow}\;\;\;
\mathcal{L}_\text{kin}=N^\dag\left(i\partial_0+\frac{\vec{\nabla}^2}{2m_\text{\tiny N}}+\mathcal{O}\left(\frac{1}{m_\text{\tiny N}^2}\right)\right)N
\end{equation}
\begin{enumerate}
\item\label{enum_L-kin-1}Fermion fields $n$(neutron or proton) with 4-momentum $p_\nu=m_\text{\tiny N}v_\nu$ do not change their velocity in the limit $m_\text{\tiny N}\to\infty$ under a momentum transfer $k_\nu$ which is small relative to $m_\text{\tiny N}$. Hence, modes with different $v_\nu$ decouple.
\item\label{enum_L-kin-2}The decoupling makes a velocity-dependent field transformation $N_v=e^{i m_\text{\tiny N}\slashed{v}v_\nu x^\nu}n_v$ convenient. The fields are split into particle
annihilation and anti-particle creation operators $N_v=N_v^++N_v^-$. The momentum transfer is assumed to be small relative to the rest mass, not providing enough energy for
the creation of particle-anti-particle pairs: $N_v^{+\dag}\slashed{\partial}N_v^-=N^{-\dag}_v\slashed{\partial}N_v^+=0$.
\item\label{enum_L-kin-3}For small momenta $q_\nu$, the Lagrangean must be invariant under a field transformation resembling a small boost, $N_v\rightarrow e^{i q_\nu x^\nu} N_v=N_{v+q/m_\text{\tiny N}}$ (only here, $N$ is a scalar field). The effective Lagrangean up to $\mathcal{O}\left(\frac{1}{m_\text{\tiny N}}\right)$ is given.
\item\label{enum_L-kin-4}The final form of the kinetic term of the effective Lagrange density, used here, is obtained in the rest frame of the nucleon after field transformations
involving $i v_\nu\partial^\nu$ and $\partial^2$. The operator $N=\left(\begin{array}{c}p\\n\end{array}\right)$ holds the neutrons and protons chosen each as a two component
Pauli spinor with mass dimension $[N]=\frac{3}{2}$. The corrections from the mass difference are assumed to be less than order $m_\text{\tiny N}^{-2}$.
\end{enumerate}
\par
The internal symmetry transformations act on the two components of the isovector $N=\left(\begin{array}{c}p\\n\end{array}\right)$. In this work, the only interaction
which breaks the isospin symmetry - \textit{i.e.}, distinguishes between two neutrons and two protons - is the Coulomb term.
\par\vskip10pt
\textbf{\large Power Counting.\hspace{.5cm}}
The EFT formalism orders contributions to an observable in powers of the typical external momentum $Q$. Initially, the ordering scheme is based on the assumption that all
coupling constants are of natural size, \textit{i.e.}:
\begin{equation}\label{eq_naturalness}
g_i\left(\Lambda\right)=\frac{c_i\left(\frac{\Lambda}{M}\right)}{M^{\delta_i-4}}\;\;\;\text{with dimensionless}\;\;\;c_i(1)\;\;\;\text{of order}\;\;\;1\;\;\;.
\end{equation}
This implies an ordering scheme with a truncated expansion $\mathcal{L}^{(\delta\leq\delta_\text{max})}$ (vertex expansion), and also a limited number of iterations
(loop expansion) of the interaction terms contained in $\mathcal{L}$. Operators of higher mass dimension or higher iterations of terms with $\delta<\delta_\text{max}$ would be of
order $\left(\frac{Q}{M}\right)^{n(\delta_\text{max})}$ and therefore suppressed. For $\delta_\text{max}=8$ the effective Lagrange density is:
{\small\begin{eqnarray}\label{eq_nlo-lagrangean}
    \mathcal{L}^{(8)}&=&N^\dagger\left(i\partial_0+\frac{\vec{\nabla}^2}{2m_\text{\tiny N}}\right)N+C^{\text{LO}}_1(N^\dagger N)(N^\dagger N)+
    C^{\text{LO}}_2(N^\dagger\sigma_i N)(N^\dagger\sigma_i N)\nonumber\\
    &&- \frac{1}{2} C^{\text{NLO}}_1  \left[  ( N^\dagger  \partial_i N )^2  +
      \left(( \partial_i N^\dagger ) N \right)^2  \right]\nonumber\\
    &&   -  \left( C^{\text{NLO}}_1 - \frac{1}{4} C^{\text{NLO}}_2 \right) ( N^\dagger \partial_i N ) \;\Big[( \partial_i N^\dagger ) N \Big]+
    \frac{1}{8} C^{\text{NLO}}_2  ( N^\dagger N ) \left[ N^\dagger\partial_i^2 N + \partial_i^2 N^\dagger N \right]\nonumber\\
    &&  - \frac{i}{8} C^{\text{NLO}}_5  \epsilon_{ijk} \bigg\{\left[ ( N^\dagger \partial_i N ) \big[( \partial_j N^\dagger ) \sigma_k N \big] +
      \big[ ( \partial_i N^\dagger ) N \big]  ( N^\dagger\sigma_j\partial_k N ) \right]\nonumber\\
    &&-( N^\dagger N )\Big[ ( \partial_i N^\dagger ) \sigma_j\partial_k N \Big]  +   ( N^\dagger \sigma_i N )\Big[ ( \partial_j N^\dagger )\partial_k N \Big] \bigg\}+ \frac{1}{4} \Big[ \left( C^{\text{NLO}
}_6 + \frac{1}{4} C^{\text{NLO}}_7  \right)    \left( \delta_{i k}\delta_{j l}+ \delta_{i l} \delta_{k j} \right)\nonumber\\
    && +\left( 2 C^{\text{NLO}}_3 + \frac{1}{2} C^{\text{NLO}}_4 \right) \delta_{i j} \delta_{k l} \Big]\,\Big[  \big[ (\partial_i \partial_j N^\dagger)\sigma_k  N \big]+
    ( N^\dagger \sigma_k \partial_i \partial_j N ) \Big] ( N^\dagger\sigma_l N )\nonumber\\
    &&  -  \frac{1}{2} \Big[ C^{\text{NLO}}_{6} \left( \delta_{i k} \delta_{jl}+ \delta_{i l} \delta_{k j} \right) +
    C^{\text{NLO}}_4 \delta_{i j} \delta_{k l} \Big]( N^\dagger \sigma_k \partial_i N ) \big[( \partial_j N^\dagger ) \sigma_l N \big]\nonumber\\
    &&- \frac{1}{8} \left( \frac{1}{2}  C^{\text{NLO}}_7\left( \delta_{i k} \delta_{j l}+ \delta_{i l} \delta_{k j}\right) -
      ( 4 C^{\text{NLO}}_3 - 3 C^{\text{NLO}}_4 ) \delta_{i j} \delta_{k l} \right)\big[  ( \partial_i N^\dagger \sigma_k\partial_j N )\nonumber\\
&& +      ( \partial_j N^\dagger\sigma_k \partial_i N ) \big] ( N^\dagger \sigma_l N )\;\;\;.
\end{eqnarray}}
The vector $\vec{\sigma}$ consists of the Pauli spin matrices, lower case Arabic letters label vector components in a Cartesian basis, and the sum convention is understood.
In leading order (LO), only operators with $\delta\leq 6$ - namely $C_{1,2}^\text{LO}$ - are included, while next-to-leading order (NLO) means also to consider $\delta = 8$.
To retain the $S$-matrix for low-energy two-nucleon reactions, the assumption that all LECs $C_i^{\text{(N)LO}}$ are of natural size has to be given up,
because the resulting perturbative treatment of all interactions cannot reproduce poles of the amplitude at momenta $Q\ll M$. The poles, if there are any,
of an amplitude derived from a theory with natural coupling constants only,
are all at momenta $Q\gtrsim M$. The peculiar analytic structure of the amplitude is resembled in the relative size of the effective range$^\text{\cite{ere-bethe}}$ parameters in both
the singlet- and triplet-neutron-proton (total spin $s=0,1$) channel (table~\ref{tab_np-para}). The scattering lengths $a_{s,t}$
are considerably larger than the effective ranges $r_{s,t}$,
which are $\mathcal{O}(1/M)$. Hence, $a_{s}$ and $a_t$ constitute additional low-momentum scales, associated with a virtual and a real bound state close to
the scattering threshold.
\begin{equation}\label{eq_np-low-scale}
\aleph\equiv\frac{1}{\left|a_{s,t}\right|}\ll M\sim\mathcal{O}(m_\pi)\;\;\;.
\end{equation}
\begin{table*}
\renewcommand{\arraystretch}{1.5}
  \caption{\label{tab_np-para}{\small Experimental$^\text{\cite{exp-anp}}$ neutron-proton effective range parameter for the spin singlet (s) and triplet (t) S-wave channel.
The conversion factor is $\hbar c=1=197.316132894~$MeV$\cdot$fm.}}
\begin{tabular}{l|cccc}
\hline
 &$a_s$&$a_t$&$r_s$&$r_t$\\
\hline\hline
 $[$fm$]$&$-23.748(10)$&$5.424(4)$&$2.75(5)$&$1.759(5)$\\
$[$MeV$^{-1}]$&$-0.12036$&$0.02749$&$0.0139$&$0.008915$\\
\hline
    \end{tabular}
\end{table*}
With the additional scale $\aleph$, a coupling constant of size $g_{1,2}\left(\Lambda\right)=\frac{c_{1,2}\left(\Lambda/M\right)}{M\aleph}$ with
$c_{1,2}\left(1\right)\sim\mathcal{O}(1)$ does no longer justify a truncation of the loop expansion. This assumption is made for $C_{1,2}^\text{LO}$.
Because of this unnatural size, an iteration of such an interaction does not change the order of the respective contribution.
Diagrammatically, this is shown in fig.~\ref{fig_eft-exp}, where all summands in the first row are of the same order with only momentum-independent vertices, while
insertions of vertices with natural sizes decrease the size of the contribution to the amplitude.
\newpage
The leading-order amplitude can be calculated analytically$^\text{\cite{eft-rev-phill}}$ in momentum space:
\begin{eqnarray}\label{eq_bubble-sum-solution}
T(p,p',E)&=&-i C+(-i C)^2\int\frac{d^3\vec{l}}{(2\pi)^3}\frac{dl_0}{2\pi}\frac{i}{l_0+E-\frac{\vec{l}^2}{2m_\text{\tiny N}}+i\epsilon}\cdot
\frac{i}{-l_0-\frac{\vec{l}^2}{2m_\text{\tiny N}}+i\epsilon}+\;\;\mathcal{O}\left(C^3\right)\nonumber\\
&=&-i C+(-i C)^2\int\frac{d^3\vec{l}}{(2\pi)^3}\frac{1}{2\pi}(2\pi i)\frac{-1}{E-\frac{\vec{l}^2}{m_\text{\tiny N}}+i\epsilon}+\;\;\mathcal{O}\left(C^3\right)\nonumber\\
&=&-i C+(-i C)^2\underbrace{\int\frac{d^3\vec{l}}{(2\pi)^3}\frac{-i}{E-\frac{\vec{l}^2}{m_\text{\tiny N}}+i\epsilon}}_{\equiv -i I_\Lambda=
(-2\pi)^{-3}\int_0^\infty dl\frac{m_\text{\tiny N}l^2}{m_\text{\tiny N}E-l^2+i\epsilon}}+\;\;\mathcal{O}\left(C^3\right)\nonumber\\
&=&C\sum_{n=0}^\infty (-i)^{2n}C^nI_\Lambda^n=\frac{1}{-\frac{1}{C}-I_\Lambda}\;\;\;.
\end{eqnarray}
The constant $C$ is a placeholder for $C_{1,2}^\text{LO}$, the propagator is derived from the kinetic part of $\mathcal{L}$, the total kinetic energy in the center of mass system is $E=\frac{\vec{p}^2}{m_\text{\tiny N}}$, and the subscript $\Lambda$ specifies a regularization scheme for the divergent integral.
\par
\input{diss_fig_bubble.tex}
\newpage
\par\vskip10pt
\textbf{\large Parameter Determination.\hspace{.5cm}}
Changes in the high-energy/short-distance structure of the theory, performed either by integrating out DoF or by the regularization, manifest themself in the values of the LECs.
At the moment, a direct calculation of the LECs from the underlying theory is not feasible. Instead, low-energy data related to coefficients of a power expansion of the amplitude
is used to determine the LECs. The expansion sketched in fig.~\ref{fig_eft-exp} is identical to the effective range expansion$^\text{\cite{ere-bethe}}$.
The latter approximates the scattering amplitude (parameterized by a \mbox{phase shift $\delta_l$}), for a relative angular momentum $l$ and center of mass momentum
$p=\sqrt{m_\text{\tiny N}E}$:
\begin{eqnarray}\label{eq_ere}
p^{2l+1}\cot\delta_l=-\frac{1}{a_l}+\frac{r_l}{2}p^2+\sum\limits^{\infty}_{n=2} v_l^{(n)}p^{2n}\nonumber\\
l=0\;(\text{S-wave}):T(p)=-\frac{4\pi}{m_\text{\tiny N}}\frac{1}{p\cot\delta_0-ip}\;\;\;.
\end{eqnarray}
This expansion is around zero momentum and introduces the coefficients $a_l$ (scattering length), $r_l$ (effective range), and $v_l$.
Before relating this expression to the amplitude from eq.~(\ref{eq_bubble-sum-solution}), a regularization scheme has to be specified. With the momentum dependence of
$I_\Lambda$ thus fixed, the result in eq.~(\ref{eq_bubble-sum-solution}) is expanded in powers of $p$ to obtain a form analogous to eq.~(\ref{eq_ere}).
Equation~\ref{eq_bubble-sum-solution} is of Lippman-Schwinger-type with a potential $V(\vec{p},\vec{p'})=-i C$. With a regulator function
\begin{equation}\label{eq_regulator-function}
f(\vec{p},\Lambda)\;\;\text{such that}\;\;{\Bigg\lbrace}\begin{array}{l}
\lim\limits_{\Lambda\to\infty}f(\vec{p},\Lambda)=1\\f(\vec{p},\Lambda)\approx 1\;\;\text{for}\;\;|\vec{p}|\;\;\text{of order}\;\;Q\;\;\text{and}\;\;\Lambda\gtrsim\mathcal{O}(M)
\end{array}\;\;\;,
\end{equation}
and using the separable form
\begin{equation}\label{eq_reg-pot}
V_r(\vec{p},\vec{p'},\Lambda)=f(\vec{p},\Lambda)(-iC)f(\vec{p'},\Lambda)
\end{equation}
in eq.~(\ref{eq_bubble-sum-solution}), yields:
\begin{equation}\label{eq_t-mat-regulated}
T(p,p',E)=\frac{f(\vec{p},\Lambda)f(\vec{p'},\Lambda)}{-\frac{1}{C}-I_\Lambda}\;\;\;\;\text{with}\;\;\;\;\;
I_\Lambda=4\pi\int_o^\infty\frac{dl}{(2\pi)^3}\frac{l^2f^2(\vec{l},\Lambda)}{E-\frac{l^2}{m_\text{\tiny N}}+i\epsilon}\;\;\;.
\end{equation}
To have a Gau\ss ian regulator in coordinate space, we chose a different nonseparable regulator which will be explained below (eq.~\ref{eq_reg-fkt-mom}).
A power expansion in $|\vec{p}|$ of the integral $I_\Lambda$ can now be related to the effective range amplitude. In leading order, EFT($\slashed{\pi}$) reproduces the
experimental scattering lengths (singlet: $S=0$, triplet: $S=1$) for
\begin{equation}\label{eq_lo-lec}
C_1^\text{LO}+\left(4S-3\right)C_2^\text{LO}=\frac{4\pi^2a_{s,t}\sqrt{2}}{m_\text{\tiny N}\pi\sqrt{2}-m_\text{\tiny N}\Lambda\sqrt{\pi}a_{s,t}}\;\;\;.
\end{equation}
Having determined the LECs $C^\text{LO}_{1,2}$, the corresponding potential can be used in the Schr\"odinger equation - equivalent to the Lippman Schwinger equation - to make predictions.
At this point, I summarize the prescription which starts from a Lagrange density $\mathcal{L}$ and results in a potential that is used in a multi-body Schr\"odinger
equation.
\begin{enumerate}
\item The first part of the power-counting scheme is a truncation of the vertex expansion of $\mathcal{L}$ based on dimensional analysis and natural coupling constants. I
will consider two orders in this work:
\begin{itemize}
\item leading order (LO): $\mathcal{L}^{(\delta_\text{max}=6)}$
\item next-to-leading order (NLO): $\mathcal{L}^{(\delta_\text{max}=8)}$
\end{itemize}
\item At each order, the vertex structure of $\mathcal{L}$ is identified with a potential, which is given in momentum space:
\begin{eqnarray}
\text{LO}:&V^\text{LO}_\text{NN}(\vec{q},\vec{k})=&C^\text{LO}_1+C^\text{LO}_2\vec{\sigma}_1\cdot\vec{\sigma}_2\;\;\;;\label{eq_lo-pot-mom-space}\\
\text{NLO}:&V^\text{NLO}(\vec{q},\vec{k})=&V^\text{LO}_\text{NN}+C^\text{NLO}_1\vec{q}^2+C^\text{NLO}_2\vec{k}^2+
\vec{\sigma}_1\cdot\vec{\sigma}_2\left(C^\text{NLO}_3\vec{q}^2+C^\text{NLO}_4\vec{k}^2\right)\label{eq_nlo-pot-mom-space}\\
&&+iC^\text{NLO}_5\frac{\vec{\sigma}_1+\vec{\sigma}_2}{2}\cdot\vec{q}\times\vec{k}+C^\text{NLO}_6\vec{q}\cdot\vec{\sigma}_1\vec{q}\cdot\vec{\sigma}_2
+C^\text{NLO}_7\vec{k}\cdot\vec{\sigma}_1\vec{k}\cdot\vec{\sigma}_2\;\;\;;\nonumber
\end{eqnarray}
Where $\vec{q}=\vec{p}-\vec{p'}$ and $\vec{k}=\frac{\vec{p}+\vec{p'}}{2}$ is used.
\item The second part of the power-counting scheme is an estimate of the relative size of terms in the loop expansion. This estimate determines whether or not
an operator in the potential can be treated as a perturbation.
\item The theory's renormalization scheme is defined by a regulator function and the low-energy data used to fit the LECs.
\item Predictions of the resulting interaction are expected to be consistent with data within the uncertainty margins:
\begin{eqnarray}\label{eq_exp-para-est}
\Delta(\text{LO})&\sim&\mathcal{O}\left(\frac{Q}{M}\right)\approx\frac{\gamma_t}{m_\pi}\approx\frac{1}{3}\;\;\;,\\
\Delta(\text{NLO})&\sim&\mathcal{O}\left(\frac{Q^2}{M^2}\right)\approx\frac{1}{9}\;\;\;,\nonumber
\end{eqnarray}
where a typical momentum of a nucleon within the deuteron of $\gamma_t=\sqrt{m_\text{\tiny N}B(d)}$ is assumed.
Larger deviations resemble either an insufficient power counting, \textit{e.g.}, LEC(s) of unnatural size, or a failure of the EFT because the typical
low-energy scale of the system is larger than expected.
\end{enumerate}
\par\vskip10pt
\textbf{\large Renormalization Scheme Dependence.\hspace{.5cm}}
I will use two related criteria to validate the power-counting scheme for a specific observable:
\begin{itemize}
\item order-by-order convergence of the prediction consistent with an expansion parameter of $\mathcal{O}\left(\frac{1}{3}\right)$;
\item renormalization scheme dependence of $\lesssim\mathcal{O}\left(\frac{1}{3}\right)$ at LO and $\lesssim\mathcal{O}\left(\frac{1}{9}\right)$ at NLO, \textit{i.e.},
an maximal deviation of approximately $30$\% ($10$\%) from the experimental value at LO (NLO).
\end{itemize}
In addition to the regulator value $\Lambda$ and a subset $\mathcal{D}'\subset\mathcal{D}$ of all low-energy ``$\mathcal{D}$(ata)'',
I define the cutoff parameter $\Lambda_\text{RGM}$ (see ch.~\ref{sec_rrgm}) as
a third parameter to completely specify the renormalization scheme. Renormalization, in general, constitutes a modification of unobservable modes, \textit{i.e.}, with
momenta outside the range of validity of the theory, while imposing low-energy constraints - renormalization conditions. The lower bound for $\Lambda$ was chosen of the order of
the breakdown scale, $\Lambda_\text{min}=150~$MeV, in order not to affect the interaction for center-of-mass momenta less than $100~$MeV.
$\Lambda_\text{RGM}$ parameterizes a change of high-energy modes
by means of adding or excluding Gau{\ss}ian basis functions with support for $|\vec{r}|\ll\frac{1}{m_\pi}$. For given $\Lambda$, $\Lambda_\text{RGM}$, and $\mathcal{D}'$, a correct
power counting predicts $o\in\mathcal{D}\setminus\mathcal{D}'$ with an uncertainty depending on the order of the calculation. A different renormalization scheme will affect those
predictions only within this uncertainty range. A larger deviation, \textit{i.e.}, larger scheme dependence, indicates an insufficient power-counting scheme. A prominent example
is the promotion of a three-body interaction to leading order in the triton channel ($s=\frac{1}{2}$, with the S-wave neutron-deuteron scattering length denoted by $a_\text{(nd)t}$).
Considering $\mathcal{D}=\lbrace a_s,a_t,B(t),a_\text{(nd)t}\rbrace$ and $\mathcal{D}'=\lbrace a_s,a_t\rbrace$ to fit $C_{1,2}^\text{LO}$, the predictions for $\mathcal{D}\setminus\mathcal{D}'$ will show
a strong dependence on $\Lambda$, known as the Phillips line$^\text{\cite{phillips-line}}$. Promoting the $\delta=9$ operator,
\begin{equation}\label{eq_tni-lagrange}
\mathcal{L}^{\delta=9}=C^{\text{LO}}_{3\text{NI}}(N^\dagger N)(N^\dagger\tau_i N)(N^\dagger\tau_i N)\;\;\;,
\end{equation}
to leading order, and using $\mathcal{D}'=\lbrace a_s,a_t,B(t)\rbrace$ yields predictions for $a_{(nd)t}$ consistent with the LO uncertainty estimate, while not affecting $A=2$
observables$^\text{\cite{tni-gang-1,tni-gang-2,tni-gang-3}}$. Accordingly, I extend the leading-order potential in eq.~(\ref{eq_lo-pot-mom-space})
for calculations in $A\geq 3$ systems by the operator:
\begin{equation}\label{eq_tni-mom-space}
V_{3\text{NI}}^\text{LO}(\vec{p}_{12},\vec{p}_{23},\vec{p'}_{12},\vec{p'}_{23})=C^{\text{LO}}_{3\text{NI}}\vec{\tau}_1\cdot\vec{\tau}_2\;\;\;.
\end{equation}
\par\vskip10pt
\textbf{\large Universality.\hspace{.5cm}}
The unnatural scaling of $C^{\text{LO}}_{3\text{NI}}$ is a consequence of a finite quantity characterizing the three-body system in the limit of a scale-free two-body subsystem in which,
\textit{e.g.}, diverging two-fermion scattering lengths, $a_{s,t}\to\infty$. This property was discovered$^\text{\cite{efimov-effect}}$ as the universal ratio of binding energies of
successive states in the spectrum of three identical bosons. In the limit $|a|\to\infty$, this spectrum consists of an infinite number of bound states with energies
$B^{(n)}(3)$ which accumulate at the lowest break-up threshold. These binding energies approach a universal ratio given by:
\begin{equation}\label{eq_efimov-ratio}
\lim_{n\to\infty}\frac{B^{(n+1)}(3)}{B^{(n)}(3)}=\frac{1}{515.03}\;\;\;\text{for}\;\;|a|\to\infty\;\;\;.
\end{equation}
Here, $n=1$ labels the largest binding energy, and \textit{universal} refers to the independence of that ratio with regards to the specific form of the two-body interaction that
generates the diverging scattering length. For instance, the LECs in eq.~(\ref{eq_lo-pot-mom-space}) can be tuned such that $|a_{s,t}|\to\infty$ for different values of $\Lambda$,
\textit{i.e.}, different shapes of the corresponding potentials. However, the ratio given in eq.~(\ref{eq_efimov-ratio}) will be independent of $\Lambda$, it is universal.
In contrast, the absolute position of the spectrum, \textit{e.g.}, fixed by
the smallest eigenvalue in the three-body spectrum, is no universal consequence of the infinite scattering length - it differs in general for different $\Lambda$.
If, however, this eigenvalue is fixed, so is
the complete spectrum, which is why in addition to $a_{s,t}$ only one additional parameter is needed to select the system that corresponds to the triton from the universality
class, associated with $a_{s,t}$. In practice, two approximations result in a finite three-body spectrum:
First, the scattering lengths are finite, yet large relative to a range set by the breakdown scale of the EFT, and second, the spectrum is calculated in a finite basis.
\par
With the extended set $\mathcal{D}'=\lbrace a_s,a_t,B(t)\rbrace$, the predictions of the EFT for $\mathcal{D}\setminus\mathcal{D}'$ are expected to by accurate up to corrections of the
order of about $30~$\%.
Strong evidence was found recently$^\text{\cite{platter-4bdy-lo}}$ that this assertion holds even if $\mathcal{D}$ includes low-energy four-body data. In contrast to the three-nucleon system,
the four-nucleon system is then at this order a universal consequence of its constituent subsystems without universal scales. An analogous promotion of a four-body counter term to
LO is therefore not necessary.
\newpage
\section{Nuclear Potentials from EFT}\label{sec_pots}
I will derive three classes of nucleon-nucleon (NN) potentials. The first class consists of the leading-order interactions, while the second and third class are both of next-to-leading
order accuracy but include different subsets of higher-order contributions. With the potentials, $A$-body observables are predicted from solutions of the Schr\"odinger equation
eq.~(\ref{eq_a-body-sgl}).
The eigenstate $\vert\phi^{(i)}\rangle$ is obtained with a variational method approximating the wave function in coordinate space. This method is based on the
idea of resonantly rearranging groups of particles within the $A$-body system and is reviewed together with the extensions for its application to the EFT($\slashed{\pi}$)
potentials in ch.~\ref{sec_rrgm}. The necessary Fourier transform of the NLO operators in eq.~(\ref{eq_nlo-pot-mom-space}) is given in app.~\ref{app_FT}, but the form of the
regulator shall be discussed here with the LO potential as an example. To minimize the numerical effort, I choose a momentum space regulator that transforms into a
Gau{\ss}ian function on the relative coordinate between the interacting particles. The conjugate momentum to this coordinate $\vec{r}$ ($\vec{r'}$) before (after) the interaction is
$\vec{p}$ ($\vec{p'}$), fig.~\ref{fig_eft-exp}. Choosing the same regulator for all vertices,
\begin{equation}\label{eq_reg-fkt-mom}
f(\vec{p},\vec{p'},\Lambda)=e^{-\left(\vec{p}-\vec{p'}\right)^2/\Lambda^2}\;\;\;\rightarrow\;\;\;V(\vec{p},\vec{p'},\Lambda)\equiv f(\vec{p},\vec{p'},\Lambda)V^\text{(N)LO}(\vec{q},\vec{k})\;\;\;,
\end{equation}
yields the convenient leading-order form:
\begin{eqnarray}\label{eq_pot-lo-coord}
\langle\vec{r}'\vert\hat{V}^\text{LO}\vert\vec{r}\rangle&=&\frac{1}{(2\pi)^6}\int d^3p'd^3p\;\langle r'\vert p'\rangle\langle p'\vert \hat{V}\vert p\rangle\langle p\vert r\rangle=
\frac{1}{(2\pi)^6}\int d^3p'd^3p\;e^{i\vec{r}'\cdot\vec{p}'}V(\vec{p}',\vec{p})e^{-i\vec{r}\cdot\vec{p}}\nonumber\\
&=&\frac{1}{(2\pi)^6}\int d^3kd^3q\;e^{i\vec{r}'\cdot(\vec{k}+\frac{\vec{q}}{2})}V(\vec{k},\vec{q})e^{-i\vec{r}\cdot(\vec{k}-\frac{\vec{q}}{2})}\nonumber\\
&=&
\frac{1}{(2\pi)^6}\int d^3kd^3q\;e^{i\vec{k}\cdot(\vec{r}'-\vec{r})}V(\vec{k},\vec{q})e^{i\vec{q}\cdot(\frac{\vec{r}+\vec{r}'}{2})}\nonumber\\
&=&\left(C^\text{LO}_1+C^\text{LO}_2\,\vec{\sigma}_1\cdot\vec{\sigma}_2\right)
\underbrace{\frac{\Lambda^3}{8\pi^{\frac{3}{2}}}e^{-\frac{\Lambda^2}{4}(\frac{\vec{r}+\vec{r}'}{2})^2}}_{\stackrel{\Lambda\to\infty}{\rightarrow}\delta\left((\vec{r}'+\vec{r})/2\right)}
\delta(\vec{r}'-\vec{r})\;\;\;.
\end{eqnarray}
The momentum-space regulator is hence not separable in $\vec{p}$ and $\vec{p'}$, a property used to derive the solution of eq.~(\ref{eq_t-mat-regulated}).
Consequently, this solution provides only a useful starting value for the numerical fit of the LO LECs, as will be shown below.
\par\vskip10pt
\newpage
\textbf{\large Leading-Order Potential.\hspace{.5cm}}
At leading order, the theory is based, in the two-nucleon sector, on $\mathcal{L}^{(6)}$. Iterations of the two vertices are all of the same order due to the additional
unnatural scales of order $Q$ in both NN S-wave channels. This power counting is therefore correctly implemented by solving the Schr\"odinger equation with a potential
\begin{equation}\label{eq_pot-coord-lo}
\hat{V}^\text{LO}_\text{NN}=
\sum\limits_{i<j}^A\left(\frac{\Lambda^3}{8\pi^{\frac{3}{2}}}e^{-\frac{\Lambda^2}{4}\vec{r}^2}\left(C^\text{LO}_1+C^\text{LO}_2\vec{\sigma}_i\cdot\vec{\sigma}_j\right)\right)\;\;\;.
\end{equation}
The two independent LECs, $C^\text{LO}_{1,2}$, are fitted to the scattering length $a_s$ in the spin-0 S-wave neutron-proton (np) channel and to either
$a_{t\text{\tiny (riplet)}}$ or the deuteron
binding energy $B(d)$ in the spin-1 S-wave np channel. $B(d)$ is identified with a pole of the amplitude, which is in LO of the effective range expansion:
\begin{equation}\label{eq_lo-bd-at}
\renewcommand{\arraystretch}{1.8}
\begin{array}{c}
-\frac{1}{a_t}-i\gamma_t=0\;\;\;\rightarrow\;\;\;B(d)=\frac{\gamma_t^2}{m_\text{\tiny N}}=-\frac{1}{m_\text{\tiny N}a_t^2}\\
\rightarrow\;\;\;B(d)\big|_{a_t=a_{t,\text{\tiny exp}}}\approx 1.41~\text{MeV}\;\;\;\text{or}\;\;\;a_t\big|_{B(d)=B(d,\text{exp})}\approx 0.0219~\text{MeV}^{-1}\;\;\;.
\end{array}
\end{equation}
Using the experimental deuteron binding energy $B(d,\text{exp})$ as input yields a scattering length that deviates from data within the LO uncertainty range. $B(d)$ is
consequently identified with a low-energy observable that can also be used to determine LO LECs. Table~\ref{tab_lo-pots} lists the leading-order potentials derived for this work,
where $B(t)$ and $B(\alpha)$ were calculated with $C_{3\text{NI}}^\text{LO}=0$. For $\Lambda\leq 800~$MeV, the dependence of the binding energies on the RGM model space was
found small: $\frac{\vert B(x,w12)-B(x,w120)\vert}{B(x,exp)}\leq 0.01$
(models spaces w12(0) are defined in table~\ref{tab_deuteron-ms} \& \ref{tab_t-ms} \& \ref{tab_he4-ms}),
while for $\Lambda=1600~$MeV only $w120$ is to be used because of the shorter range of this interaction.
\begin{table*}
\renewcommand{\arraystretch}{1.5}
  \caption{\label{tab_lo-pots}{\small Parameters of leading-order potentials of the form eq.~(\ref{eq_pot-lo-coord}). If unspecified, quantities are given in MeV.}}
\footnotesize
\begin{tabular}{cl|lllcc|c|ccc}
\hline
$\Lambda$ [MeV]&$\frac{\Lambda^2}{4}$ [fm$^{-2}$]&$8^{-1}\pi^{-\frac{3}{2}}\Lambda^3C^\text{LO}_1$&
$8^{-1}\pi^{-\frac{3}{2}}\Lambda^3C^\text{LO}_2$&$|C_S|$&$|C_T|$&$\frac{\Lambda^6}{2^6\pi^3}C^\text{LO}_{3\text{NI}}$&$\mathcal{D}'$&$B(d)$&$B(t)$&$B(\alpha)$\\
\hline\hline
$200$&$0.25682$&$-41.549$&$-5.1372$&$0.31$&$2.6$&$0.56$&$a_{s,t}$    &$2.5315$&$9.1529$&$31.975$\\
            &$0.25685$&$-40.592$&$-4.8183$&$0.38$&$2.5$&$0.028$&$a_s,B(d)$&$2.2245$&$8.5156$&$30.371$\\
$300$&$0.57784$&$-80.727$&$-6.7272$&$0.55$&$1.4$&$5.83$&$a_{s,t}$     &$2.0587$&$13.047$&$53.308$\\
            &$0.57791$&$-81.480$&$-6.9786$&$0.54$&$1.5$&$6.47$&$a_s,B(d)$&$2.2246$&$13.476$&$54.459$\\
$800$&$4.1091$&$-493.38$&$-15.467$&$0.34$&$0.44$&$190.0$&$a_{s,t}$    &$1.6168$&$54.285$&$262.59$\\
            &$4.1096$&$-500.34$&$-17.789$&$0.33$&$0.45$&$221.5$&$a_s,B(d)$&$2.2246$&$57.539$&$271.47$\\
$1600$&$16.436$&$-1897.8$&$-29.731$&$0.18$&$0.21$&$2792.0$&$a_{s,t}$    &$1.5145$&$187.288$&$374.508$\\
              &$16.438$&$-1913.8$&$-35.091$&$0.18$&$0.21$&$4551.0$&$a_s,B(d)$&$2.2246$&$194.24$  &$382.603$\\
\hline
    \end{tabular}
\end{table*}
In practice, the scattering lengths are calculated from the phase shift $\delta$ in the respective channel at an energy $E_0=0.4~$keV via
$a_{s,t}=\frac{1}{\sqrt{m_\text{\tiny N}E_0}}\tan\delta_{s,t}$.
In the two-nucleon sector, the phase shifts were calculated by numerically integrating the Schr\"odinger equation using \texttt{Mathematica} and alternatively with the
Resonating Group Method (RGM). The former uses a
steepest descent optimization method to fit the LEC, while the latter employs a genetic algorithm (app.~\ref{app_gen-alg}). The value predicted by eq.~(\ref{eq_lo-lec})
for the separable regulator is used either as a starting value (steepest descent) or center of the search interval (genetic algorithm) in the two approaches. The LECs obtained
with both methods differ by $\ll 1\%$. This difference transients to a numerical uncertainty for predictions of observables which is insignificant relative to the expected
systematic error at leading order.
\par
The leading-order LECs we found are of a size consistent with a basic EFT assumption (discussion of eq.~(\ref{eq_np-low-scale})). Specifically,
one$^\text{\cite{eft-rev-vK}}$ estimates,
\begin{equation}\label{eq_lo-coeff-nat-1}
C_1^\text{LO}+(4S-3)C_2^\text{LO}\approx\frac{4\pi}{m_\text{\tiny N}\aleph}\underbrace{\gamma}_{\sim\mathcal{O}(1)}\approx 10^{-3}~\text{MeV}^{-2}\;\;\;,
\end{equation}
where I set $\aleph=\frac{|a_s|+|a_t|}{2}$ and the exact value of $\gamma$ depends on the chosen renormalization scheme, \textit{e.g.}, parameterized by $\Lambda$.
The value of eq.~\ref{eq_lo-coeff-nat-1} is not exceeded by the numbers for $C_{S(T)}=10^4\cdot 8\pi^\frac{3}{2}\Lambda^{-3}(C_1^\text{LO}+(4S-3)C_2^\text{LO})$
as given in table~\ref{tab_lo-pots} which are consitently smaller than $10^{-4}~\text{MeV}^{-2}$.
\par
The two-body interaction thus determined, without proper renormalization through a three-body vertex, predicts a triton binding energy $B(t)$ that increases with $\Lambda$, \textit{i.e.}, with decreasing range of the potential at constant $B(d)$. This behavior is known as the Thomas effect$^\text{\cite{thomas-effect}}$. This divergence disappears naturally
with an additional renormalization constraint at LO on a three-body observable.
I choose a three-body interaction of the form in eq.~(\ref{eq_tni-lagrange}), regularized analogous
to the two-nucleon interaction to yield an operator in coordinate representation of the form:
\begin{equation}\label{eq_tni-pot-coord}
\hat{V}^\text{LO}_\text{3NI}=\sum\limits_{\stackrel{i<j<k}{\text{cyclic}}}^A
\left(\frac{\Lambda^3}{8\pi^{\frac{3}{2}}}e^{-\frac{\Lambda^2}{4}\vec{r}_{ij}^2}\right)\cdot
\left(\frac{\Lambda^3}{8\pi^{\frac{3}{2}}}e^{-\frac{\Lambda^2}{4}\vec{r}_{jk}^2}\right)
C^{\text{LO}}_{3\text{NI}}\,\vec{\tau}_i\cdot\vec{\tau}_j\;\;\;.
\end{equation}
In addition to the the physical constraint put on the minimal value for $\Lambda$, practical calculations were facilitated by also imposing an upper bound.
The increasing $B(t)$ requires a correspondingly stronger 3NI to fit data and poses an inconvenience regarding the numerical technique (see ch.~\ref{sec_rrgm} for terminology).
Specifically, the variational basis has to expand two wave functions which differ widely in their spacial extension, with the more deeply bound system naturally confined to a less
extended region in space (see discussion of the triton charge radius in ch.~\ref{subsec_r3-ch}). Hence, the basis has to expand a wave function with small support - corresponding
to the LO state without 3NI - relative to the support of the renormalized state. By using cutoff values in the range $300~\text{MeV}\lesssim\Lambda\lesssim800~\text{MeV}$
for which $B(t)$ at LO without 3NI does not exceed 60~MeV, a single triton basis could be used for all cutoffs. This interval does not present a principal limitation on the
regulator imposed by the RGM. For $\Lambda$s exceeding the boundaries, the basis must be augmented with sufficiently narrow width parameters to account for
the structure of the wave function at distances $r\sim\Lambda^{-1}$. Albeit constraining the cutoff to this interval, the reduction of the basis size for a triton substructure within
$A>3$ trial wave functions has to be performed for each $\Lambda$ separately (see discussion of the $^4$He core as a subsystem of $^6$He in ch.~\ref{sec_a6}
for an analogous case).
\par
An elegant way of dealing with larger $\Lambda$s is provided by the fact that the number of bound states increases with the cutoff. This opens the possibility to
fit not the deep ground state to the triton but rather an excited state - which is already closer to the datum -. The advantage is a relatively small contribution from the 3NI at
the expense of an assessment of the convergence of the excited state within the RGM model space. Although this convergence is guaranteed (see Mini-Max theorem, \textit{e.g.},
\cite{varga}), in practice, a larger basis are necessary. For this reason I do not draw this option and choose $\Lambda$ from the above defined interval, where for all practical
bases only one bound three-nucleon state exists. In ch.~\ref{sec_a6}, this identification of a physical nucleon with an excited state in the spectrum of EFT($\slashed{\pi}$) is discussed further. There, in the six-nucleon system, we will find multiple bound states. If any of them corresponds to the physical 6-helium ground state or if the
the theory does not sustain a shallow bound state at all at the respective order will be a central problem in that analysis.
Here, a strong cutoff dependence observed in NLO corrections for those deep states is suggested as only one discriminating factor between states in- and
outside the range of validity of EFT($\slashed{\pi}$).
\par\vskip10pt
\textbf{\large Next-to-Leading-Order Potential.\hspace{.5cm}}
Truncating the vertex expansion at $\delta_\text{max}=8$ with a promoted three-nucleon interaction (3NI), the numerical uncertainty in the amplitude - calculated
as sketched in fig.~\ref{fig_eft-exp} - is estimated to decrease by a relative factor of $\frac{Q}{M}$. The effective range expansion provides two additional parameters of
$\mathcal{O}(M^{-1})$. Specifically, the NLO input data is taken as a 4-element subset of $\mathcal{D}=\lbrace a_s,a_t,r_s,r_t,B(d),B(t)\rbrace$.
The tree-level amplitude of $\mathcal{L}^{(8)}$ is given by eq.~(\ref{eq_nlo-pot-mom-space}), and transforms (app.~\ref{app_FT}) with a regulator function of the form
eq.~(\ref{eq_reg-fkt-mom}) into a coordinate space function:
{\small
\begin{eqnarray}\label{eq_nlo-pot-coord-space}
\hat{V}^\text{NLO}&=&
\sum\limits_{i<j}^AI_0\left(\Lambda,r\right)\left(A_1+A_2\vec{\sigma}_i\cdot\vec{\sigma}_j\right)+\left(A_3+A_4\vec{\sigma}_i\cdot\vec{\sigma}_j\right)\Big\lbrace I_0\left(\Lambda,r\right),\vec{\nabla}^2\Big\rbrace\nonumber\\
&&+I_0\left(\Lambda,r\right)\left(A_5+A_6\vec{\sigma}_i\cdot\vec{\sigma}_j\right)\vec{r}^2+I_0\left(\Lambda,r\right)A_7\vec{L}\cdot\vec{S}+
I_0\left(\Lambda,r\right)A_8\left(\vec{\sigma}_i\cdot\vec{r}\vec{\sigma}_j\cdot\vec{r}-\frac{1}{3}\vec{r}^2\vec{\sigma}_i\cdot\vec{\sigma}_j\right)\nonumber\\
&&-A_9\Bigg\lbrace I_0\left(\Lambda,r\right),\Big[\big[\partial^r\otimes\partial^s\big]^{2}\otimes\big[\sigma_1^p\otimes\sigma_2^q\big]^{2}\Big]^0\Bigg\rbrace\;;\\&&\nonumber\\
&&\;\;\;\text{with}\;\;\;{\vec{L}=-i\vec{r}\times\vec{\nabla}}\;\;,\;\;\vec{S}=\frac{1}{2}(\vec{\sigma}_i+\vec{\sigma}_j)\;\;,\;\;
I_0\left(\Lambda,r\right)=e^{-\frac{\Lambda^2}{4}\vec{r}^2}\;\;\;,\nonumber
  \end{eqnarray}
}
and nine redefined LECs $A_i$ which are functions of the original $C^\text{(N)LO}_i$. The explicit form of this dependence on $C^\text{(N)LO}_i$
is given in table~\ref{tab_pot-coord-lec-spect} in the appendix. Contributions from EFT($\slashed{\pi}$)
to effective range parameters for $l=1$ (P-wave) and the transition $l=0\rightarrow l=2$ (SD-mixing) are suppressed relative to contributions to the effective ranges $r_{s,t}$
by at least another factor
of $\frac{Q}{\Lambda}$. This allows for an arbitrary constraint on the predictions of $\hat{V}^\text{NLO}$ in these partial waves, because only the next orders will introduce
LECs that have to be fitted to physical values. I will use two types of constraints, depending on whether the NLO operators are iterated (nonperturbative treatment) or not.
The specific constraints are given in table~\ref{tab_p-wave-constr}. We chose the values for the phase shifts $\delta_l$ and $\epsilon_1$ as small fractions of the Nijmegen
partial wave data$^\text{\cite{pwa-online}}$ (PWA) because then LECs are of magnitudes that do not pose numerical problems.
\begin{table*}
\renewcommand{\arraystretch}{1.5}
  \caption{\label{tab_p-wave-constr}{\small Next-to-leading order constraints on observables receiving dominant contributions from terms of higher order in
the vertex expansion.}}
\begin{tabular}{c|cc}
\hline
&nonpert.&pert.\\
\hline\hline
$\delta_1\left(\langle^{2s+1}P_J|\hat{V}^\text{NLO}|^{2s+1}P_J\rangle\right)$&$\mathcal{O}\left(0.1\cdot\delta_1^\text{PWA}\right)$&0\\
$\epsilon_1\left(\langle^3S_1|\hat{V}^\text{NLO}|^3D_1\rangle\right)$&$\mathcal{O}\left(0.1\cdot\epsilon_1^\text{PWA}\right)$&0\\
\hline
\end{tabular}
\end{table*}
\par\vskip10pt
\newpage
\textbf{\normalsize Nonperturbative (nNLO).}\\
The first scheme I used, identifies the functions given in eq.~(\ref{eq_nlo-pot-coord-space}) \& (\ref{eq_tni-pot-coord}) with a potential, from which predictions are
made by solving the corresponding Schr\"odinger equation. This equation of motion for the wave function can be transformed into a Lippman Schwinger type
integral equation like eq.~(\ref{eq_bubble-sum-solution}), and by that it can be seen that its solution is found by iterating all operators in the potential. The
amplitude, thus obtained, includes in addition to the LO and NLO diagrams shown in fig.~\ref{fig_eft-exp} contributions of $\mathcal{O}\left(\frac{1}{M^{n\geq 3}}\right)$.
The effect is analogous to that of the regulator function. The resulting amplitude can be cast into the form
\begin{equation}\label{eq_amplitude-weinberg}
T(|\vec{p}|)\propto\frac{1}{\frac{1}{C^\text{LO}}+\sum\limits_{i=0}^{\infty}\left(d_i+f_i\right)p^{2i}+i p}\;\;\;,
\end{equation}
with expansion coefficients $d_i\neq 0$ even for $i>1$ resulting from the iterated NLO vertices. The $f_i$'s parameterize the expansion of the regularization
analogous to $I_\Lambda$ in eq.~(\ref{eq_t-mat-regulated}) at LO. The cutoff dependence leaking through non-zero coefficients with $i>1$ has to be
$\lesssim\mathcal{O}\left(\frac{Q}{M}\right)^2$ for this power-counting scheme to be useful. The $d_i$ will be a linear combination of LECs from
eq.~(\ref{eq_nlo-pot-mom-space}), and therefore, this criterion could be met by a cancellation of large coefficients. Although this is acceptable from the EFT
point of view, the associated difference in the relative magnitude of matrix elements causes numerical uncertainties which increase significantly beyond $A=4$.
For $A\leq 4$, however, explicit calculations will provide evidence that the criterion is met, and the numerical treatment is consistent in this respect
with calculations using different renormalization schemes for which analytical solutions exist even at NLO.
\par
The redefined LECs are chosen with the genetic algorithm to minimize the weight function
\begin{equation}\label{eq_nlo-fit-para}
W\left(\lbrace A_i\rbrace\right)=\sum\limits_{o\in\mathcal{D}'}\frac{|o(\text{exp})-o(\text{trial})|}{|o(\text{exp})|}\\
\end{equation}
{\small$$\text{ with}\;\;\;\mathcal{D}'=\Big\lbrace\left(\text{Re,Im}\lbrace e^{2i m(E)}\rbrace,m\in\lbrace\delta\left(^{1,3}S_{0,1},^{1,3}P_{0,1,2},\epsilon_1\right)\rbrace,E\in]0~\text{MeV},0.35~\text{MeV}]\right),B(d)\Big\rbrace$$}
In order to assess convergence and accuracy, potentials with different short distance structure - parameterized by the regulator cutoff $\Lambda$ and the input data - are derived
(table~\ref{tab_nlo-pots}).
\begin{figure}
  \includegraphics[width=.7\textwidth]{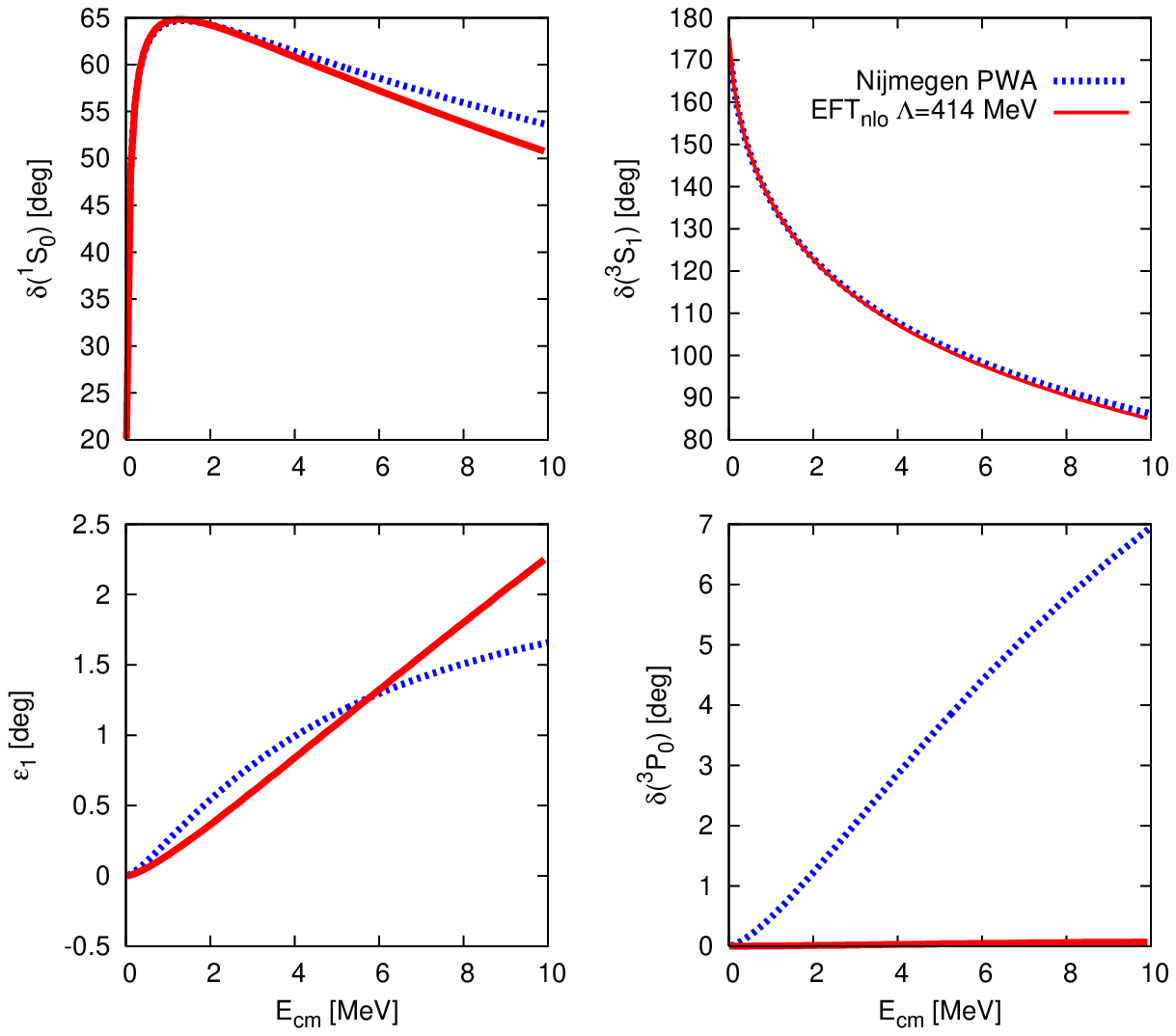}
  \caption{\label{fig_nn-phases}\small Selected neutron-proton scattering phase shifts
    for the $^1V_\slashed{\pi}$ potential and values from the Nijmegen partial
    wave analysis$^\text{\cite{pwa-online}}$ which were used to fit the LECs.}
\end{figure}
The predicted NN phase shifts, shown in fig.~\ref{fig_nn-phases}, are consistent with expectations regarding LO to NLO convergence and accuracy
at the respective order. In particular, the validity of the constraint on the P-wave phase shifts is apparently satisfied and the predictions for energies larger than the fitting interval
remain small. We iterate that those constraints reduce the number of independent LECs in eq.~(\ref{eq_nlo-pot-coord-space}) to five.
Also, as argued above, they are consistent with the EFT philosophy that higher order
interactions can only induce higher order corrections in observables.
Different P-wave constraints, corresponding to either option in table~\ref{tab_p-wave-constr},
allow for such a controlled modification of short-distance structure.
Enforcing zero P-wave phase shifts provides no significant
gain in computer time in calculations in $A\leq 4$ systems. We therefore exploit the
additional handle on higher order effects provided by the first option which allows non-zero values
to gauge the accuracy of a NLO calculation. A consequence of this approach are non-vanishing
rank one and two interactions corresponding to LECs $A_7$ and $A_{8,9}$, respectively.
These tensor structures are also found when one implements the SD-interactions, which
enter beyond NLO. We choose to include these as representations of higher-order effects
and to constrain their parameters by the SD-mixing angle.
\setlength{\tabcolsep}{0.2em}
\begin{table*}
  \caption{\label{tab_nlo-pots}\small Numerical values of the cutoff $\Lambda$ and the LECs of the NLO EFT($\slashed{\pi}$) potentials.
    The LECs were fitted to $B(d,\text{exp})$~\cite{exp-deut} and NN phase shifts.}
\footnotesize
    \begin{tabular}{cc|ccccccccc}
\hline
      &$\Lambda$&$A_1$&$A_2$&$A_3$&$A_4$&$A_5$&$A_6$&$A_7$&$A_8$&$A_9$\\
      &\tiny{[MeV]}&\tiny{[MeV]}&\tiny{[MeV]}&\tiny{[MeV$\cdot$fm$^2$]}&\tiny{[MeV$\cdot$fm$^2$]}&\tiny{[MeV$\cdot$fm$^{-2}$]}&
      \tiny{[MeV$\cdot$fm$^{-2}$]}&\tiny{[MeV]}&\tiny{[MeV$\cdot$fm$^{-2}$]}&\tiny{[MeV$\cdot$fm$^{2}$]}\\
\hline\hline
      $^1V_\slashed{\pi}$&414&-143&464&-57.0&68.8&-52.4&-346&-3.97&-62.1&-0.111\\
      $^2V_\slashed{\pi}$&432&-612&944&-157&110&351&-723&66.0&-168&-0.137\\
      $^3V_\slashed{\pi}$&544&-1224&1036&-432&336&1704&-1851&-78.9&-378&-0.130\\
      $^4V_\slashed{\pi}$&544&-1625&-89.6&-298&49.8&1870&34.3&-990&-734&$2.13\cdot 10^{-3}$\\
      $^5V_\slashed{\pi}$&544&-405&317&-125&108&287&-493&-271&-360&$-9.88\cdot 10^{-4}$\\
      $^6V_\slashed{\pi}$&648&84.9&-324&-888&404&4845&-2064&-342&-2319&-0.716\\
      $^7V_\slashed{\pi}$&648&-1316&1039&-530&431&3775&-3666&-75.4&-1052&-0.178\\
      $^8V_\slashed{\pi}$&672&301&-143&-708&480&3527&-3450&-192&-1613&-0.257\\
      $^9V_\slashed{\pi}$&672&-158&-201&-362&275&2001&-1659&79.7&-1270&-0.0850\\
\hline
    \end{tabular}
\end{table*}
\par\vskip10pt
\newpage
\textbf{\normalsize Perturbative (pNLO).}\\
Although the partial inclusion of higher-order contributions through an iteration of NLO operators in the course of solving the Schr\"odinger equation does not result in
a cutoff dependence inconsistent with the expected NLO uncertainty, \textit{i.e.}, the uncertainty and convergence estimates are still valid, a perturbative treatment is expected to have
two main advantages:
\begin{enumerate}
\item The numerical problems related to the cancellation of large numbers to yield relatively small ones can be avoided. In the denominator of the amplitude in
eq.~(\ref{eq_amplitude-weinberg}) only the sum of the expansion coefficients is constrained, $d_i+f_i\lesssim\mathcal{O}(M^{-2i})$. Large LECs matching this condition pose
a numerical problem in the course of diagonalising the Hamilton matrix for $A>4$ systems.
\item The computational cost is significantly reduced because of the scalar operator structure of the interaction.
\end{enumerate}
Projecting the potential in partial waves, the LECs $A_i$ can be expressed in terms of spectroscopic coefficients
$C^\text{(N)LO}\left(^{2s+1}L_J\right)$ (table~\ref{tab_pot-coord-lec-spect} in app.~\ref{app_FT}). Imposing the constraints in table~\ref{tab_p-wave-constr}, \textit{i.e.}, set
$C_{\left(^{2s+1}P_J\right)}^\text{\tiny NLO}=0=C_{\left(\text{SD}\right)}^\text{\tiny NLO}$, results in an interaction where the leading-order part
$\hat{V}^\text{LO}=\hat{V}^\text{LO}_\text{NN}+\hat{V}^\text{LO}_{3\text{NI}}$ - given in
eq.~(\ref{eq_pot-coord-lo}) \& (\ref{eq_tni-pot-coord}) - is still iterated, but the NLO part contributes only perturbatively. The latter is now of the form,
\begin{eqnarray}\label{eq_nlo-pot-coord-pert}
\hat{V}^\text{NLO}_\text{\tiny pert}&=&\sum\limits_{i<j}^A\Big(e^{-\frac{\Lambda^2}{4}\vec{r}_{ij}^2}\left(A'_1+A'_2\vec{\sigma}_i\cdot\vec{\sigma}_j\right)\nonumber\\
&&+\left(A'_3+A'_4\vec{\sigma}_i\cdot\vec{\sigma}_j\right)\Big\lbrace e^{-\frac{\Lambda^2}{4}\vec{r}_{ij}^2},\vec{\nabla}^2\Big\rbrace+
\frac{3}{4}\Lambda^4e^{-\frac{\Lambda^2}{4}\vec{r}_{ij}^2}\left(A'_3-A'_4\vec{\sigma}_i\cdot\vec{\sigma}_j\right)\vec{r}_{ij}^2\Big)\nonumber\\
&&+\sum\limits_{\stackrel{i<j<k}{\text{cyclic}}}^Ae^{-\frac{\Lambda^2}{4}\vec{r}_{ij}^2}e^{-\frac{\Lambda^2}{4}\vec{r}_{jk}^2}A'_{3\text{NI}}\,\vec{\tau}_i\cdot\vec{\tau}_j\;\;\;.
\end{eqnarray}
following a prescription given by \cite{kaplan-pnlo}. The three coefficients corresponding to momentum independent operators are functions of the two independent NLO LECs
and the three LO LECs, $A'_{3\text{NI},1,2}=F\left(A'_{3,4},A_{3\text{NI},1,2}\right)$.
This relation enforces the constraint $\lim\limits_{E\to 0}T^\text{LO}(E)=\lim\limits_{E\to 0}T^\text{NLO}(E)$ which is not satisfied automatically if a cutoff regulator is used.
Explicitly, in the limit of taking the energy to zero, the NLO calculation has to yield the LO values for the scattering lengths and binding energies of the deuteron
and triton. This is guaranteed, first, by an expansion of the LECs in the low-momentum scale in addition to that of the operators,
\begin{equation}
A_i\equiv\sum_{k=-1}^\infty A_i^{[k]}\left(\lambda\right)\;\;\;,
\end{equation}
where $\lambda$ parameterizes the regularization scheme. Here, this expansion is resembled in the $A'_{1,2}$ terms. Second, one has to match the LECs to an appropriately expanded amplitude, where my choice is listed below.
\par
Prior to this work, the strictly perturbative treatment of EFT interactions was sucessfully applied to three and four fermions in a harmonic trap$^\text{\cite{stetcu-harm-trapper}}$
using short-range interactions similar to
the ones here. To results for three-nucleon scattering, analyzed with this approach$^\text{\cite{nd-ham-meh}}$, we compare in ch.~\ref{subsec_nd-phase}.
Furthermore, calculations$^\text{\cite{lee-lat}}$ in a discretized space (lattice) were conducted for ground-state energies of $A=3,4,6,12$ nuclei with a
chiral effective field theory which also treats NLO corrections perturbatively.
\setlength{\tabcolsep}{0.2em}
\begin{table*}
  \caption{\label{tab_nlo-pots-pert}\small  Numerical values of the cutoff $\Lambda$ and the LECs of the pNLO EFT($\slashed{\pi}$) potentials.}
\footnotesize
    \begin{tabular}{cc|cc|ccccc}
\hline
$\Lambda$&$\mathcal{D}'$&$\frac{\Lambda^3}{8\pi^{\frac{3}{2}}}C_1^\text{LO}$&$\frac{\Lambda^3}{8\pi^{\frac{3}{2}}}C_1^\text{LO}$&$A'_1$&$A'_2$&$A'_3$&$A'_4$&$A'_{3\text{NI}}$\\
\small{[MeV]}&&\small{[MeV]}&\small{[MeV]}&\small{[MeV$\cdot$fm$^2$]}&\small{[MeV$\cdot$fm$^2$]}&\small{[MeV$\cdot$fm$^{-2}$]}&\small{[MeV$\cdot$fm$^{-2}$]}&\small{[MeV]}\\
\hline\hline
300&$a_s\,,\,B(d)$&$-81.480$&$-6.9786$&$-43.198$&$-60.593$&$-10.062$&$-7.1631$&$-3.0509$\\
400&$a_s\,,\,B(d)$&$-136.66$&$-9.1401$&$-129.41$&$-190.07$&$-18.536$&$-12.634$&$2.0703$\\
500&$a_s\,,\,B(d)$&$-81.505$&$-6.9786$&$22.954$&$9.6193$&$1.2529$&$0.83394$&$-35.996$\\
\hline
    \end{tabular}
\end{table*}
\par
For this work, the interaction is derived for the analysis of the 5- and 6-helium system,
and proceeds in two steps. The LO LECs are determined to yield the unperturbed states which are
then used to calculate the NLO matrix elements:
\newpage
\begin{enumerate}
\item Iteration of LO vertices is encoded in two- and three-body eigenstates with $\mathcal{D}'=\lbrace a_s,B(d),B(t)\rbrace$
and scattering states with $E=0.4~$keV:
\begin{eqnarray}
C_{1,2}^\text{LO}&:&\underbrace{\left(\hat{T}+\hat{V}^\text{LO}\right)}_{\equiv \hat{H}^\text{LO}}\big|\psi\big\rangle=E\big|\psi\big\rangle\stackrel{a_{s},B(d)}{\longrightarrow}
\Bigg\lbrace\begin{array}{l}
E=0.4~\text{keV}:\,\text{LO scattering states}\;\big|\psi_0^{(\pm)}\big\rangle\\
E=-B(d):\;\text{deuteron ground state}\;\big|d\big\rangle
\end{array}\nonumber\\
C_{3\text{NI}}^\text{LO}&:&\hat{H}^\text{LO}\big|\psi\big\rangle=B(t)\big|\psi\big\rangle\;\rightarrow\;\text{triton ground state}\;\big|t\big\rangle\nonumber\\
&&\Rightarrow\;\;\hat{H}^\text{LO}\big|\psi\big\rangle=E\big|\psi\big\rangle\;\rightarrow\;B(\alpha,\text{LO}),\big|\alpha\big\rangle
\end{eqnarray}
The ground state wave function and the energy of the four-nucleon system are predictions at LO.
\item Leading-order wave functions are perturbed by single NLO interactions with LECs $A'_{1,2,3\text{NI},3,4}$ fixed by the constraints:
\begin{enumerate}
\item[(i)] $\big\langle\psi_0^{(-)}\left(^1S_0\right)\big|\hat{V}^\text{NLO}_\text{\tiny pert}\big|\psi_0^{(+)}\left(^1S_0\right)\big\rangle=0$
\item[(ii)] $\big\langle\psi_0^{(-)}\left(^3S_1\right)\big|\hat{V}^\text{NLO}_\text{\tiny pert}\big|\psi_0^{(+)}\left(^3S_1\right)\big\rangle=
\frac{4\pi}{m_\text{\tiny N}}\left(\frac{1}{\left(m_\text{\tiny N}B(d)\right)^{-1/2}-ip}-\frac{1}{-\frac{1}{a_t}-ip}\right)$
\item[(iii)] $\big\langle d\big|\hat{V}^\text{NLO}_\text{\tiny pert}\big|d\big\rangle=0$
\item[(iv)] $\big\langle t\big|\hat{V}^\text{NLO}_\text{\tiny pert}\big|t\big\rangle=0$
\item[(v)] $\big\langle\alpha\big|\hat{V}^\text{NLO}_\text{\tiny pert}\big|\alpha\big\rangle=-B(\alpha,\text{LO})+B(\alpha)$
\end{enumerate}
This means using the np scattering lengths, and the binding energies of the deuteron, triton, and $\alpha$ as input. We will also use interactions
which fit to $B(\alpha)$ at LO. For those, constraints (iv) and (v) are changed accordingly.
\end{enumerate}
\par
The reasoning behind the unusual choice of input data to fit the NLO LECs is given in app.~\ref{app_rgm-pnlo}, after technical details of the numerical method
and findings in the n-d scattering system, necessary for the understanding, have been discussed. Here, we stress that a four-body datum instead
of an effective range is used to determine NLO LECs.
The second condition includes the scattering length in the deuteron channel in the input data. At LO, $B(d)$ is used to determine the LEC in this channel, resulting
in a $^3S_1$ scattering length of $a_t(\text{LO})=\left(m_\text{\tiny N}B(d)\right)^{-1/2}$. Through condition (ii), the experimental $a_t$ is reproduced at pNLO.
We derived three potentials (table~\ref{tab_nlo-pots-pert}) and applied them to the five- and six-body system
(fig.~\ref{fig_he5-spect} \& \ref{fig_6he-spect}).
All calculations so far, if not stated otherwise, utilize the variational method which I will introduce in the following chapter.
\section{The Resonating Group Method}\label{sec_rrgm}
For the approximate solution of the few-body problem, I chose a variational approach in which the total wave function is sought as a linear combination of products
of bound subsystems. This decomposition into fragments is based on the idea of nucleons jumping \textit{resonantly} from one defined \textit{group} within a
nucleus to the other$^\text{\cite{rgm-a-wheeler,rgm-b-wheeler}}$.
This ansatz is expected intuitively to be particularly handy for systems where the jumps occur in short time intervals relative
to the resting time in a certain configuration. If in addition the state of interest has one dominating component, in which it spends most of its life, this approach allows
for an efficient yet accurate description. For this work, the method was chosen for its following properties:
\begin{itemize}
\item Calculation of nuclear bound- and scattering states in principle for any number of particles $A$ in one framework.
\item Direct access to an approximation of the totally antisymmetric wave function is provided.
\item Individual basis states can be classified as describing high- or low-energy modes, and hence renormalization arguments provide a
guideline for a systematic refinement of the model spaces.
\item Correct treatment of long-range forces, \textit{e.g.}, Coulomb interactions between charged constituents.
\item Multi-channel framework allows the treatment of breakup reactions (limited to two-fragment asymptotic states).
\end{itemize}
I will review the calculation of observables in this framework, while details about the evaluation of coordinate space matrix elements and model spaces are
dealt with in app.~\ref{app_rgm-coord-me} and during the discussion of the results for the respective systems.
\par
In our incarnation$^\text{\cite{hmh-rrgm}}$, the Resonating Group Method (RGM) treats the scattering of two - in general composite - nuclei and decomposes the translationally invariant Hamilton operator - the
center of mass motion is separated off - accordingly into fragment-internal, short- and long-range inter-fragment parts (relate to fig.~\ref{fig_rgm-cluster} for an example and
notations):
\begin{equation}\label{eq_rgm-hamiltonian-allg}
\hat{H}=\hat{H}_{\text{f}1}+\hat{H}_{\text{f}2}+\hat{T}_{\text{rel}}+\hat{V}_{\text{Coul,rel}}+
\underbrace{\sum_{i,j,k\in \text{f}1\vee\text{f}2}\hat{V}^\text{(N)LO}-\hat{V}_{\text{Coul,rel}}}_{\equiv\hat{V}_\text{short}}\;\;\;.
\end{equation}
This motivates a factorization of relative motion from the channel function in the ansatz for the total wave function:
\begin{equation}\label{eq_rgm-wfkt-ansatz}
\psi^m=\mathcal{A}\sum_n^{N(\text{ch})}\psi_\text{ch}^n\psi_\text{rel}^{mn}\;\;\;,\;\;\;
\begin{array}{c}\mathcal{A}=\sum\limits_{\mathcal{P}\in S(A)}(-1)^\mathcal{P}\mathcal{P}\\N(\text{ch}):\;\;\text{number of channels}\end{array}
\end{equation}
The wave function is antisymmetrized ($\mathcal{A}$) with a symmetric group $S(A)$ for $A$ particles decomposed into double cosets$^\text{\cite{selig-cosets}}$.
This decomposition reduces the number of spacial matrix elements to be calculated to one representative for each double coset instead of one for each permutation $\mathcal{P}$.
The superscript $m$ labels the boundary condition imposed on the relative motion wave function:
\begin{equation}\label{eq_rgm-rel-wfkt}
\psi_\text{rel}^{mn}=\delta^{mn}F^n+a^{mn}G^n+\sum_\nu b_\nu^{mn}e^{-\gamma_{n\nu}\vec{R}^2}\;\;\;,
\;\;\;\begin{array}{c}\vec{R}\propto\vec{r}_\text{cm}(\text{f1})-\vec{r}_\text{cm}(\text{f2})\\a^{mn},b_\nu^{mn}:\;\;\text{variational parameters}\end{array}\;\;\;,
\end{equation}
with
\begin{equation}\label{eq_frag-int-eigenfkts}
\lim_{|\vec{R}|\to\infty}\hat{H}\psi^m=\left(\hat{H}_{\text{f}1}+\hat{H}_{\text{f}2}\right)\psi^m+E_\text{rel}\psi^m\;\;\;,
\end{equation}
\textit{i.e}, $F^n$ ($G^n$) is the regular (irregular) Coulomb function$^\text{\cite{coul-abram}}$ for the relative motion of the two fragments with a kinetic energy
$E_\text{rel}=\frac{\vec{k}^2}{2m_\text{red}}$. In its current implementation, the RGM imposes the boundary condition
\begin{equation}\label{eq_asymp-wfkt}
u_l(kR)=\frac{i}{2}\sqrt{\frac{k}{m_\text{red}}}\left(h_l^-(kR)-e^{2i\delta_l}h_l^+(kR)\right)\;\;\;,\;\;\;\text{with}\;\;\;h_l^\pm(kR)=G_l(kR)\pm i F_l(kR)\;\;\;,
\end{equation}
to adopt the convention of~\cite{newton} for the scattering matrix.
The boundary condition allows regular waves in channel $n$ only, while admixtures of other channels, and the corrections for relative separation $|\vec{R}|$
of the fragments where $V_\text{short}(R)\neq 0$, are encoded in $a^{mn}$ and $b_\nu^{mn}$, respectively. The variational parameters are determined by the
extremal condition (Kohn variation$^\text{\cite{kohn-variation}}$):
\begin{equation}\label{eq_kohn-variation}
\delta\left(\big\langle\psi^m\big|\hat{H}-E\big|\psi^m\big\rangle-\frac{1}{2}a^{mn}\right)\stackrel{!}{=}0\;\;\;.
\end{equation}
In general, the variational approach yields an asymmetric reactance matrix, an artefact eliminated by a so called Kato correcture$^\text{\cite{kato}}$, after which the
$S$-matrix is derived:
\begin{equation}\label{eq_s-from-a}
S=\frac{\mathbb{1}+i a}{\mathbb{1}-i a}\;\;\;.
\end{equation}
The divergence of $G_l(\rho)$ at $\rho=0$ is regularized here by using
\begin{equation}\label{eq_reg-irreg-fkt}
\tilde{G}_l^n(kR)=\left(\sum_{j=2l(n)+1}^\infty\frac{\left(\beta R\right)^j}{j!}e^{-\beta R}\right)\cdot G_l^n(R)\;\;\;,
\end{equation}
instead of $G^n_l$ in a partial wave expanded eq.~(\ref{eq_rgm-rel-wfkt}).
This approximation, although negligible in the determination of $a^{mn}$, has a significant effect on matrix
elements involved in the fit of perturbative NLO LECs. Specifically,
$\lim_{\Lambda\to\infty}\big\langle\psi_0^{(-)}\big|\hat{V}^\text{LO}_\text{\tiny pert}\big|\psi_0^{(+)}\big\rangle$ receives no contribution from $\tilde{G}$ for values
of the regulator parameter $\beta$ that suppress the function over the range of support of the potential (see also app.~\ref{app_rgm-pnlo}).
Increasing $\beta$ to mitigate the effect of the regulator
for potentials of about the same range requires in turn a larger model space to approximate the $\tilde{G}$. Therefore, I choose relatively small cutoff values
$\Lambda$ corresponding to longer-range potentials, and used available independent bound-state data to determine the LECs for the potential
$\hat{V}_\text{\tiny pert}^\text{NLO}$ of eq.~(\ref{eq_nlo-pot-coord-pert}).
\par
Like the Coulomb functions as solutions to the relative motion part, the channel wave function - containing the bound states of possibly composite fragments -
is expanded in a finite basis in order to render the method practical.
Our implementation employs a Gau{\ss}ian basis. Another approach, recently developed$^\text{\cite{rgm-nav-qual}}$, uses a harmonic-oszillator basis and therefore
orthogonal basis vectors - an advantage over the Gau{\ss}ian basis.
I distinguish between three types of expansions:
\begin{enumerate}
\item \textit{Cluster expansion}: Of all possible groupings $\mathcal{G}$, only a subset $\mathcal{G}'$ is considered in the ansatz, \textit{e.g.}, from
$$\mathcal{G}(^6\text{He})=\big\lbrace^4\text{He-2n},\text{t-t},^3\text{He-3n},\text{\mbox{d\hspace{-.55em}$^-$}-\mbox{d\hspace{-.55em}$^-$}-2n},\text{(pp)-2n-2n}\big\rbrace$$
the 6-helium ground state was found$^\text{\cite{wurzer-halo}}$ to be dominated by a single configuration $\mathcal{G}'=\big\lbrace^4\text{He-2n}\big\rbrace$
(see fig.~\ref{fig_6he-fragm} for a sketch). In a scattering calculation at energies below a
certain threshold, closed channels contribute in the form of so called distortion channels. Such channels do not have an asymptotic tail and are used to enlarge the basis
for a better approximation of the wave function in the region of configuration space where the colliding fragments interact.
Therefore, those channels are included to increase to variational space although they do not correspond to, \textit{e.g.}, possible breakup of a composed system.
\item \textit{Partial wave expansion}: Orbital angular momenta up to $l_\text{m,rel(int)}$ are considered for the relative (fragment-internal) wave functions.
For the bound states of the fragments, I include all coupling schemes of fragment-internal angular momenta $l$ consistent with its total orbital angular momentum $L$.
For an $A$-nucleon fragment there are $A-1$ $l$'s and $A-2$ intermediate couplings $l'$ to specify.
Likewise, all coupling schemes of the individual spins of the fragment's constituents to yield a good total spin quantum number $S$, that can be coupled with $L$
to good $J$, are included in the wave function ansatz. Only the $A-2$ intermediate spin coupling values have to be specified because of the conserved nucleon spin of $\frac{1}{2}$.
\item \textit{Gau{\ss}ianization}: The radial parts of the partial-wave-expanded functions are expanded in a Gau{\ss}ian basis parameterized by a finite set of width parameters,
$\mathbf{\gamma}$.
\end{enumerate}
The regularization parameter $\Lambda_\text{RGM}$, introduced in sec.~\ref{sec_eft}, is then a function of: $\mathcal{G}'$,
a set of possibly different $l_\text{max}$, and $\mathbf{\gamma}$.
The convergence of the calculation was assessed with respect to all three expansions. The decomposition in two fragments, and the angular momentum coupling
scheme are manifest in the ansatz for the channel wave function:\\
\begin{equation}\label{eq_rgm-ch-wfkt}
\psi_\text{ch}=\Bigg[\frac{1}{R}Y_{L_\text{\tiny rel}}(\hat{\vec{R}})\otimes\Big[\phi^{J_{f1}}\otimes \phi^{J_{f2}}\Big]^{S_{c}}\Bigg]^J\;\;\;,
\end{equation}
where the abreviation for the coupling of two spherical tensors,
\begin{equation}\label{eq_angl-mom-coupl}
\left[S^{m_s}\otimes L^{m_l}\right]^{JM}=\sum_{m_{s,l}}(Sm_sLm_l|JM)S^{m_s}L^{m_l}\;\;\;,
\end{equation}
is used. The channel spin $S_c$ is  coupled from the total spins $J_{fi}$ of the two fragments whose wave functions are denoted $\phi^{J_{fi}}$.
Appropriate labels for components of the spherical tensors are understood. The implied coupling scheme - relative orbital angular momentum $L_\text{rel}$ with $S_c$ to
a total $J$ - differs from the fragment-internal scheme for $J_{fi}$. The ansatz for the fragment wave function is:\\
\begin{equation}\label{eq_rgm-int-wfkt}
\phi^{J}=\mathcal{A}\sum_{\stackrel{f\in\mathcal{G}'}{j}}\sum_{\stackrel{\mathbf{l}_j,\mathbf{s}_j}{\mathbf{\gamma}_j}}c(j)\Bigg[\Big[\prod_{n=1}^{A-1}e^{-\gamma_{n}(j)\vec{\rho}_n^2}
\mathcal{Y}_{l_{n}(j)}\left(\vec{\rho}_n\right)\Big]^{L(j)}\otimes\Xi^{S(j)}\Bigg]^J\;\;\;,
\end{equation}
\begin{minipage}[c]{.4\textwidth}
with\hspace{.2cm}
\footnotesize
$
\begin{array}{l}
m\in\lbrace 1,A-1\rbrace\;\;,\;\;n\in\lbrace 1,A-2\rbrace\\
\mathbf{l}_j=\big\lbrace (l_m(j),l'_n(j)),L(j)\big\rbrace\\
\end{array}
$
\par\vspace{0pt}
\end{minipage}
\begin{minipage}[c]{.4\textwidth}
,\hspace{.3cm}and\hspace{.2cm}
\footnotesize
$
\begin{array}{l}
\mathbf{s}_j=\big\lbrace s_n(j),S(j)\big\rbrace\\
\mathbf{\gamma}_j=\big\lbrace\gamma_{m}(j)\big\rbrace\\
\mathcal{Y}_{l}\left(\vec{r}\right)=|\vec{r}|^lY_{l}\left(\hat{\vec{r}}\right)
\end{array}
$\hspace{.2cm}.
\par\vspace{0pt}
\end{minipage}\\
\input{diss_rgm-cluster.tex}
The two fragments of the example in fig.~\ref{fig_rgm-cluster} also represent two different values of $f$, \textit{i.e.}, substructures within a fragment. The superposition
coefficients $c$ are determined by solving the generalized eigenvalue problem because of the nonorthonormal nature of the basis:
\begin{equation}\label{eq_gen-ev-prob}
\big\langle\mathcal{A}\phi_m^J\big|\hat{H}\big|\mathcal{A}\phi_n^J\big\rangle =
E\underbrace{\big\langle\mathcal{A}\phi_m^J\big|\mathcal{A}\phi_n^J\big\rangle}_{\equiv N\stackrel{\text{\tiny in gen.}}{\neq}\mathbb{1}}\;\;\;,
\;\;\;\text{with}
\;\;\;m,n=\lbrace f,j,\mathbf{l}_j,\mathbf{s}_j,\mathbf{\gamma}_j\rbrace\;\;\;\text{defining a basis vector.}
\end{equation}
The eigensystem is solved by diagonalizing $\hat{H}'$ defined as
\begin{equation}\label{eq_eigensystem}
\underbrace{N^{-\frac{1}{2}}\hat{H}N^{-\frac{1}{2}}}_{\hat{H}'}-E\mathbb{1}=0\;\;\;.
\end{equation}
The inversion of the norm matrix $N$ increases the numerical uncertainty and renders the method useless if $N$ has eigenvalues of order $10^{-12}$ or smaller.
Eigenvalues of this magnitude can result from basis states which are linearly depended up to differences which cannot be resolved by the numerical precision, and have
to be excluded. In app.~\ref{app_rgm-wfkt}, a three-nucleon system is considered as an example for the construction of the fragment wave function eq.~(\ref{eq_rgm-int-wfkt})
and the following determination of the ground-state wave function.
\par
With $f_l$ and $\tilde{g}_l$ expanded in Gau{\ss}ians - in practice, a basis parameterized by one of the sets $\mathbf{\gamma}_j$ for the states in
eq.~(\ref{eq_rgm-int-wfkt}) is used - the matrix elements necessary to determine the reactance matrix $a^{mn}$ in the channel basis require a basis transformation
resembling the different angular momentum coupling schemes and hence can be written for a generic operator $\hat{o}$ as a linear combination of elements of the form:
\begin{equation}\label{eq_bv-me}
M_{mn}=\langle\phi_m^{J'M'}\vert\mathcal{A}\hat{o}\vert\phi_n^{JM}\rangle\;\;\;\;\;\;\;\;\;\text{(notation def. in eq.~(\ref{eq_gen-ev-prob}))}.
\end{equation}
The spacetime symmetry of the theory (ch.~\ref{sec_eft}) allows only spherical tensor operators $\hat{o}(r,m_r)$ of rank $r=0$ which makes the following separation
in coordinate- ($\hat{o}_o$) and spin part ($\hat{o}_s$) possible:
\begin{equation}\label{eq_sph-tens-op}
\hat{o}(0,0)=\sum_{m_q}\hat{o}_o(q,m_q)\Bigg[\hat{o}_s(q,m_q)\underbrace{\frac{(-1)^q}{\sqrt{2q+1}}\Bigg](-1)^{m_q}}_{=(qm_qq-m_q|00)}\;\;\;.
\end{equation}
The operator structure for the EFT potentials in eq.~(\ref{eq_nlo-pot-coord-space}) \& (\ref{eq_tni-pot-coord}) matches this form if the tensor operator is transformed,
$\vec{\sigma}_i\cdot\vec{r}\vec{\sigma}_j\cdot\vec{r}-\frac{1}{3}\vec{r}^2\vec{\sigma}_i\cdot\vec{\sigma}_j=
\sqrt{5}[\left[\sigma_i\otimes\sigma_j\right]^2\otimes\left[r\otimes r\right]^2]^0$.
Using Racah algebra$^\text{\cite{edmonds}}$, the matrix element can be written in the form
{\small\arraycolsep.3pt\renewcommand{\arraystretch}{.5}
\begin{equation}\label{eq_op-me-prod-red-me}
M_{mn}=\delta_{JJ'}\delta_{MM'}\sum_{\mathcal{P}\in S(A)}(-1)^\mathcal{P}c^*(j)c(j)(-1)^{L+2S+S'-J}\left\lbrace
\begin{array}{ccc}
S&L&J\\L'&S'&r
\end{array}
\right\rbrace
\langle L'\vert\vert \mathcal{P}\hat{o}_o(r)\vert\vert L\rangle\langle S'\vert\vert \mathcal{P}\hat{o}_s(r)\vert\vert S\rangle\;\;\;.
\end{equation}}
The Wigner-Eckart theorem defines the reduced matrix elements
\begin{equation}\label{eq_red-me-spin}
\langle S'\vert\vert \mathcal{P}\hat{o}_s(r)\vert\vert S\rangle =\frac{\sqrt{2S'+1}}{(SSrS'-S\vert S'S')}\langle S'S'\vert \mathcal{P}\hat{o}_s(r,S'-S)\vert SS\rangle\;\;\;.
\end{equation}
The magnetic quantum labels are conventionally chosen to be maximal. Eq.~\ref{eq_red-me-spin} holds analogously for the reduced coordinate space matrix element.
\par\vskip10pt
\newpage
\textbf{\large Evaluation of the spatial matrix elements.\hspace{.5cm}}
The basic structure of all coordinate-space matrix elements necessary to obtain $M_{mn}$ in eq.~(\ref{eq_bv-me}) for a given pair (triplet) of interacting particles $ij(k)$ is:
\begin{equation}\label{spac_me_2}
J(\mathcal{P};l_1,m_1,\ldots,l_z,m_z)=\int d\vec{s}_1...d\vec{s}_{A-1}e^{-\sum_{\mu\mu'}^{A-1}\rho^{ij(k)}_{\mu\mu'}(\mathcal{P})\vec{s}_\mu
\vec{s}_{\mu'}}\cdot\prod_{n=1}^z\mathcal{Y}_{l_nm_n}\Bigg(\underbrace{\sum_{\mu=1}^{A-1}\xi_{\mu n}^{ij(k)}(\mathcal{P})\vec{s}_\mu}_{\equiv\vec{Q}_n}\Bigg)\;\;.
\end{equation}
This expression is valid if the following conditions are satisfied:
\begin{itemize}
\item The single particle coordinates $\vec{r}_i$ are transformed in Jacobi coordinates
$\vec{s}_i$ via an orthogonal matrix in \textit{bra} and \textit{ket}.
\item The permutation $\mathcal{P}$ is applied to the right, \textit{i.e.}, to the coordinates of the potential operators and the \textit{ket}.
\item The part of an operator acting in coordinate space, $\hat{o}_o$, is expressed in terms of Gau{\ss}ians and spherical harmonics, as in eq.~(\ref{eq_nlo-pot-coord-space}).
The choice for the regulator results in Gau{\ss}ian radial potential functions, and therefore, no additional expansion is necessary.
\item The Jacobi coordinates of the \textit{bra} are to be chosen as the independent variables, \textit{i.e.}, all other coordinates are expressed in terms of them.
\end{itemize}
These operations define the transformation matrices $\rho$ and $\xi$ (eq.~(\ref{spac_me_2})), both depending on the coordinates of the interacting particles and the permutation.
The $\vec{Q}_n$ are hence, in general, linear combinations of the independent variables $\vec{s}_n$. $z$ equals the total number of inter-cluster
coordinates in \textit{bra} and \textit{ket} plus the number of spherical harmonics from the potential operator. For example, the operator $\propto A_8$ in
eq.~(\ref{eq_nlo-pot-coord-space}) adds 2 to $z$. The two major simplifications which lead to an analytic expression for $J(\mathcal{P})$ are:
\begin{itemize}
\item the diagonalization of $\rho_{\mu\mu'}$ using a transformation matrix $T_{\lambda\lambda'}$ with
$T_{\lambda\lambda}=1$ and $T_{\lambda\lambda'}=0$ for $\lambda >\lambda'$,
\item expressing the spherical harmonics in terms of their generating function$^\text{\cite{edmonds}}$,
\begin{equation}\label{eq_gen-fkt}
(\vec{b}\cdot\vec{r})^L=b^L\sum_{m=-L}^LC_{Lm}b^{-m}\mathcal{Y}_{Lm}(\vec{r})\;\;,\;\;
\end{equation}
with $\;\;\vec{b}=(1-b^2,i(1+b^2),-2b)\;\;$, and $C_{Lm}=(-2)^LL!\sqrt{\frac{4\pi}{(2L+1)(L-m)!(L+m)!}}\;\;\;$.
\end{itemize}
An example of this prescription is given in app.~\ref{app_rgm-coord-me} for the calculation of the spacial part of the operator $\propto A_9$ -
as the only structure not present in previously considered potentials, like \texttt{AV18}$^\text{\cite{av18}}$, CD-Bonn$^\text{\cite{bonn}}$ - in eq.~(\ref{eq_nlo-pot-coord-space}).
\par\vskip10pt
\textbf{\large Evaluation of the spin matrix elements.\hspace{.5cm}}
A generic spin operator with rank $r$ in eq.~(\ref{eq_nlo-pot-coord-space}) acting on particles $i$ and $j$ is given as:
\begin{equation}\label{eq_spin_op}
\hat{o}_s(ij,r)=\sum_mC_m\hat{o}_s^m(i)\hat{o}_s(j)^m\;\;\;o_s^m(i)\in\lbrace \mathbb{1}_s(i),\vec{\sigma}_{i}\rbrace\;\;\;.
\end{equation}
The constants $C_m$ represent the coupling of spherical tensors of rank $1\;(\vec{\sigma})$ and scalars
$(\mathbb{1})$ to a spherical tensor of rank $r$. The transition from the spherical components
$\sigma^{\pm 1}$ to raising and lowering operators $\sigma^{\pm}$ is done via
$\sigma^{\pm}=\mp\frac{1}{\sqrt{2}}\sigma^{\pm 1}$, with known action on spin eigenfunctions. Accordingly, the state $\vert SS\rangle$ is expanded in
a linear combination of products of single particle spin eigenfunctions
\begin{equation}
\vert SS\rangle=\sum_kC_k\vert\frac{1}{2}m_1\rangle\cdot ...\cdot\vert\frac{1}{2}m_\text{\tiny N}\rangle\;\;\;.
\end{equation}
The $C_k$ contain the Clebsch-Gordan coefficients of all couplings. The matrix elements $\langle S'S'\vert \mathcal{P}\hat{o}_s(ij,r)\vert SS\rangle$
are calculated by applying $\hat{o}_s(ij,r)$ to the right, yielding another product function. Now all permutations
$\mathcal{P}$ are determined which result in non-zero matrix elements whose value depends on the specific operator coefficients.
\section{A=2}\label{sec_a2}
\begin{figure}
\includegraphics[width=.6\columnwidth]{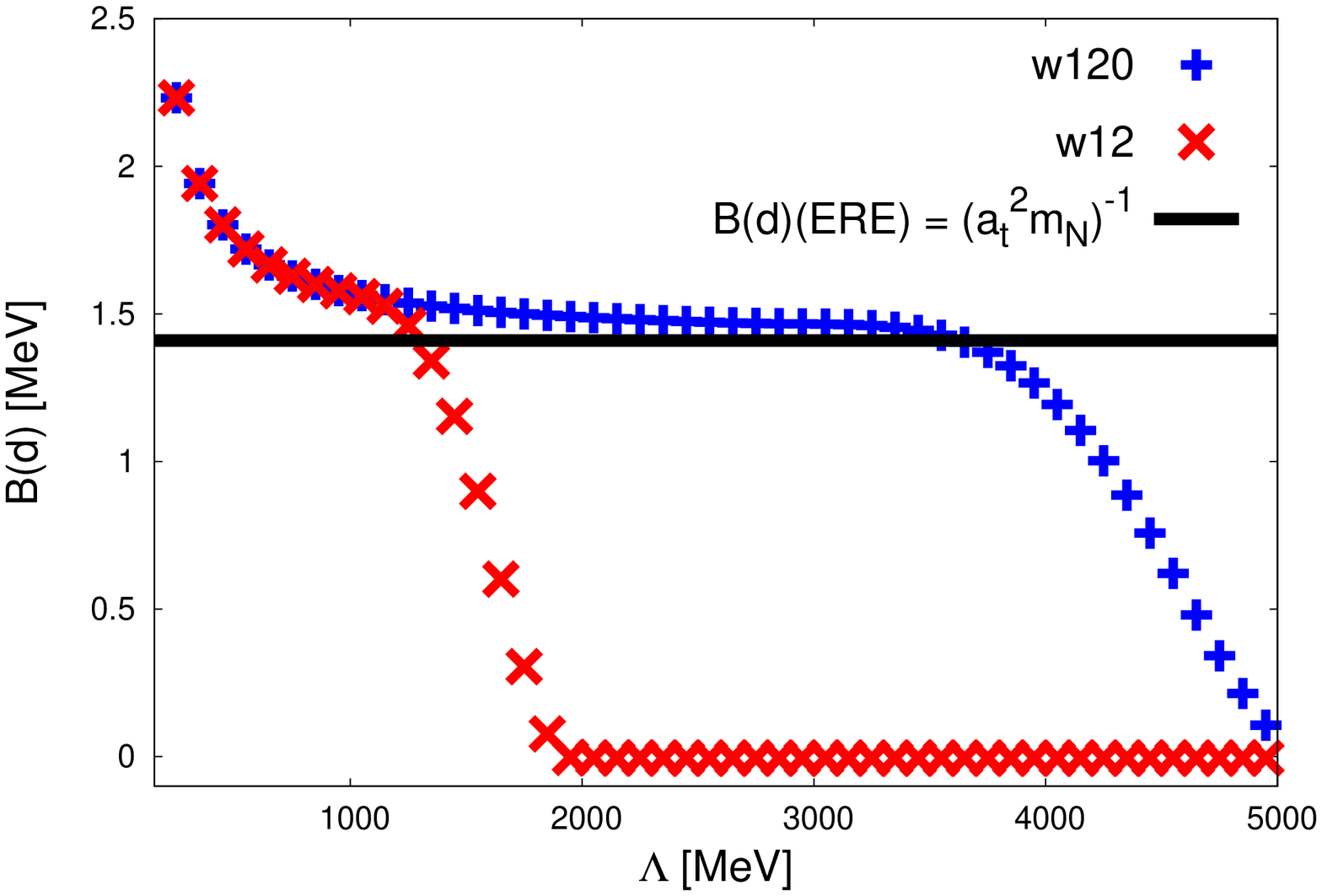}
  \caption{\label{fig_bd-cutoff}\small Binding energy $B(d)$ of the deuteron in
    leading-order EFT($\slashed{\pi}$) as a function of the Gau{\ss}ian momentum
    cutoff $\Lambda$ for two different RGM model spaces. The space $w120$
    contains narrower width parameters, and hence can be used for the
    potentials corresponding to larger values of $\Lambda$. Using the triplet
    neutron-proton scattering length $a_t$ as experimental input, the effective
    range formula predicts $B(d)=1.41~$MeV.}
\end{figure}

The two-nucleon system is relevant for this work because, first, it provides most of the data used to determine the LECs, and second, the sensitivity of observables
to $\Lambda_\text{RGM}$ can be assessed by a comparison to predictions by other numerical and analytical calculations.
The parameter $\Lambda_\text{RGM}$
is defined in ch.~\ref{sec_rrgm} as a set of fragmentations $\mathcal{G}'$, a set of maximal orbital angular momenta with corresponding coupling schemes, and a set
of width parameters covering an interval which is cut off at short and long distances. Although $\Lambda_\text{RGM}$ is a renormalization parameter in the same sense as the
regulator width $\Lambda$, probing for inconsistencies in the power counting is less practical with it because it is not straightforward how to change $\Lambda_\text{RGM}$ such
that short-distance physics only is affected, especially for $A>3$ systems. The theoretical uncertainty in an observable at order $n$ must not
significantly exceed $\left(\frac{Q}{M}\right)^n$, \textit{i.e.}, $\forall\Lambda\gtrsim M$ predictions have to agree within this error margin. For $\Lambda_\text{RGM}$,
there are corresponding inequalities, one for the angular momenta $l_m$, and another for the width parameters (assuming $\mathcal{G}'=\mathcal{G}$).
Analogous to the $\Lambda$ dependence, the soundness of the power counting guarantees that under a modification of the model space, \textit{i.e.},
an addition or removal of basis states of type eq.~(\ref{eq_rgm-int-wfkt}) with $l_n(j)>l_\text{max}$ and
$\gamma_n(j)\gtrless\gamma_{\stackrel{\text{\tiny max}}{\text{\tiny min}}}$ - corresponding to high energy modes - predictions for low-energy observables change
only at order $\left(\frac{Q}{M}\right)^n$. Choosing a complete basis is equivalent to $\Lambda\to\infty$. For two reasons I favor to operate in a numerically complete
basis, which does not exclude short-distance modes from the start:
\begin{itemize}
\item In practice it is more convenient to vary $\Lambda$ once a model space with $\Lambda_\text{RGM}$ is chosen.
\item It is unclear how a certain $\Lambda_\text{RGM}$, imposed in the $A=2$ sector, is consistently translated to $A>2$ systems, \textit{e.g.}, what
triton model space is to be used given a deuteron model space to yield the same $\Lambda_\text{RGM}$? Therefore, a change in the renormalization
scheme is expected from $A=2$ to any larger system. Harmless in the case of an approximately complete basis, a smaller model space could result
in a significant dependence of the LECs on $\Lambda_\text{RGM}$ in addition to that on $\Lambda$. Hence, LECs determined from $A=2,3$ data and
used to predict $A=4$ observables would falsely suggest a failure of the power counting because of the change in $\Lambda_\text{RGM}$.
\end{itemize}
The two-nucleon model space, as specified in table~\ref{tab_deuteron-ms}, consists of a set of $20$ width parameters and a pure S-wave ($l_1=0$) for
the LO and perturbative NLO calculation, while admixtures of D-waves ($l_1=2$) are possible in the nonperturbative NLO approach.
\begin{table*}
\renewcommand{\arraystretch}{1.5}
  \caption{\label{tab_deuteron-ms}{\small Two-nucleon RGM basis used for the determination of LECs and predictions. The values were optimized for a
potential model calculation and the basis is labeled $w120$. $w12$ refers to a basis with widths $\tilde{\gamma}_1(j)=0.1\cdot\gamma_1(j)$. The
notation is defined in eq.~(\ref{eq_rgm-int-wfkt}). The deuteron state is specified by spin $s=1$, a singlet deuteron by $s=0$, while the neutron-neutron and
proton-proton systems have $s=0$, too. The total isospin is fixed by: $l+s+t\stackrel{!}{=}~$odd.}}
\footnotesize
\begin{tabular}{l|ll|c}
\hline
&widths $[\text{fm}^{-2}]$&&dim\\
\hline\hline
LO,pNLO&
$\gamma_1(j)\in\mathcal{W}_{120}=\left\lbrace
\begin{array}{l}
129.567,51.3467,29.4729,13.42339,8.214456,\\
4.447413,2.939,1.6901745,1.185236,0.84300,\\
0.50011,0.257369,0.13852,0.071429,0.03852,\\
0.01857,0.009726,0.005619,0.00277,0.00101
\end{array}\right\rbrace$&
$l_1(j)=0\;\;\;\forall j$&$20$\\
nNLO&$\gamma_1(j)\in\mathcal{W}_{120}$&$l_1(j)=\Bigg\lbrace\begin{array}{l}0\;\;;\;\;j\leq 20\\2\;\;;\;\;j>20\end{array}
$&40\\
\hline
    \end{tabular}
\end{table*}
The small values for $\gamma_1$ are included to expand asymptotic states $h_l^\pm$ (eq.~(\ref{eq_asymp-wfkt})) while the large widths are needed for the bound
state wave functions, for which structure at distances $\propto\Lambda^{-2}$ is relevant. The LO wave function is subject to two boundary conditions: a rapid decrease
towards zero, $r\to 0$, starting at $|\vec{r}|\approx\Lambda^{-2}$ and a certain asymptotic decay for $|\vec{r}|\gg M^{-1}$, independent of $\Lambda$.
With a basis like $w120$, the second condition can be met
for binding energies of the order of the physical $B(d)$, but any such basis, with a maximal width $\gamma_\text{max}$, will not be able to simultaneously
reproduce the falloff for $\Lambda\gtrsim\sqrt{\gamma_\text{max}}$ (eq.~(\ref{eq_pot-lo-coord})). This last assertion is verified by checking the consistent prediction for the deuteron
binding energy in two calculations:
\newpage
\begin{enumerate}
\item The LO LEC in the deuteron channel, $C_T=C_1^\text{LO}+C_2^\text{LO}$, is fitted to $a_t$ with a standard numerical integration, using \texttt{Mathematica}
and not the RGM, of the Schr\"odinger equation for $M<\Lambda\to\infty$.
\item With this $C_T$ and the RGM model spaces $w12$ and $w120$, an RGM deuteron binding energy $B(d,\Lambda)$ is calculated.
\end{enumerate}
Figure~\ref{fig_bd-cutoff} shows consistent predictions for $B(d)$ in $w12$ and $w120$ up to a threshold value $\Lambda_t^{w12}\approx 1.4~$GeV. For
$\Lambda\gtrsim\Lambda_t^{w12}$, $w12$ fails in expanding a bound state, while $w120$ does so above $\Lambda_t^{w120}\approx 3.6~$GeV.
Furthermore, the predicted $B(d)$ converges as expected for $\Lambda\to\infty$ to the analytical prediction given by eq.~(\ref{eq_lo-bd-at}).
\par
\begin{figure}
\includegraphics[width=0.6\columnwidth]{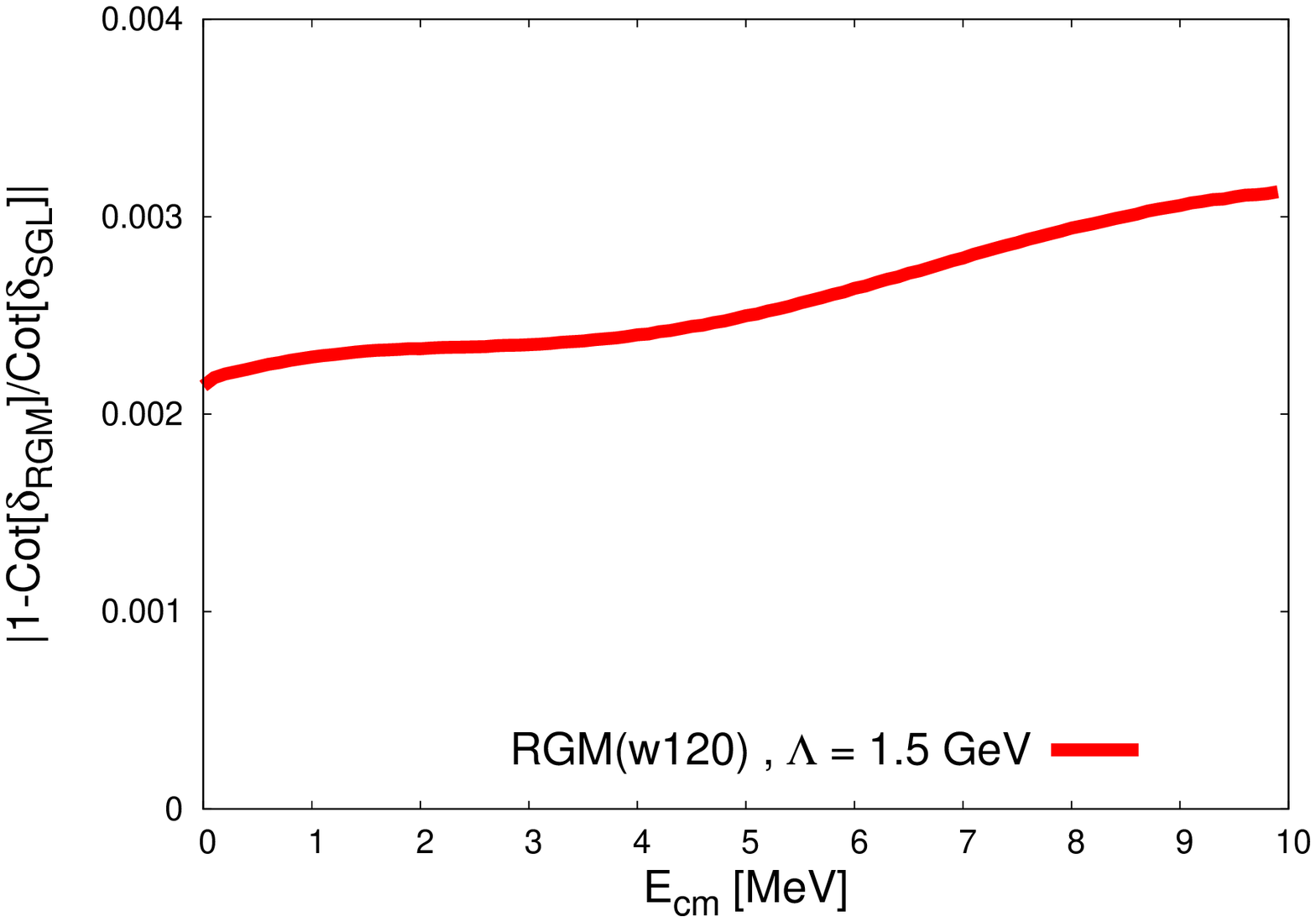}
  \caption{\label{fig_sgl-kett}\small Relative deviation between the $^3S_1$ np phase shift results obtained
with the RGM and a numerical integration of the Schr\"odinger equation. A leading-order potential with one parameter fit to $a_t$ and a
cutoff $\Lambda=1.5~$GeV was used. Notice the scale on the ordinate.}
\end{figure}

The accurate expansion of scattering states for leading-order potentials and the converged variational determination of scattering phase shifts is demonstrated
by comparison to predictions by a direct integration of the Schr\"odinger equation with \texttt{Mathematica} for the neutron-proton $^3S_1$-phase shifts in
fig.~\ref{fig_sgl-kett}.
\par
I conclude from these analyes that the dependence of two-body LECs, obtained with a basis $w120$ for $\Lambda\lesssim 2.5~$GeV, on $\Lambda_\text{RGM}$
is insignificant relative to the systematic uncertainty of a NLO calculation with EFT($\slashed{\pi}$). With this basis and the limitation on the cutoff, the RGM is
an accurate numerical method for $A=2$. For the next larger system, the triton, the basis $w120$ is used to expand the deuteron fragment
and a part of the width set $\mathcal{W}_{120}$ to expand the radial function of the second neutron's coordinate in order to translate
this accuracy to the $A=3$ system.
\newpage
\section{A=3}\label{sec_a3}
While two-nucleon observables at momenta $p\ll\frac{1}{|a_{s,t}|}$ are parameterized accurately by the two scattering lengths only, an additional parameter is
needed for the three-nucleon amplitude (see discussion of eq.~(\ref{eq_efimov-ratio})). The dependence of observables on the renormalization parameter - manifest,
\textit{e.g.}, in the Phillips line, the universal ratio eq.~(\ref{eq_efimov-ratio}), or the Thomas effect - expresses this feature of the nuclear interaction. The estimate for the
expansion parameter in eq.~(\ref{eq_exp-para-est}) would be false if a similar promotion of a higher dimensional operator occurs - \textit{i.e.}, a LEC violating
eq.~(\ref{eq_naturalness}) with $\delta >8$. The cutoff dependence of three three-body observables is investigated here at LO and NLO. The single
additional parameter allows for a tangible graphical representation of this dependence by plotting the prediction for an observable as a function of another for
varying renormalization parameters. A renormalized order $n$ interaction corresponds to a locus of points bounded by a rectangle around the datum. The
size of this bounding box resembles the theoretical uncertainty with respect to different unresolved microscopic structures and relating these dimensions
at successive orders assesses the expansion parameter of the effective field theory. This is how the na\"ive estimate of eq.~(\ref{eq_exp-para-est}) is tested here.
All three-body observables were obtained with the RGM with variational basis defined in table~\ref{tab_t-ms}. In addition, we show in app.~\ref{app_rgm-wfkt} with the example
of a smaller basis explicit formulas for the triton trial wave function.
\begin{table*}
\renewcommand{\arraystretch}{1.5}
\caption{\label{tab_t-ms}{\small Triton and 3-helium RGM model spaces. Quantum numbers, which are not specified, run over all values compatible with the
total angular momentum and parity, $J^\pi=\frac{1}{2}^+$, and coupling rules, \textit{e.g.}, the $l_{1,2}=2$ set contains basis vectors with total orbital angular
momentum $L=0,1,2$. The nNLO basis includes all possible combinations of $l_1$, $l_2$, and $s$.
The sets $\mathcal{W}^{(1,2)}_{120}$ are subsets of $\mathcal{W}_{120}$ containing every other value starting with the first (second) width.}}
\footnotesize
\begin{tabular}{l|ccc|c|c}
\hline
&$l_1$&$l_2$&$s$&width sets&dim\\
\hline\hline
LO,pNLO&$0$&$0$&$1$&$\gamma_{2}(j)\in\mathcal{W}_{120}$,$\gamma_{1}(j)\in\mathcal{W}^{(1)}_{120}$&200\\
&$0$&$0$&$0$&$\gamma_{2}(j)\in\mathcal{W}_{120}$,$\gamma_{1}(j)\in\mathcal{W}^{(2)}_{120}$&200\\
nNLO&$\lbrace 0,1,2\rbrace$&$\lbrace 0,1,2\rbrace$&$\lbrace 0,1\rbrace$&$\gamma_{1,2}$ given in \cite{mythesis}&220\\
\hline
    \end{tabular}
\end{table*}
\subsection{Triton charge radius}\label{subsec_r3-ch}
\begin{figure}
\includegraphics[width=.75\textwidth]{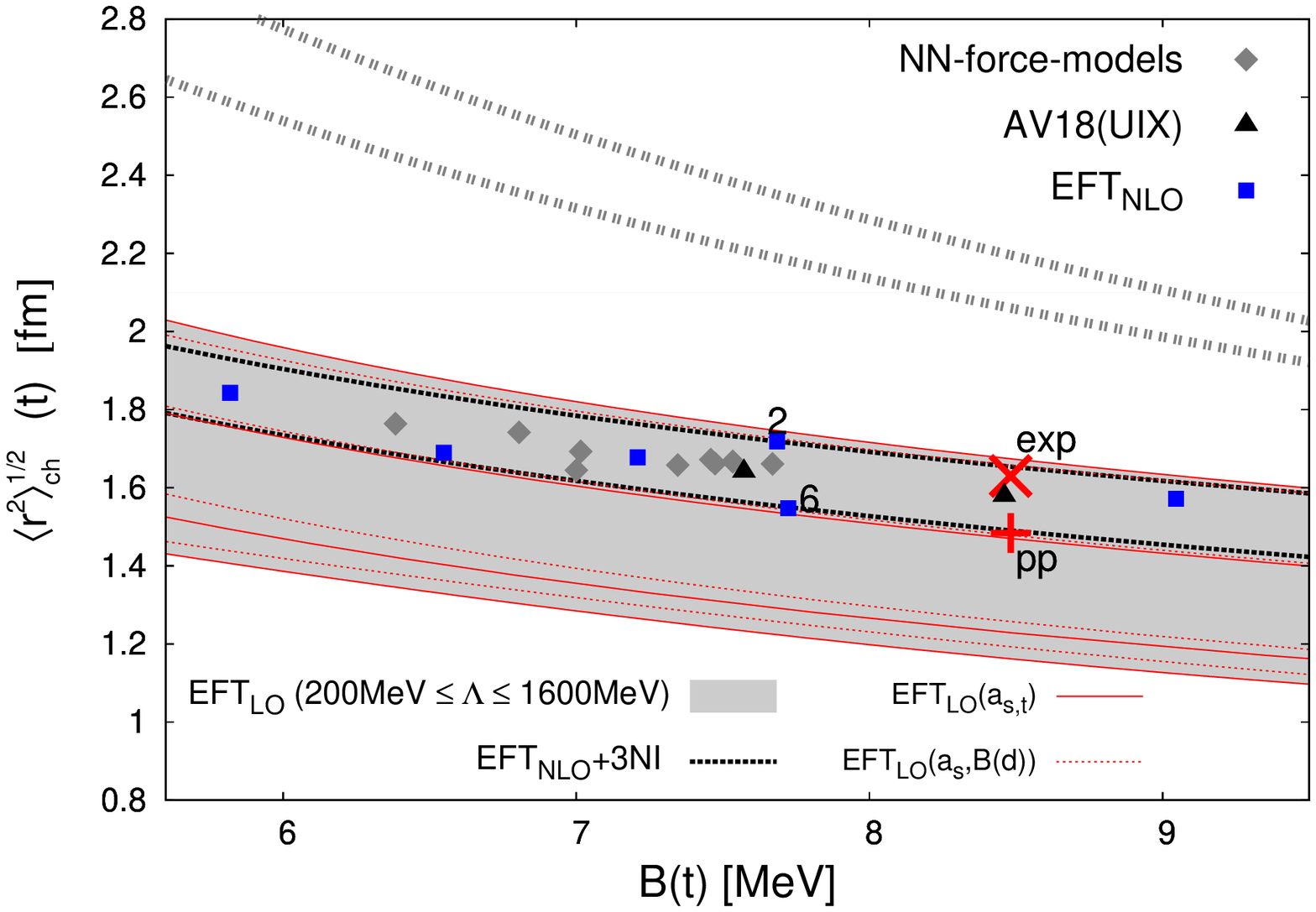}
  \caption{\label{fig_rch-t}\small The correlation between the triton charge radius and its binding energy.
The narrower band mapped out by the EFT($\slashed{\pi}$) NLO results from the RGM (blue squares, black dashed lines) as described in the text,
compared to our RGM LO results (gray shaded area) and a LO calculation by~\cite{platter-rch} (dashed gray lines), demonstrates convergence,
also to the datum (e\textcolor{red}{x}p. value) and a customary finite-nucleon-size-correction of it (pp).
The values from \texttt{AV18(+UIX)} (black triangles, RGM calculation for this work) and a variety of other
potential models (gray diamonds~\cite{rch-pots-1,rch-pots-2}) are consistent with our proposed EFT($\slashed{\pi}$) NLO correlation band.}
\end{figure}

The variational RGM approximates the ground-state wave function $\langle\vec{r}\vert t\rangle$ of the triton as the eigenfunction corresponding to $B(t)$. Hence,
a numerically complete variational basis and therefore a converged $B(t)$ transients into convergence for all other observables derived from the ground-state
wave function. In particular, we consider the average charge radius which is calculated from the wave function using:
\begin{equation}\label{eq_rch-t}
\langle r^2\rangle_{\text{\tiny ch}}^{1/2}=\left(\langle t|\sum_{i=1}^3\frac{1}{2}\vec{r}^2_i\left(1+\tau^3_{i}\right)|t\rangle\right)^{\frac{1}{2}}\;\;\;,
\end{equation}
where $\vec{r}_i$ is the position and $(1+\tau^3_{i})$ the charge operator of the $i$-th nucleon. An analysis of the correlation between $B(t)$ and
$\langle r^2\rangle_{\text{\tiny ch}}^{1/2}$, as motivated above, will provide information about the convergence properties of EFT($\slashed{\pi}$), and is
presented in fig.~\ref{fig_rch-t}.
Predictions are shown for the LO and nNLO EFT($\slashed{\pi}$) interactions of this work, a LO EFT($\slashed{\pi}$) calculation$^\text{\cite{platter-rch}}$
extracting observables by solving Faddeev integral equations, various nuclear force models$^\text{\cite{rch-pots-1,rch-pots-2}}$, and two RGM calculations
for the \texttt{AV18}$^\text{\cite{av18}}$ two-nucleon- and the \texttt{AV18+UIX}$^\text{\cite{uix}}$ two- and three-nucleon interaction model.
For the discussion of this graph, I define two terms:
\begin{description}
\item[correlation line:]$\left\lbrace\left(\,B(t)\left(C_{3\text{NI}}\right),o\left(C_{3\text{NI}}\right)\,\right)\,,\,-\infty <C_{3\text{NI}}<\infty\,\right\rbrace\;\;\;,$\\
\textit{i.e.}, fixed $\Lambda$, while $\Lambda_\text{RGM}$ needs to be adjusted for shallow or deep bound states relative to the state at $C_{3\text{NI}}=0$.
In this section, $\langle r^2\rangle_{\text{\tiny ch}}^{1/2}$ is considered as the correlated observable $o$.
\item[order $n$ correlation band:]$\left\lbrace\left(\,B(t)\left(\Lambda,\Lambda_\text{RGM},C_{3\text{NI}},\mathcal{D}'\right),
o\left(\Lambda,\Lambda_\text{RGM},C_{3\text{NI}},\mathcal{D}'\right)\,\right)\,\right\rbrace\;\;\;,$\\
\textit{i.e.}, all renormalization parameters $\Lambda,\Lambda_\text{RGM},C_{3\text{NI}}$ and conditions at $n$-th order, $\mathcal{D}'$, accessible within this RGM approach
are varied.
\end{description}
The EFT formalism identifies the band as the area in which the predictions of any two-body interaction - parameterized by $a_{s,t}$ or $a_s,B(d)$ - are predicted
to be. In that sense, the bands are the universal three-body consequences of specific two-body interactions. By definition, the bands are unbounded, resembling the
nonperturbative dependence on the renormalization parameters. Renormalizing the theory with an additional constraint on the three-body spectrum, enforcing a ``triton''
energy in an interval around the datum of width set by the uncertainty of the considered order, automatically constrains all other low-energy observables derived
from the wave function corresponding to this bound state, or scattering observables with one nucleon impinging on the deuteron.
\par
Turning now to the results in fig.~\ref{fig_rch-t}, one first confirms the expected behavior of an increasing $\langle r^2\rangle_{\text{\tiny ch}}^{1/2}$
for more loosely bound systems. Results of two EFT($\slashed{\pi}$) LO calculations are shown.
The RGM LO potentials map out a band (gray shaded area) which includes the datum. In addition to the experimental value,
$\langle r^2\rangle_{\text{\tiny pp}}^{1/2}=\left(\langle r^2\rangle_{\text{\tiny ch}}-\langle R^2_{\text{\tiny p}}\rangle-\frac{3}{4m_\text{\tiny N}^2}-\frac{N}{Z}\langle R_n^2\rangle\right)^{1/2}$
is given as a measure which takes into account the finite extension of the nucleons$^\text{\cite{r2-n-p}}$,
$\langle R^2_{\text{\tiny p}}\rangle=0.769(12)~\text{fm}^2$ and $\langle R^2_{\text{\tiny n}}\rangle=-0.1161(22)~\text{fm}^2$, plus the Darwin-Foldy term$^\text{\cite{darwin-foldy}}$
(red pp cross in fig.~\ref{fig_rch-t}).
The corrections constitute a partial removal of higher-order effects. Hence, it has to be an element of the LO band. The important reference point here is the datum itself,
because all finite-size and relativistic corrections are expected to be included order by order in the EFT expansion.
\par
The lower and upper LO line from~\cite{platter-rch}  (dashed gray lines in fig.~\ref{fig_rch-t}) result from a fit of the LO LECs $C_{1,2}^\text{LO}$
to either $a_{s,t}$ or $B(d),a_s$, respectively, providing some measure of higher-order effects. For the RGM LO calculation, we used $a_{s,t}$ (red solid lines)
and $B(d),a_s$ (red dotted lines, with the closest solid red line corresponding to the same $\Lambda$) to determine the LECs,
and varied the cutoff from $200~\text{MeV}$ (top edge) to $1.6~\text{GeV}$ (lower edge) in steps of $100~\text{MeV}$. The dependence on $\mathcal{D}'$, \text{i.e.},
using $B(d)$ or $a_t$ as input, measured in terms of relative shift of the correlation line, becomes comparable to that observed by doubling $\Lambda=800~$MeV
to $1.6~$GeV
but is found negligible for a doubling of $\Lambda=200~$MeV. We choose the lower bound for $\Lambda$ to be in the region of the pion mass. The upper
bound is set by the observation that the band width is increased little by increasing $\Lambda$ from $800~$MeV to $1.6~$GeV compared to the broadening
from $200~$MeV to $400~$MeV. The band appears to be saturated, with the bulk
of its width coming from the region $\Lambda\in[200;800]~$MeV.  For fixed
$\Lambda$, $C_{3\text{NI}}^\text{LO}$ was varied to generate the correlation lines.
\par
The quasi-exact Faddeev calculation of~\cite{platter-rch} and the RGM
results do not overlap. With fig.~\ref{fig_bd-cutoff} \& \ref{fig_sgl-kett} I demonstrated that numerical inaccuracies
of the RGM are negligible, and conjecture that the non-overlap can be traced to the
differences between the regularization schemes employed. The Faddeev
calculation uses the separable cutoff function $f_\kappa(\vec{p},\vec{p}')=\text{exp}\left(-\vec{p}^2/\kappa^2\right)\text{exp}\left(-\vec{p}'^2/\kappa^2\right)$
with $\kappa\geq 1600~\text{MeV}$, while the RGM uses
the non-separable regulator of eq.~(\ref{eq_reg-fkt-mom}) plus an implicit regulator imposed by
the finite number of width parameters. Therefore, both methods combined can be
viewed as providing a check of residual regularization-scheme dependence.  A
conservative estimate of LO effects is thus the range of results covared by the
combination of both methods. The LO accuracy at the physical triton binding
energy is thus $\pm0.6~\text{fm}$, and the measured charge radius happens to lie
right in the middle of the LO band. Additional regularization schemes,
\textit{e.g.}, with a range of cutoff values in the Faddeev approach, are thus
conjectured to lead to an overlap between the two bands.
\par
At NLO, the potentials without 3NI map out a correlation band which is more
narrow than its LO counterpart, and again contains the datum. The width of the
band can be estimated by considering the difference between the results using
the extreme cases provided by potentials $^2V_\slashed{\pi}$ and $^6V_\slashed{\pi}$.
They produce the same NN scattering lengths and nearly the same triton binding energy, and their deuteron
binding energies differ by less than $10$\%, while their triton charge radii differ by
about $10$\%, consistent with the expectation of a NLO calculation. Finally, they
are based on two significantly different cutoff values, $\Lambda(^2V_\slashed{\pi})\approx
400~$MeV and $\Lambda(^6V_\slashed{\pi})\approx 650~$MeV. We therefore can base an estimate of
the NLO band on the range mapped out by varying the 3NI for these two
potentials. These lines are included in fig.~\ref{fig_rch-t}.
\par
Variation of the cutoff $\Lambda$, of the fitting-input $\mathcal{D}'$, and of the 3NI lead therefore all
to similar assessments of the uncertainty of the theory at NLO. At fixed $B(t)$,
the charge radius varies from LO to NLO by $\lesssim 30$\%, in agreement with the
power counting which estimates the correction according to eq.~(\ref{eq_exp-para-est}) to be approximately $\frac{1}{3}$.
\par
All three values, namely, $\pm 0.2~$fm from the na\"ive estimate $Q\approx\frac{1}{3}$
and the observed convergence from LO to NLO, and $\pm 0.1~$fm from the above mentioned
difference between $^2V_\slashed{\pi}$ and $^6V_\slashed{\pi}$, would be equally valid
estimates for the theoretical uncertainty of this NLO calculation.
Using the more conservative error estimate and the assumption for the NLO error band center
at $1.6~$fm at the experimental $B(t)$, EFT($\slashed{\pi}$) predicts a value
\begin{equation}\label{eq_nlo-rch}
\langle r^2\rangle^{1/2}_{\text{\tiny ch}}(t,\text{NLO})=(1.6\pm 0.2)~\text{fm}
\end{equation}
within the band of the leading-order value $\langle r^2\rangle^{1/2}_{\text{\tiny ch}}(t,\text{LO})=(2.1\pm 0.6)~$fm as quoted from~\cite{platter-rch}.
The NLO value is found in good agreement with experiment$^\text{\cite{exp-t_rch}}$, $\langle r^2\rangle^{1/2}_{\text{\tiny ch}}(t,\text{exp})=(1.63\pm 0.03)~$fm.
\par
Another argument in favor of our definition of the correlation band is provided by the
results of the two phenomenological models, \texttt{AV18}$^\text{\cite{av18}}$ and \texttt{AV18+UIX}$^\text{\cite{uix}}$. In general, EFT predicts that
the results of a potential which reproduces or shares input observables, at least to the accuracy
required at the considered order, deviate from the results of an appropriate EFT
potential by less than the theoretical uncertainty of the EFT values at this order in the
applicability range of the EFT.
This criterion is easily met by \texttt{AV18(+UIX)}, and hence its predictions have to be
consistent with the proposed correlation band. The two-body potential \texttt{AV18}, reproducing the
Nijmegen phases much more accurately than required to fall into this category of potentials,
is expected to yield a value at a position within the $(\langle r^2\rangle_{\text{\tiny ch}}(t)-B(t))$-band. The prediction
for the triton charge radius of \texttt{AV18+UIX} is expected to deviate less than $10$\% from
the experimental datum within the error band, because this model has a three-body interaction added to reproduce the experimental $B(t)$.
Both expectations are consistent with the results shown in fig.~\ref{fig_rch-t}.
\par
In conclusion, the results for these three-nucleon observables show that although the
potentials are approximately NN phase-shift equivalent, they differ
in their predictions of three-body observables. As mentioned above,
a proper renormalization of the theory requires therefore one three-nucleon contact interaction.
Setting this 3NI to zero in the potentials $^{1-9}V_\slashed{\pi}$,
the expected dependency  of observables in $A>2$ systems on how the unobservable short-distance physics is modeled is observed.
Different short-distance physics is modeled by the potentials not only by varying cutoff values but also by
differing sets of LECs for the same cutoff, while the scheme- and
regulator-dependent three-body interaction parameter is chosen as
zero. With this parameter fitted to the triton binding energy, the prediction for the triton charge radius is consistent
with experiment within the expected uncertainty range. A significant convergence from LO to NLO is observed.
\subsection{Trinucleon binding energy splitting}\label{subsec_tri-nucl-split}
\begin{figure}
\includegraphics[width=.7\textwidth]{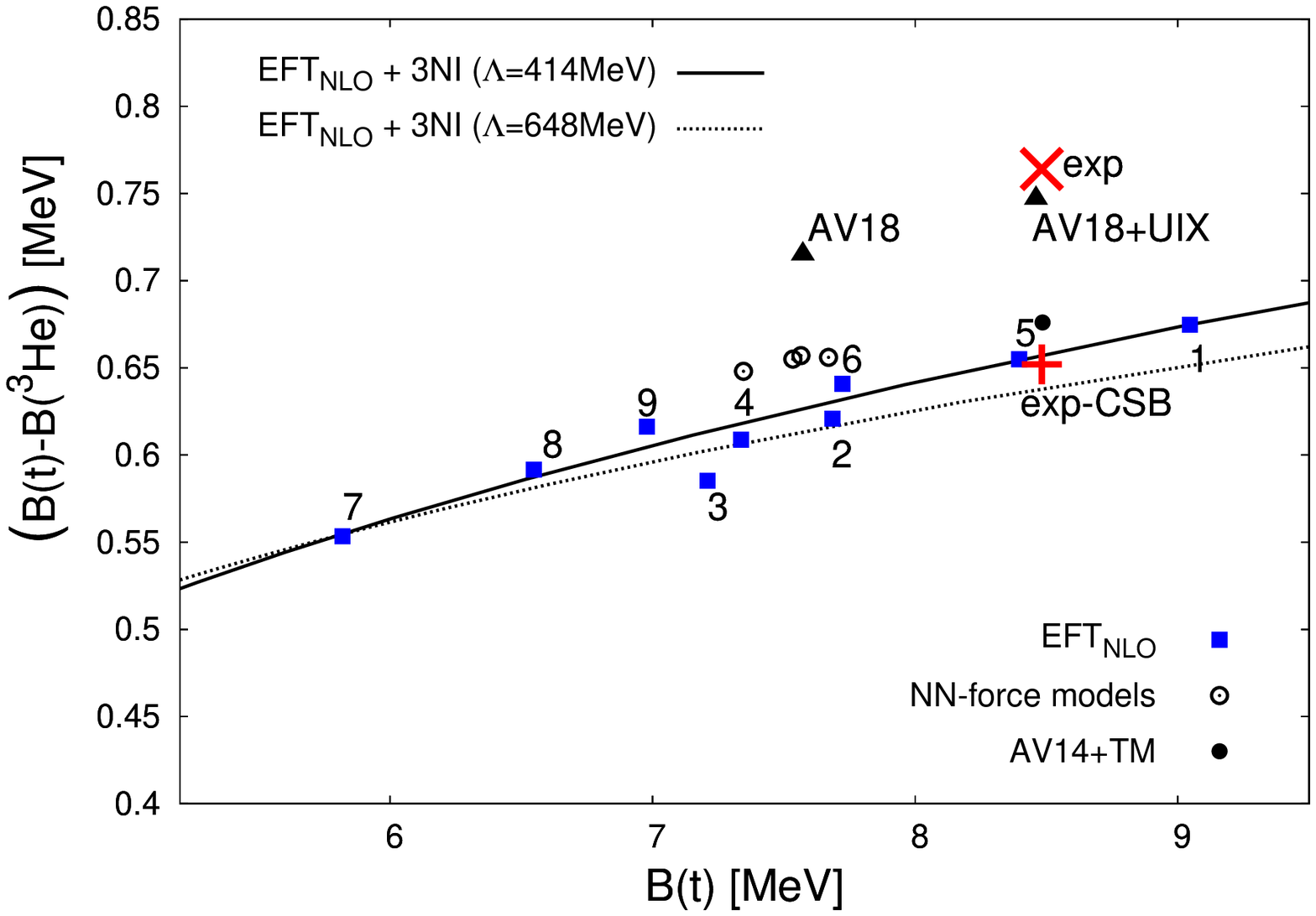}
\caption{\label{fig_trinucl-splitt}\small Binding energy difference between $^3$He
  and $^3$H for EFT($\slashed{\pi}$) potentials (filled squares) compared to
  various isospin-invariant potential models
  calculations$^\text{\cite{rch-pots-2,pest-pot}}$ (circles) and RGM values for
  \texttt{AV18(+UIX)} (triangles) which contain charge symmetry breaking terms.
  For the potentials $^{1}V_\slashed{\pi}$ (solid) and $^{7}V_\slashed{\pi}$ (dashed),
  a smooth variation of the 3NI leads to the two correlation lines.
  The upright cross is the experimental value without the contribution from CSB
  terms$^\text{ \cite{csb-tni}}$.}
\end{figure}
In fig.~\ref{fig_trinucl-splitt}, the RGM results for the splitting between
the binding energies of \mbox{3-helium} and the triton are shown for the potentials
$^{1-9}V_\slashed{\pi}$. At NLO in the pion-less EFT, the strong interaction
is isoscalar and hence does not break charge symmetry.  Therefore, charge
symmetry breaking (CSB) comes in our RGM calculation only from including the
Coulomb interaction between the protons in $^3$He. The EFT($\slashed{\pi}$)
results, hence, only show the model-independent contribution of Coulomb interactions
to the trinucleon binding energy splitting.
The correlation band is mapped out by the EFT($\slashed{\pi}$) potentials
with zero 3NI and by smoothly varying this
three-body parameter for potentials
$^{1,7}V_\slashed{\pi}$. Both approaches result in correlations consistent with each other
and with the results of the other NLO potentials.
At the experimental triton binding energy, this leads to predicting
\begin{equation}\label{eq_nlo-splitt}
  \left(B(t)-B({^3\text{\small He}})\right)=\left( 0.66\pm0.03\right)~\text{MeV}\;\;\;.
\end{equation}
Here, we estimate the theoretical uncertainty by the spread of the
phase-equivalent NN potentials as in the previous section, e.g.~comparing
$^{3,9}V_\slashed{\pi}$. The \textit{a priori} error estimate at NLO of
$\lesssim10$\% gives a larger uncertainty of $\pm0.07~\text{MeV}$. However,
one should keep in mind that including isoscalar strong interactions by
higher-order terms of the effective-range expansion has identical effects on
the strong interactions inside the triton and $^3$He. These effects cancel
out in the difference and only survive indirectly, as the strength of the
Coulomb interaction in a system is also correlated to its size. We therefore
quote the width of the correlation band as the error estimate of our calculation.
\par
This value deviates by about $0.1~$MeV from the experimental value$^\text{\cite{opper-csb}}$ of
$0.764~$MeV. In line with the argumentation
above, we attribute this difference to isospin-breaking charge-independence breaking (CIB) and CSB interactions
coming from the explicitly broken chiral symmetry in the strong and
electro-weak sector from which only the parts resulting in the Coulomb force
have been considered in this calculation. They enter in EFT($\slashed{\pi})$
only at higher order. To support this assertion, results of the potential
models \texttt{AV18(+UIX)}, which contain CSB interactions, are included in
fig.~\ref{fig_trinucl-splitt}.  Both potentials are not elements of the
correlation band suggested by the EFT($\slashed{\pi}$) points but agree with a
shifted band, centered around the datum. In contrast, the values from charge-symmetric
potential models in fig.~\ref{fig_trinucl-splitt} lie within the NLO
EFT($\slashed{\pi}$) band. This leads us to predict a model-independent CSB/CIB
contribution to the binding difference in NLO EFT($\slashed{\pi}$) at the
experimental triton binding energy of
\begin{equation}\label{eq_nlo-CSB}
  \left(B(t)-B({^3\text{\small He}})\right)^\text{CSB/CIB}=\left( 0.10\mp0.03\right)~\text{MeV}\;\;\;,
\end{equation}
anti-correlated with the Coulomb contribution to give the experimentally
established difference. This is to be compared with the contributions from
two- and three-nucleon CSB interactions which stem from Chiral Effective Field
Theory, Breit and vacuum polarization corrections, and from corrections to the
kinetic energy operator. These were calculated$^\text{\cite{csb-tni}}$ to sum
up to $(0.112\mp0.022)~$MeV, leaving about $(0.652\pm0.022)~$MeV for the soft
photon effects, dominated by the Coulomb interaction. This is in perfect
agreement with the EFT($\slashed{\pi})$ result.
\subsection{Neutron-deuteron scattering}\label{subsec_nd-phase}
The following analysis of the neutron-deuteron (n-d) scattering system in the triton (total spin-$\frac{1}{2}$) channel serves as the first example of a scattering calculation
in the RGM framework with leading-order EFT($\slashed{\pi}$) potentials. The RGM results and their comparison to previous calculations motivate the peculiar choice for the
input data $\mathcal{D}'$ for the determination of the pNLO coefficients (see ch.~\ref{sec_pots}).
\par
The effective range expansion, as introduced in eq.~(\ref{eq_ere}), expands the amplitude around zero momentum and yields a deuteron binding energy which differs
by about $0.6~$MeV from the datum at LO. Effective range corrections to $B(d)$ are then large relative to the ones in an expansion around the deuteron binding momentum
$\gamma_t$. The latter expansion, by construction, reproduces the experimental $B(d)$ already at LO. In \cite{hgrie-zpara} it was shown how the convergence rate of EFT($\slashed{\pi}$)
renormalized by a matching to the Z-parameterized two-nucleon amplitude is improved for three-nucleon scattering observables compared to an equally admissible
theory with LECs fitted to the ERE amplitude. The former parameterization uses the deuteron pole position $\gamma_t$ and its residue as input, while the latter employs
$a_t$ and $r_t$. The intuitive expectation that accuracy in the reproduction of the tail of the deuteron wave function transients to a better description of low-energy
elastic neutron scattering off the deuteron was confirmed$^\text{\cite{hgrie-zpara}}$ and is demonstrated by the faster convergence of EFT($\slashed{\pi}$) from LO to NLO.
Na\"ively, I expect a faster convergence in the $^6$He system with all of its subsystems bound close to data in analogy. Following this intuition, $B(d,t,\alpha)$ are used as input and not, \textit{e.g.}, effective ranges from the two-nucleon sector. The improved accuracy is also expected in low-energy $\alpha$-n scattering
if $B(\alpha,\text{exp})$, \textit{i.e.}, the target, is reproduced accurately at LO with respect to its binding energy. Important differences between the n-d and n-$\alpha$ reactions which
could account for a different result are:
\begin{itemize}
\item The $\alpha$ state is bound deeply below the lowest breakup threshold compared to the deuteron, and relative to the low-momentum scale of EFT($\slashed{\pi}$).
Neutron scattering on an $\alpha$ target should therefore be independent of breakup effects over a larger range of energies.
\item The $\alpha$'s coordinate-space wave function is not totally symmetric due to a reduction of the RGM basis, while the model space for the n-d
reaction was sufficiently large to assume the symmetry of the deuteron's wave function.
\end{itemize}
\par
In this section, the findings of \cite{hgrie-zpara,nd-ham-meh} for the three-nucleon system are confirmed in the sense that LO EFT($\slashed{\pi}$) predictions for n-d
phase shifts with the RGM are found to be more accurate, \textit{i.e.}, less sensitive to cutoff and input variations, for interactions that reproduce $B(d,\text{exp})$. Specifically, this is
shown for the correlation between the binding energy and the n-d scattering length in the triton channel - this is known as the Phillips line and was discovered by
\cite{phil-line} and explained by \cite{efimov-line} - and
the n-d S-wave phase shifts below the deuteron breakup threshold. The analysis extends the existing EFT($\slashed{\pi}$) (N)LO calculations insofar as it provides a
more detailed assessment of the uncertainty at LO. In contrast to \cite{hgrie-zpara}, the LECs are determined here from the ERE as given in eq.~(\ref{eq_ere}),
\textit{i.e.}, an expansion around zero momentum.
\par
\begin{figure}
\includegraphics[width=\columnwidth]{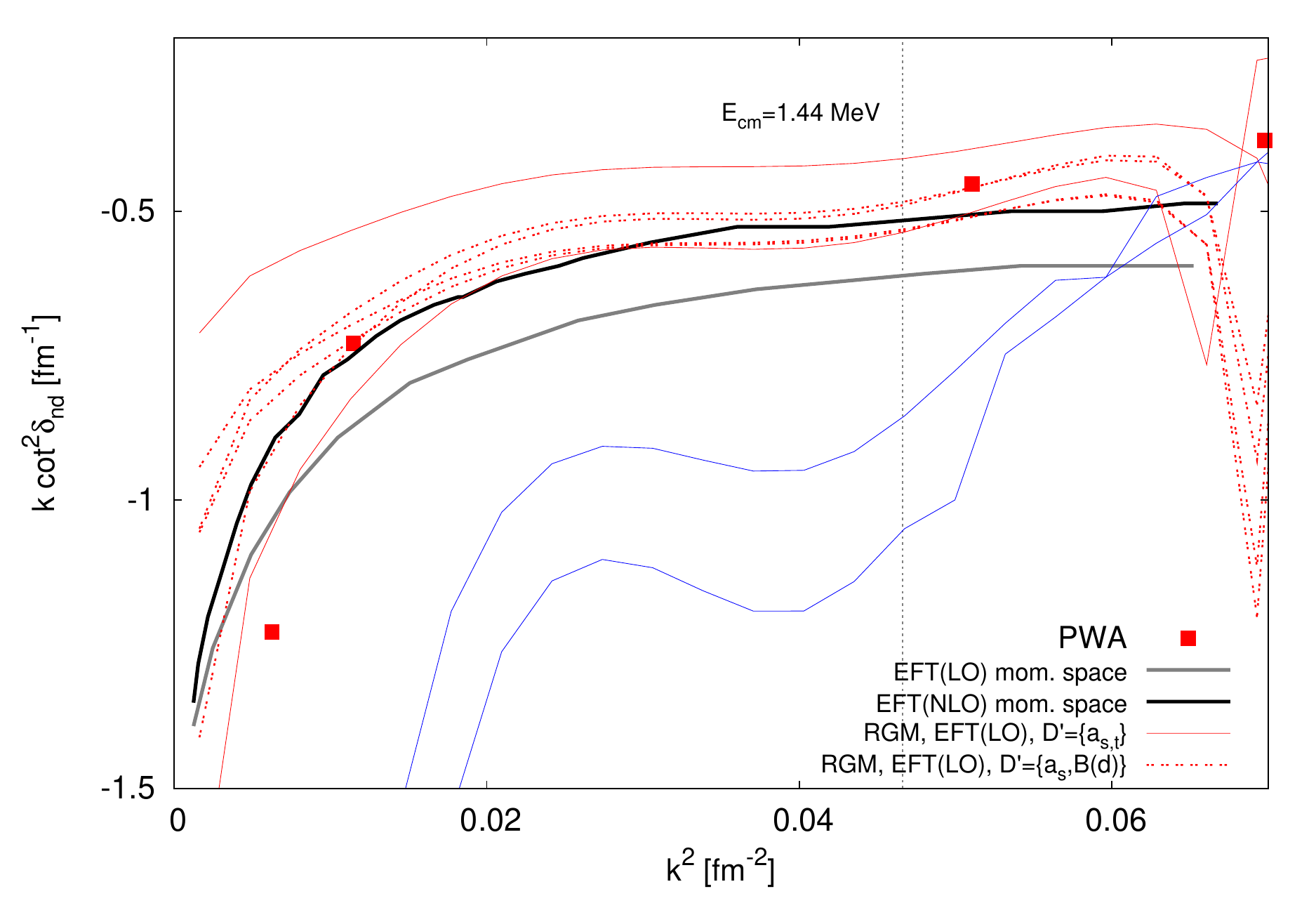}
\caption{\label{fig_d-n-ph}\small
Results for the Doublet S-wave phase shift $\delta_\text{\tiny nd}$ for neutron-deuteron scattering expressed in $k\cot^2\delta_\text{\tiny nd}$ with center of mass momentum $k$.
The vertical line marks the deuteron breakup threshold at the LO $B(d)$. The RGM LO predictions (red and blue lines) are compared to a phase shift analysis$^\text{\cite{nd-exp}}$
and LO (NLO) momentum space calculations$^\text{\cite{nd-ham-meh}}$ (gray (black) line). Predictions shown as solid lines use $a_t$ instead of $B(d)$ (dashed) as input.
The blue lines highlight the specific RGM results for $\Lambda=800,1600~$MeV.}
\end{figure}

First, I discuss the n-d phase shifts in the spin-$\frac{1}{2}$ channel shown in fig.~\ref{fig_d-n-ph}.
The RGM model space uses the $w120$ basis (\ref{tab_deuteron-ms}) for the deuteron fragment and the singlet-deuteron and nn distortion channels.
The width set $\mathcal{W}_{120}$ defines the basis for the radial fragment-relative function.
RGM results are displayed for LO EFT($\slashed{\pi}$) potentials with $200~\text{MeV}\leq\Lambda\leq1.6~\text{GeV}$
(blue and red lines). Two sets of input data $\mathcal{D}'$ are used for each $\Lambda$ to determine the three LO LECs (eq.~(\ref{eq_pot-coord-lo}) \& (\ref{eq_tni-pot-coord})):
$\mathcal{D}'_1=\lbrace a_{s,t},B(t)\rbrace$  and $\mathcal{D}'_2=\lbrace a_{s},B(d),B(t)\rbrace$. Values for $B(d)$ of potentials fitted to $\mathcal{D}'_1$ range from
$\sim 2.5~$MeV ($\Lambda=200~$MeV) to $\sim 1.5~$MeV ($\Lambda=1600~$MeV). The predicted phase shifts as shown in fig.~\ref{fig_d-n-ph} are found much more
accurate for potentials whose LECs are either determined via $\mathcal{D}'_2$ or their $\Lambda$ is such that the experimental $B(d)$
is approximately reproduced even if $\mathcal{D}'_1$ serves as input data.
\par
In general, the uncertainty at a given order is assessed by the sensitivity of predictions to modification of unobservable short-distance physics. Here,
if a change in the short-distance structure is induced solely by a change in $\Lambda$, the uncertainty in $\delta_\text{\tiny nd}$, measured by the width of the band defined by
phases whose corresponding potentials use $B(d)$ as input (red dashed lines), is smaller compared to the uncertainty found if short distance structure is also varied by
using differrent input data, namely $\mathcal{D}'_1$. The phases $\delta_\text{\tiny nd}$ deviate most from the phase-shift analysis data (red squares) for potentials which
use $\mathcal{D}'_1$ and whose cutoff is large, \textit{i.e.}, $\Lambda=800,1600~$MeV (blue lines). The LO uncertainty of EFT($\slashed{\pi}$) is estimated by the width
of the band bounded by the predictions of the potential with $\Lambda=1600~$MeV (lower blue line) and that with $\Lambda=200~$MeV (top solid red line). This band is not
expected to widen any further, because, first, the maximal deviation from $B(d,\text{exp})$ is almost reached with $\Lambda=1.6~$GeV
(see discussion of fig.~\ref{fig_bd-cutoff}). Second, lowering $\Lambda$ below $200~$MeV will cut off not only unobservable high-energy modes but increase the
distortion of observable low-energy modes. The instability of the phases predicted by the RGM in fig.~\ref{fig_d-n-ph} for energies $\gtrsim 1.5~$MeV resemble the
crossing of the deuteron breakup threshold. The position of this threshold is set by $B(d)$ and therefore depends on $\Lambda$ and $\mathcal{D}'$.
For the potential with $\Lambda=1.6~$GeV and $\mathcal{D}'_1$, for instance, the collision becomes inelastic at $E_\text{cm}\gtrsim 1.5~$MeV
(vertical line in fig.~\ref{fig_d-n-ph}) compared to a breakup of the deuteron for $E_\text{cm}\gtrsim 2.6~$MeV for
LO interactions with $\Lambda=200~$MeV and $\mathcal{D}'_1$ in EFT($\slashed{\pi}$).
Threshold positions are variable within uncertainty margins of a given order. The values of $\Lambda$ and the input set
$\mathcal{D}'$, for example, constrain the deuteron binding energy to $1.4~\text{MeV}\lesssim B(d)\lesssim 2.7~$MeV. This fact can be utilized as follows:
\par
The numerical effort for multichannel scattering calculations can be reduced by widening the threshold separation without simultaneously losing accuracy in
phase shifts at energies below the lowest breakup threshold. I consider the four-body problem of the $^4$He $0^+$-channel as an example.
In this system, the two-channel calculation, with only t-p and $^3$He-n open, is expected to be predicted accurately like the n-d reaction of this chapter
if the renormalization scheme is chosen wisely.
This choice is conjectured to set the scheme parameters $\Lambda$ and $\mathcal{D}'$, such that, on the one hand, the two lowest thresholds - t-p and $^3$He-n -
are accurately reproduced, \textit{e.g.}, by including $B(t)$ in $\mathcal{D}'$, and on the other hand, the d-d threshold is lowered, \textit{e.g.}, using $\Lambda=1.6~$GeV and $\mathcal{D}'_1$.
The benefit of the larger gap between the $^3$He-n and d-d thresholds is a reduction of the basis size.
\par
Concluding the discussion of $\delta_\text{\tiny nd}$, the RGM uncertainty band calculated at LO in EFT($\slashed{\pi}$) contains the (N)LO predictions of
\cite{nd-ham-meh} ((black) gray solid in fig.~\ref{fig_d-n-ph}) as well as the partial wave analysis data$^\text{\cite{nd-exp}}$.
\par
\begin{figure}
\includegraphics[width=0.9\columnwidth]{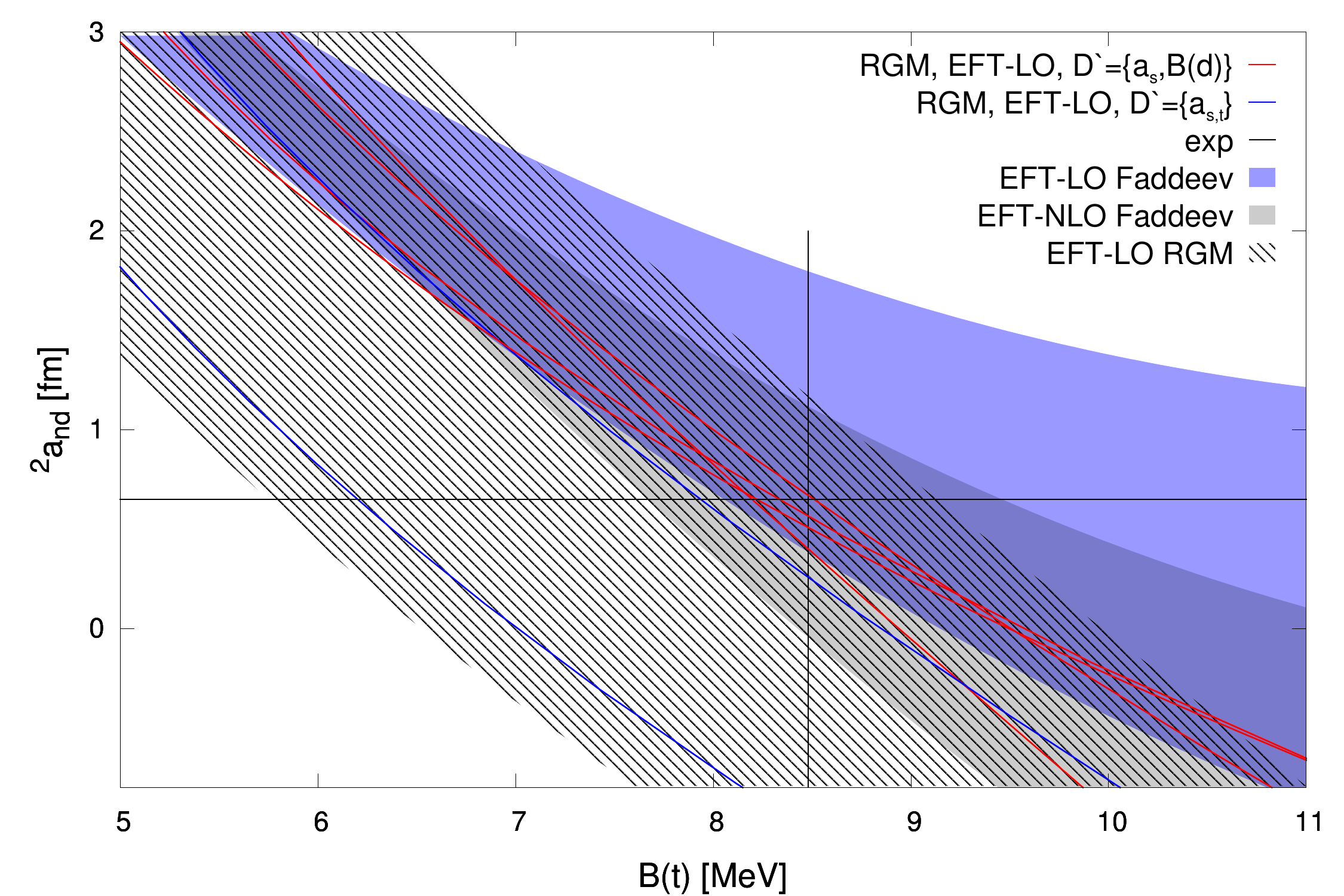}
\caption{\label{fig_phillips}\small
Correlation between $B(t)$ and the doublet S-wave scattering length $^2a_\text{\tiny nd}$ for the neutron-deuteron collision (Phillips line). Experimental values are shown as solid black lines.
The (gray) blue shaded area is the (N)LO EFT($\slashed{\pi}$) uncertainty band from \cite{hgrie-zpara}. The hatched area is the LO band obtained here with the RGM.
(Blue) Red correlation lines mark RGM LO EFT($\slashed{\pi}$) predictions for cutoffs $200~\text{MeV}\leq\Lambda\leq 1.6~$GeV and $\mathcal{D}'_{(1)2}$ as input.}
\end{figure}

By considering the n-d scattering length $^2a_\text{\tiny nd}$ in the triton channel correlated to $B(t)$, I confirm the conclusions drawn from the analysis of $\delta_\text{\tiny nd}$.
Furthermore, the RGM uncertainty assessment derived from a change in the cutoff $\Lambda$ and input data, can be compared to \cite{hgrie-zpara} at LO in EFT($\slashed{\pi}$),
where only the regulator parameter was varied. The results are displayed in fig.~\ref{fig_phillips}, where $^2a_\text{\tiny nd}$ was derived from the phase shift at
$E_\text{cm}=4~$keV, calculated with the RGM and the same LO EFT($\slashed{\pi}$) potentials used for the $\delta_\text{\tiny nd}$ predictions. The correlation lines were
obtained by fitting the two-body LECs to $\mathcal{D}'_{(1)2}\setminus\lbrace B(t)\rbrace$ ((blue) red lines) with fixed $\Lambda$ and a variation of the three-body
parameter $A_{3\text{NI}}$.
\par
As for $\delta_\text{\tiny nd}$, the uncertainty assessed by potentials based on $\mathcal{D}'_2$ with their precise reproduction of $B(d)$ is smaller compared to
that which includes $\mathcal{D}'_{1}$ predictions. In more detail, the width of the band mapped out by $\mathcal{D}'_{2}$ interactions alone (red lines) resembles the
sensitivity to short distance physics parameterized by a regulator value of $200~\text{MeV}\leq\Lambda\leq 1.6~$GeV. Hence, a sole change in $\Lambda$
would suggest a LO EFT($\slashed{\pi}$) uncertainty of
\begin{equation}\label{eq_a-nd-1}
^2a_\text{\tiny nd}=0.67\pm0.19~\text{fm}\;\;\;,
\end{equation}
with $A_{3\text{NI}}$ fitted to $B(t)$. A further probe into the short-distance structure of the theory, by including interactions using $a_t$ instead of $B(d)$ as input ($\mathcal{D}'_1$),
maps out a broader band (hatched area in fig.~\ref{fig_phillips} with its (upper) lower bound set by a potential with $\Lambda=(200)1600~$MeV and LECs fitted to
$\mathcal{D}'_{1}\setminus\lbrace B(t)\rbrace$), suggesting the less accurate LO prediction:
\begin{equation}\label{eq_a-nd-2}
^2a_\text{\tiny nd}=0.67\pm1.5~\text{fm}\;\;\;.
\end{equation}
Increasing $\Lambda$ beyond $1.6~$GeV is not expected to widen the band significantly, because of the relatively small width increase induced by increasing
$\Lambda$ from $800~$MeV (lower solid blue line in fig.~\ref{fig_phillips}) to $1600~$MeV (lower edge of hatched area), compared to the increase from $400~$MeV
(upper blue solid line) to $800~$MeV.
We have thus demonstrated how the limitation to a certain set of input data reduces the uncertainty of an EFT significantly. A cutoff variation as the only probe to dependence
on high-energy modes does not reveal the overall uncertainty which is found considerably larger, \textit{i.e.}, the width of the hatched area, representing the correlation for
$\Lambda$ and $\mathcal{D}'$, is wider than the band mapped out by the potentials using $\mathcal{D}'_1$ (red lines).
\par
For the range of triton binding energies considered here, the RGM LO EFT($\slashed{\pi}$) correlation band overlaps with the (N)LO bands
((gray) blue shaded in fig.~\ref{fig_phillips}) calculated also with EFT($\slashed{\pi}$) but using a Z-parameterized amplitude to fit LECs and which were obtained by calculating the
three-body observables from Faddeev integral equations solved in momentum space$^\text{\cite{tar-stern}}$. The broader RGM band is explained by the more
detailed probe of short-distance structure. While in \cite{hgrie-zpara} the band spread is induced by a cutoff variation only, for the RGM band, input data
is varied additionally. Taking the overlap of both calculations as the LO correlation band, all NLO predictions are elements of this broader LO band -
resolving this issue arising in \cite{hgrie-zpara}.
\subsection{Conclusions from A=3}\label{subsec_a3-summary}
The analysis of correlated observables is a useful tool worthy to be applied in larger systems because:\vspace{-.2cm}
\begin{itemize}
\item The shape of the area mapped out by varying the renormalization scheme assesses a bound for the expansion parameter(s) of the theory. An unbounded area
indicates an inconsistency in the power counting.\vspace{-.3cm}
\item A discrimination between insufficient interaction models and numerical effects can be made. Consider, for instance, a potential tuned accurately to two-body
data but missing the triton binding energy significantly (\textit{e.g.}, \texttt{AV18} fig.~\ref{fig_rch-t}). Also, the position and width of a correlation band for a scattering
observable - not one derived from a bound state, as then, a different, in most cases more tedious, numerical algorithm is required - is known from a relatively simple
EFT($\slashed{\pi}$) LO calculation. A prediction of the model inconsistent with this band is then only related to numerical uncertainties. This understanding will be
employed in the four-nucleon scattering calculations presented below.\vspace{-.3cm}
\item A guideline for theoretical model building is given. A modification of an interaction to improve the prediction for a selected low-energy observable must simultaneously
impact correlated observables in a way consistent with the accuracy of the interdependence.\vspace{-.3cm}
\item Conflicting measurements - nonoverlapping error intervals - can be categorized by falling in or out of the correlation band at sufficiently high order. A comparison
of the limitations of the theory, \textit{e.g.}, neglected weak interactions or three-body break-up channels, and of the experimental setup, suggest an interpretation
of the discrepancy in the data.
\end{itemize}
The RGM results presented here confirmed for the n-d scattering length and the phase shift in the S-wave spin-$\frac{1}{2}$ channel how a certain
choice of NN data for the determination of the LECs reduces LO uncertainty. The precisely reproduced deuteron is to be favored over, \textit{e.g.}, an accurate
fit of neutron-proton scattering phase shifts at $E_\text{cm}\to 0~$MeV, if the LO uncertainty for n-d scattering observables is sought to be minimized.
In general, the input set $\mathcal{D}'$ which will minimize the LO uncertainty will be a function of the predicted observable. The accuracy of predictions
for the triton charge radius, for example, did not increase (see solid and dashed red lines in fig.~\ref{fig_rch-t}) when $B(d)$ instead of $a_t$ was
included in $\mathcal{D}'$.
We will, nevertheless, translate the approach to the larger systems (ch.~\ref{sec_a5} \& \ref{sec_a6}) in which $^4$He will play the role of the deuteron as the dominant
substructure in the $^6$He ground state, and as the target in elastic $\alpha$-n scattering. Explicitly, LO results for the $^6$He bound-state energy and $^5$He properties
are expected to be less sensitive to $\Lambda$ and $\Lambda_\text{RGM}$ if $B(\alpha,\text{exp})$ is used instead of $B(t)$ to fit the LO three-body parameter $A_{3\text{NI}}$.
By reducing the sensitivity to $\Lambda_\text{RGM}$, I expect to obtain converged results, with respect to the basis size, at lower dimensionality and by
that increase the flexibility of the RGM in $A>4$ calculations. Although, the focus is on a real bound state in the six-nucleon and a virtual one in the five-nucleon system,
even unimproved convergence will contribute to the understanding of the different behavior of an EFT for bound- and scattering systems with regard to
the position of the bound state in the spectrum relative to the nearest breakup threshold. To be specific, it is of interest to investigate what effect the relatively large gap
between $B(d)$ and $B(t)$ relative to that between $B(^6\text{He})$ and $B(\alpha)$ has on EFT convergence.
\section{A=4}\label{sec_a4}
An investigation of systems of $N+1$ particles with an interaction derived in the $N'\leq N$ body sector is presented here for $N=3$. The foregoing analyses of the
three-nucleon system with EFT($\slashed{\pi}$) confirmed at NLO that neither triton nor low-energy neutron-deutron scattering observables are universal consequences
of the large two-nucleon scattering lengths. By increasing the number of particles by one proton, the analogous problem asks: Is the 4-helium ground state
and/or low-energy four-body two-fragment scattering observables on the ground states of the subsystems, \textit{i.e.}, deuteron, triton, and 3-helium, universal for
interactions parameterized by, \textit{e.g.}, $a_{s,t},B(t)$?
\subsection{Tjon correlation}\label{subsec_tjon}
First, in analogy to the triton, correlations amongst bound-state observables are investigated, specifically, the Tjon line$^\text{\cite{tjon}}$
which relates the ground-state energies of the triton and \mbox{4-helium}.
By doing that, it implies the existence of a four-fermion bound state for all short-range interactions parameterized by two-fermion scattering lengths $a_{s,t}$ and the 
smallest three-fermion binding energy. While this is a non-trivial consequence itself, the different characeteristics of this stated compared to those of its bound
two- and three-body fragments make it even more interesting. Specifically, it is bound by $\sim 20~$MeV below the lowest breakup threshold, triton-proton, compared to
only $\sim 6~$MeV for the triton below d-n, and $\sim 2~$MeV in the deuteron relative to n-p. Previous LO EFT($\slashed{\pi}$) calculations$^\text{\cite{platter-tjon}}$
suggested that this large gap does indeed follow universally with a relatively small uncertainty, if $\mathcal{D}'=\lbrace a_s,a_t/B(d),B(t)\rbrace$ is used to fix the
relative positions of $B(d)$ and $B(t)$. The assessment of the uncertainty was limited there to a change in input data, changing $a_t$ to $B(d)$, which, in the light of
the results shown in fig.~\ref{fig_d-n-ph} \& \ref{fig_phillips} could underestimate the error. In this work, we will show that not only the bound structure, but also the
large gap to threshold can be explained, counterintuitively as it suggests large typical momenta, with EFT($\slashed{\pi}$).
For this purpose, I extend the LO analysis$^\text{\cite{platter-tjon}}$ to NLO in the EFT($\slashed{\pi}$) framework, and include the Coulomb interaction in the potential.
\begin{table*}
\renewcommand{\arraystretch}{1.5}
  \caption{\label{tab_he4-ms}{\small 4-helium Gau{\ss}ian width parameters $\gamma_n$ for the RGM basis in LO and pNLO calculations. The $\gamma$ subscripts
label the Jacobi coordinate $\vec{s}$ with $\vec{s}_\text{rel}\equiv\vec{s}_3$}, and $\vec{s}_{12}=\vec{s}_1+\vec{s}_2$ is the intermediate spin coupled from particles
1 and 2 - a neutron and a proton - within the triton state. The $81$ triton, $81$ $^3$He, and $41$ d-d states form a $1206$-dimensional basis with the $6$ relative
width parameters. The width sets were derived from $\mathcal{W}_{120}$ and optimized with the genetic algorithm (ch.~\ref{app_gen-alg}) for a converged
$B(d,t)$ in this smaller model space.}
\footnotesize
\begin{tabular}{c|l|l}
\hline
$f\in\mathcal{G}'$&widths [fm$^{-2}$]&dim\\
\hline\hline
$\begin{array}{c}\text{t-p} \\\text{\&}\\ ^3\text{He-n}\end{array}$&\scriptsize
$s_{12}=1\;\;:\;\;\gamma_1(j)\in\mathcal{W}_9=\left\lbrace
\begin{array}{l}
49.6384602,26.3058574,11.7732549,\\
4.47594489,1.65292568,0.81911105,\\
0.33017059,0.09380507,0.02205199
\end{array}\right\rbrace$\;\;\;,\;\;\;
$\gamma_2(j)\in\left\lbrace \begin{array}{l}26.30585735,\\4.47594489,\\0.81911105,\\0.093805066\end{array}\right\rbrace$
&36\\
&
$s_{12}=0\;\;:\;\;\gamma_1(j)\in\mathcal{W}_9$\;\;\;,\;\;\;
$\gamma_2(j)\in\left\lbrace \begin{array}{l}49.6384602,11.7732549,1.65292568,\\0.330170586,0.022051993\end{array}\right\rbrace$
&45\\
d-d&
$\gamma_{1,2}(j)\in\left\lbrace \begin{array}{l}42.88895245,15.27652652,4.040844294,\\0.790390928,0.138826844,0.026116458\end{array}\right\rbrace$
&21\\
\hline
\multicolumn{3}{c}{$\gamma_3(j)\in\mathcal{W}_{120}$, elements 3, 5, 7, 9, 11, 13 are used $\forall f$}\\
\hline
    \end{tabular}
\end{table*}
The variational basis used for the LO and (p,n)NLO calculations includes all two-fragment groupings with bound fragments:
\begin{equation}\label{eq_4he-frags}
\mathcal{G}'=\left\lbrace\text{triton-proton (t-p) , 3-helium-neutron ($^3$He-n) , deuteron-deuteron (d-d)}\\
\right\rbrace\;\;\;.
\end{equation}
Singlet-deuteron-singlet-deuteron (\mbox{d\hspace{-.55em}$^-$}-\mbox{d\hspace{-.55em}$^-$}) and di-neutron-di-proton (nn-pp) fragments are
not included in the basis as they were found to contribute insignificantly to the
bound state in contrast to their role in the scattering basis (see below). For those unphysical objects, the isospin
invariance of the two-nucleon EFT($\slashed{\pi}$) up to NLO implies identical Hamiltonian eigenvalues for nn and \mbox{d\hspace{-.55em}$^-$} states.
Both states have a total spin $s=0$, an
orbital angular momentum of $l=0$, and an isospin of $t=1$.
As the total angular momentum and parity of the fragments in a certain grouping is set, it suffices to specify the maximal orbital angular momentum on each
coordinate to define the spin and orbital angular momentum structure of the model space if all coupling schemes to a total
angular momentum state $J^\pi=0^+$ are allowed. For the LO and pNLO basis, S-waves were used on all coordinates except the fragment-relative coordinate between two deuterons
on which, in addtition, a D-wave was allowed, $l_\text{rel(d-d)}=0,2$.
This limitation in $l$ is motivated by the analogy to a rectangular-well potential. Analytic bounds for the formation of a bound state on the depth
$C_\text{LO}$ and width $\frac{\Lambda^2}{4}$ of the well can be derived$^\text{\cite{schiff-angl}}$ which would require, \textit{e.g.}, for $l=1$:
$C_\text{LO}\frac{\Lambda^4}{16}\leq\frac{\pi^2\hbar^2}{2m_\text{\tiny N}}$. Analogously, for the Gau\ss ian EFT($\slashed{\pi}$) potentials with fixed LECs and cutoff, \textit{i.e.},
depth and width, particles in partial waves with higher angular momentum will eventually not bind together. Hence, such states will not overlap with a four-body bound
state. Furthermore, the strength of the interaction in higher partial waves is by construction small in the EFT framework as shown in fig.~\ref{fig_nn-phases}.
We verified the assertion by comparing results for $B(\alpha)$ obtained in this EFT model space and one which was used for \texttt{AV18(+UIX)} predictions.
The latter also includes up to three relative D-waves coupled to a total $l=0$. The difference was found insignificant relative to LO uncertainty from the interaction.
The basis definition is completed by the width sets given in table~\ref{tab_he4-ms}.
The nNLO basis is truncated by $l_n\leq 2\;\forall n$, and the width sets resemble those
defined for the RGM \texttt{AV18(+UIX)} calculations$^\text{\cite{mythesis}}$ regarding their dimensionality with values optimized by the genetic algorithm (app.~\ref{app_gen-alg})
for each cutoff.
\par
To assess if the reduction in the number of triton components in the 4-helium basis relative to its size in a three-body basis (ch.~\ref{sec_a3}) is
still guaranteed for an almost complete \mbox{4-helium} model space, the binding
energy $B(\alpha)$ thus obtained was compared to one calculated in the much
larger scattering model space.
The two values differed by less than 50~keV. This model space, used for the $0^+-$channel
scattering calculation, is defined below and is spanned by more than 7000 basis vectors.
This study of reducing the dimension of the 3-nucleon fragments in the $\alpha$ particle
without deviating significantly from the assumed converged value is crucial for
applications of the method to $A>4$ systems regarding practicality in terms of computing cost and accuracy in predicted observables.
\par
\begin{figure}
  \includegraphics[width=0.95\textwidth]{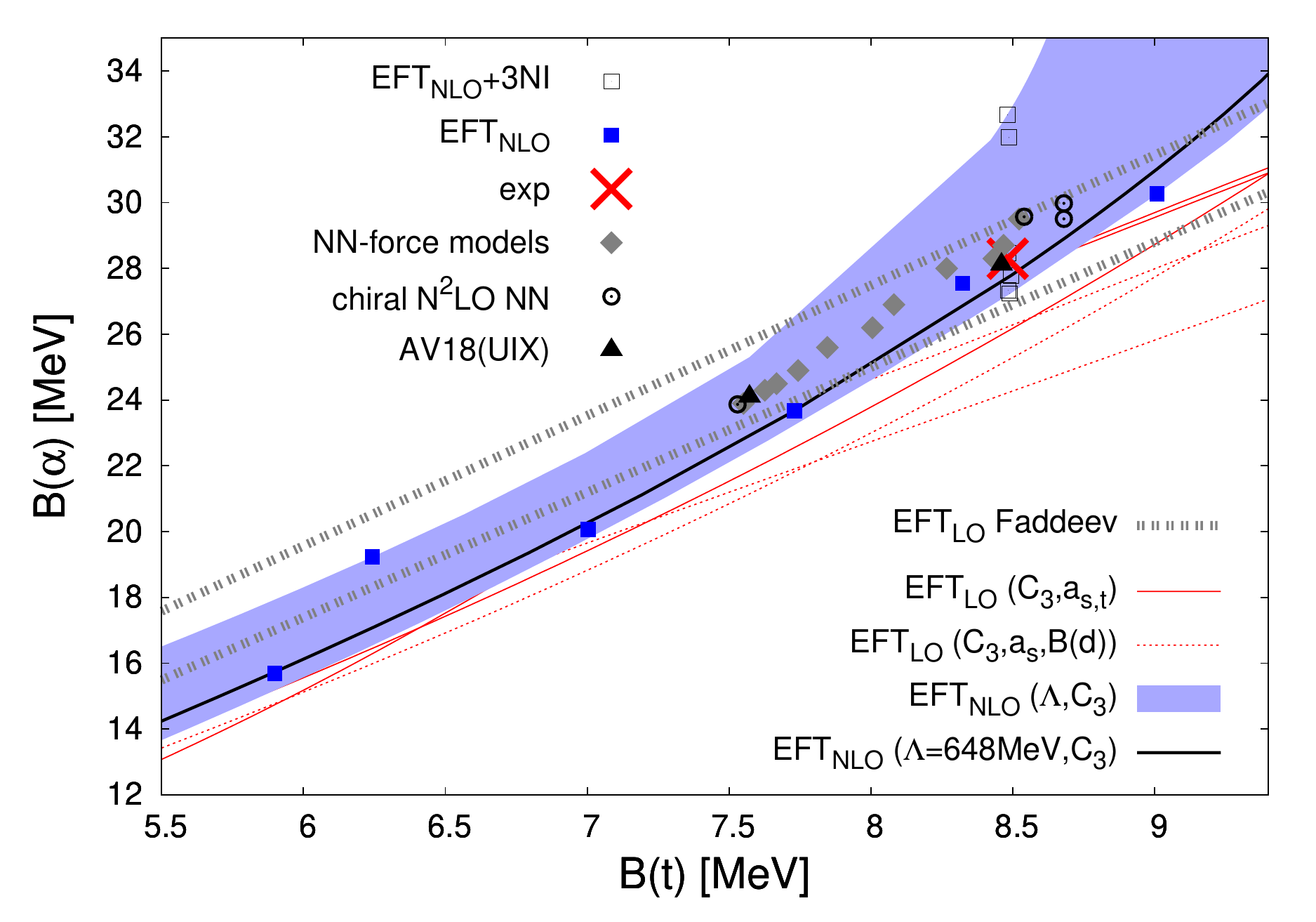}
  \caption{\label{fig_tjon}\small Correlation between the triton and $^4$He
    binding energies (Tjon line). The filled squares are the results using NLO
    EFT($\slashed{\pi}$) NN potentials from table~\ref{tab_nlo-pots} with no 3NI, and the empty squares represent
    the predictions for the 3NI fitted to $B(t)$. The shaded NLO band results from
    a variation of the 3NI for those potentials.
    We compare to LO results from~\cite{platter-tjon}, where the upper (lower) dashed line was calculated with LEC fitted to $a_{s,t}$ ($B(d),a_s$), and to our
    LO RGM calculation with potentials from table~\ref{tab_lo-pots}.
    $B(\alpha,\text{exp})$ is taken from~\cite{exp-ba}, the values for
    \texttt{AV18(+UIX)} are reported in~\cite{mythesis}, and those using a chiral expansion to NLO and N$^2$LO
    in~\cite{xpt-epelbaum}.}
\end{figure}

In fig.~\ref{fig_tjon}, the results are compared to a LO band calculated by solving Faddeev/Yakubovski integral equations quasi-exactly by $\text{\cite{platter-tjon}}$
(referred to as FY results).
The spread of the NLO values is not in conflict with EFT($\slashed{\pi}$) which allows a $\lesssim 10$\% uncertainty at NLO.
Like for the triton charge radius, the upper (lower) boundary of the FY results was obtained by choosing different NN
observables, $a_{s,t}(B(d),a_s)$, to fit the LO LECs for cutoffs high enough so that $B(\alpha)$ did not change when further changing the cutoff.
Figures~\ref{fig_rch-t} \& \ref{fig_tjon} confirm that the LO accuracy would be overestimated with this method.
The RGM LO calculation uses the same interactions as in sect.~\ref{sec_a3}, with cutoff values from $400~\text{MeV}$
(lowest red dotted line setting the lower bound for the RGM LO band in fig.~\ref{fig_tjon}) to $1.6~\text{GeV}$ (top solid red line setting the upper bound).
In contrast to the RGM LO correlation between $B(t)$ and the triton charge radius, the width of the RGM LO Tjon band
is however not converged. The shifts between the positions of the correlation lines corresponding to cutoffs of $400~\text{MeV}$,
$700~\text{MeV}$ (middle red solid line), and $1.6~\text{GeV}$ indicate that a variation
of $\Lambda$ even beyond $1.6~\text{GeV}$ would be necessary to assess the LO uncertainty from cutoff variations only.
However, elaborate technical modifications are required for
the RGM calculation at higher cutoffs. Analogous to the conservative estimate for the LO uncertainty in fig.~\ref{fig_rch-t},
the LO Tjon correlation band is thus mapped out by both the RGM and the Faddeev/Yakubovsky
(thick dashed lines) results, which overlap nicely in this plot. This combined correlation band includes the datum
and the narrower NLO band, which, from fig.~\ref{fig_tjon}, has at the experimental $B(t)$ a width of about $5~$MeV centered around $28~$MeV and
results in a prediction of
\begin{equation}\label{eq_b4he-nlo}
B(\alpha,\text{NLO})=(28\pm 2.5)~\text{MeV}\;\;\;,
\end{equation}
which is consistent with the expected NLO uncertainty of about $10$\% and with experiment.
Comparing the NLO uncertainty to LO accuracy, which we estimate by the extremal predictions at $B(t,\text{exp})$ for $B(\alpha)$,
\textit{i.e.} LO $\Lambda=400~$MeV (lowest red dotted line) and the upper edge of the NLO band, to be about $\pm 6~$MeV,
would be consistent with an expansion parameter of the EFT of the same order as the na\"ive estimate of $\sim \frac{1}{3}$.
Again, the results of the \texttt{AV18(+UIX)} models lie within the proposed band as it is expected
of all interaction models of at least NLO. The observed broadening of the correlation band
is a manifestation of the momentum dependence of the EFT expansion. The accuracy decreases
with increasing typical momentum, eventually leading to a breakdown of the expansion.
\par
From the fact that there is still a one-parameter correlation, it is concluded that no
four-nucleon contact interaction is required to renormalize the theory at
NLO. One three-body parameter fitted to data suffices to yield
proper NLO predictions for four-body observables within the theoretical
accuracy (empty squares in fig.~\ref{fig_tjon}).
\subsection{Neutron-3-helium scattering}\label{subsec_a3he-n}
\begin{figure}
  \includegraphics[width=0.7\textwidth]{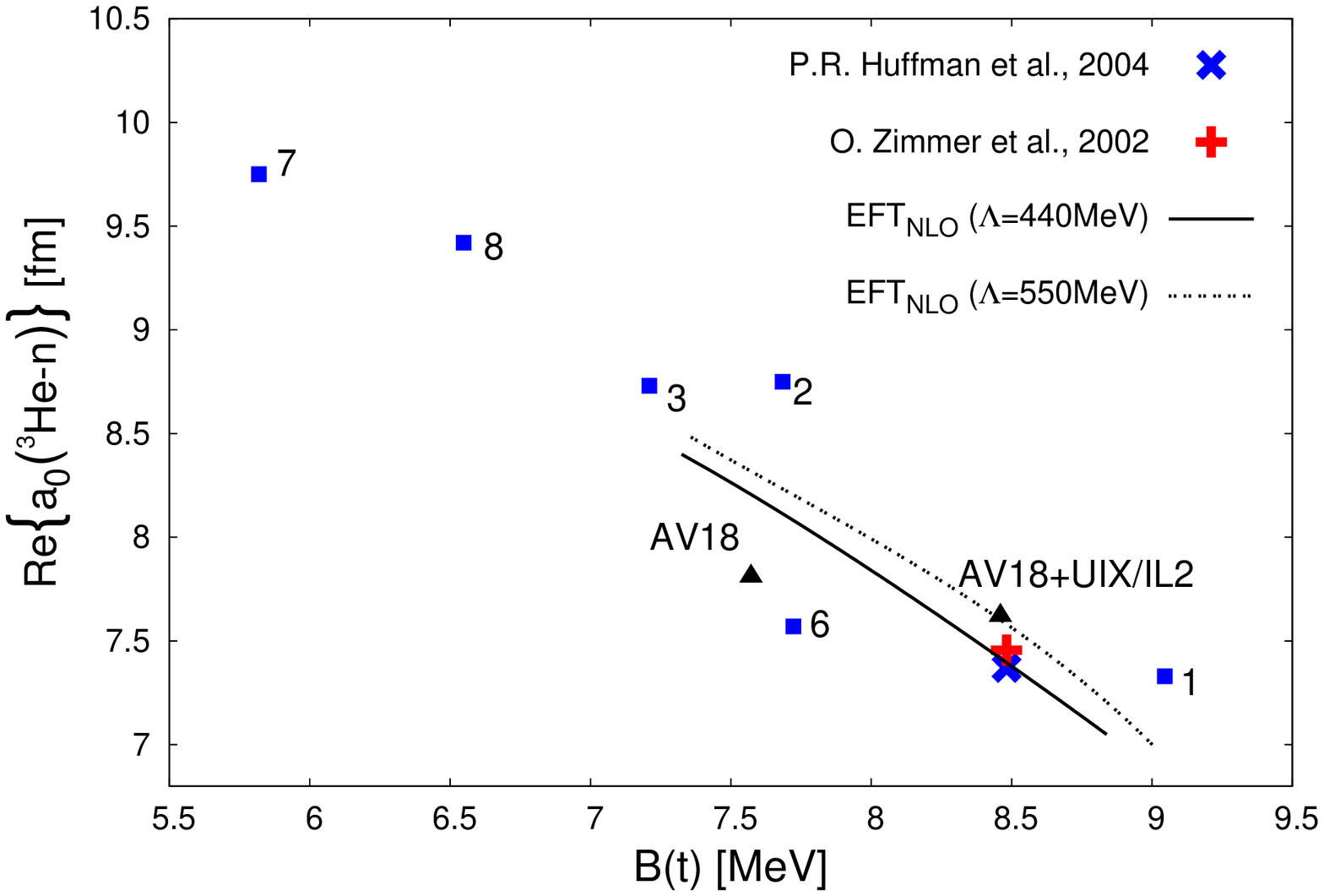}
  \caption{\label{fig_a0}\small The correlation between the triton binding energy
    and the real part of the spin singlet, S-wave scattering length
    $a_0(^3\text{{\small He-n}})$ for elastic $^3$He-n scattering. Blue squares
    represent the values of the NLO EFT($\slashed{\pi}$) potentials with a
    numerical uncertainty of $\mathcal{O}(0.02~\text{fm})$. A variation of the 3NI
    yields the continuous correlation lines for two selected NLO potentials with a fixed
    cutoff.
}
\end{figure}

As in the $A=3$ system, a low-energy scattering observable should solidify the assertions about the power counting
in $A=4$ which were based in the preceding section on bound-state properties.
In principle, all low-energy observables should be correlated with the triton
binding energy. The recent results$^\text{\cite{delt-port-n3h}}$ for
the singlet and triplet $n-^3$H scattering lengths using
three potential models and a N$^3$LO chiral potential are
evidence for this assertion in the four-nucleon scattering
system. Here, the real part of the S-wave spin singlet
scattering length $a_0(^3\text{{\small He-n}})$ for elastic
\mbox{3-helium-neutron} scattering is investigated. In fig.~\ref{fig_a0},
its value is shown as a function of $B(t)$ for six potentials
$V_\slashed{\pi}$ including the Coulomb interaction.
For two NLO potentials with $\Lambda=440~\text{MeV}$ (solid line)
and $\Lambda=550~\text{MeV}$(dashed line), with two-nucleon LECs fixed,
we also show the effect of a smooth variation of the three-nucleon interaction parameter.
\par
To extract $a_0(^3\text{{\small He-n}})$, six two-fragment channels,
$^3$He-n, t-p, d-d ($l_\text{rel}=0,2$), (nn)-(pp), and \mbox{d\hspace{-.55em}$^-$}-\mbox{d\hspace{-.55em}$^-$} are included.  The latter two consist of
unbound fragments and usually model possible three- and four-body breakup
reactions.  They, as well as the two d-d channels, are however for this
calculation only needed to provide configurations for distortion channels
since only the t-p channel is open a few eV above the $^3$He-n threshold.  For
the fragment wave functions $\phi^{J_{fi}}$ (see eq.~(\ref{eq_rgm-rel-wfkt})), a 224 dimensional basis
was used for the triton and \mbox{3-helium}, and a 9 dimensional one for the
deuteron, whose six $l_n=0$ vectors built the nn, pp, and \mbox{d\hspace{-.55em}$^-$} states.  For
these six channels, the 20 width parameters $w_{12}$ were used for the
$\gamma_n(j)$ in eq.~(\ref{eq_rgm-rel-wfkt}).  Almost all configurations included to
build those physical channels could be recycled as distortion channels to
allow for more freedom in the minimization of the variational functional.
Less than ten configurations had to be excluded to avoid numerical linear
dependences.  In each distortion channel, four to six relative width
parameters $\gamma_n(j)$, taken from $w_{12}$ with
$\gamma_n(j)>0.02~\text{fm}^{-2}$, were used.  Numerical stability and
convergence of $a_0(^3\text{{\small He-n}})$ were assessed by increasing the
number of included relative widths $\gamma_n(j)$ by one for each distortion
channel, yielding changes in $a_0(^3\text{{\small He-n}})$ of the order of
the numerical uncertainties, given that the initial $\gamma_n(j)$ were chosen
appropriately.  The lowest eigenvalue of the Hamiltonian is equal to the ground-state energy of \mbox{4-helium}
in this model space and was allowed to change in this process by not more than 10~keV.
Significantly larger changes in this eigenvalue which lead to a result not of
the order of magnitude suggested by the LO Tjon band signal numerical
linear dependences. If the model space is too small, or if the width parameters for
the relative wave function were chosen inappropriately, changes of the order of
$100~$keV up to a few MeV are expected.
The \mbox{3-helium-neutron} scattering length $a_0(^3\text{{\small He-n}})$ was
calculated from the corresponding diagonal element $S_{22}$ of the S-matrix at
a center of mass energy of $E_\text{cm}<10~$eV above the $^3$He-n threshold$^\text{\cite{hmh-4he}}$:
\begin{equation}\label{eq_scatt-length}
  a_0(E_\text{cm})=\frac{1-S_{22}(E_\text{cm})}{i\sqrt{\frac{3}{4}ME_\text{cm}}\big(1+S_{22}(E_\text{cm})\big)}\;\;\;.
\end{equation}
A fraction of the flux is diverted into the open triton-proton channel,
resulting in a nonzero $S_{12}$ S-matrix element and hence a nonzero imaginary
part of $a_0(^3\text{{\small He-n}})$.
The 3NI variation yields an almost linearly increasing imaginary part with increasing $B(t)$ from about
$\text{Im}\lbrace a_0(^3\text{{\small He-n}})\rbrace\approx -7.5~\text{fm}$ at
$B(t)\approx 5.5~\text{MeV}$ to $\text{Im}\lbrace a_0(^3\text{{\small He-n}})\rbrace\approx -2.0~\text{fm}$
at $B(t)\approx 9.1~\text{MeV}$. This qualitative observation is consistent with the fact that the
splitting of trinucleon binding energies decreases with decreasing triton binding energy (see sect.~\ref{subsec_tri-nucl-split}), which results in a
smaller separation between the respective thresholds in four-nucleon scattering.
This handle on the threshold separation suggests an approach to circumvent numerical
problems associated with the proximity of thresholds by extrapolating results for their physical
values from calculations performed at more deeply bound 3-nucleon states. This flexibility is conjectured to
facilitate future calculations.
\par
In fig.~\ref{fig_a0}, the predictions of the NLO potentials with and without 3NI
for $\text{Re}\lbrace a_0(^3\text{{\small He-n}})\rbrace$ decrease with increasing
triton binding energy. They map out a band which includes the datum.  A comparison
to hard sphere scattering qualitatively explains this behavior. A higher \mbox{3-helium} binding
energy $B({^3\text{\small He}})$ corresponds to a smaller nucleus analogous to the
triton as shown in fig.~\ref{fig_rch-t}.  As the scattering length is
proportional to the radius of the hard sphere, $a_0(^3\text{{\small He-n}})$
is expected to decrease for increasing $B({^3\text{\small He}})$.
We define the NLO correlation band to be centered around the values predicted
by the potentials with zero 3NI.
The change in $a_0(^3\text{{\small He-n}})$ observed between $^2V_\slashed{\pi}$
and $^6V_\slashed{\pi}$ serves as an error estimate for the band centered around
$7.5~$fm. Again, the assumption of the center of the band at the experimental $B(t)$
follows from the values in fig.~\ref{fig_a0}.
Thus, EFT($\slashed{\pi}$) reports at NLO:
\begin{equation}\label{eq_a0-value-nlo}
\text{Re}\lbrace a_0(^3\text{{\small He-n}})\rbrace=\left(7.5\pm 0.6\right)~\text{fm}\;\;\;.
\end{equation}
The 3NI was varied only over the depicted range of trition binding
energies, where one and the same fixed RGM variational space can be used.
The fragment model space for the two interactions was optimized
for the 3NI parameter fitted to $B(t)$. Only the number of included distortion channels
was changed to reach convergence within this space for each value of the 3NI.
The band mapped out by the two 3NI lines for
$\left(7.5~\text{MeV}~\lesssim B(t)\lesssim 9~\text{MeV}\right)$ includes the datum, and its slope is
consistent with the one indicated by the 6 NLO potentials without 3NI.
This explicitly demonstrates that a variation of the 3NI has the same effect as
varying the short-distance part of the NN interaction.
The NN model \texttt{AV18} yields a value within the error band. Adding the UIX three-body
interaction moves this point into the $10$\% NLO uncertainty radius around the datum.
\par
\begin{figure}
  \includegraphics[width=0.8\columnwidth]{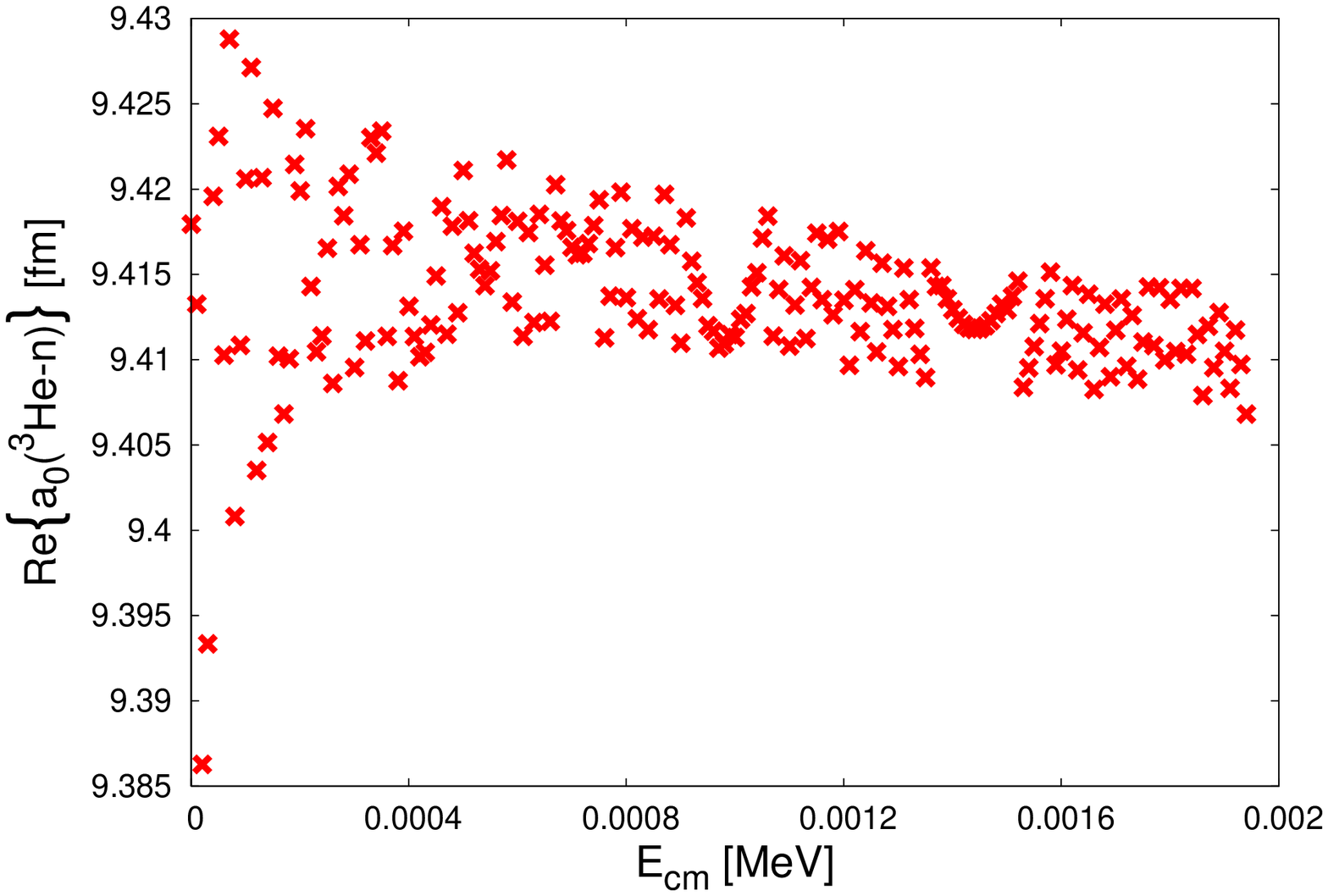}
  \caption{\label{fig_a0ofe}\small The real part of the spin singlet, S-wave
    scattering length for elastic $^3$He-n scattering as a function of the
    center of mass energy above the $^3$He-n threshold at which
    $a_0(^3\text{{\small He-n}})$ is calculated from
    eq.~(\ref{eq_scatt-length}). Notice the scale on the vertical axis. Taking
    the central value of the band as the actual value, its decrease from
    200~eV to 2~keV is due to effective range corrections, while the band is a
    result of numerical noise. $^8V_\slashed{\pi}$ was used for this plot,
    with the other $V_\slashed{\pi}$ potentials showing similar behavior
    with respect to the size of the higher order corrections and numerical
    fluctuations.}
\end{figure}
To gauge the size of the error introduced by numerical uncertainties
from diagonalizing the Hamiltonian for the calculations with zero 3NI, $a_0(^3\text{{\small He-n}})$ was
calculated over a wider range of energies, $E_\text{cm}\leq 2~$keV.
Figure~\ref{fig_a0ofe} shows $\text{Re}\lbrace a_0(^3\text{{\small He-n}})\rbrace$
as a function of the matching energy at which it is
calculated for the potential $^8V_\slashed{\pi}$.  The fluctuations resulting
in the band-like shape with a width of about $0.015$~fm out of
  $a_0(^3\text{{\small He-n}})\approx9.415$~fm are due to uncertainties
associated with the diagonalization of numerically singular matrices.  The
error due to effective range corrections can be estimated by observing the
difference,
\begin{equation}\label{eq_a0corr}
  |\text{Re}\lbrace a_0(200~\text{eV})\rbrace-\text{Re}\lbrace a_0(2~\text{keV})\rbrace|\approx 0.01~\text{fm}\;\;\;.
\end{equation}
We therefore conclude that the numerical precision of our result is about
$1\%$ and is considerably smaller than the theoretical accuracy of $\sim10\%$
attributed to a NLO calculation. For \texttt{AV18}, the same analysis produced stable
results with respect to numerical fluctuations, \textit{i.e.} no visible band
as in fig.~\ref{fig_a0ofe} is found. This is attributed to the model
space used for \texttt{AV18}, which was optimized for this potential and purged of
states with large but mutually canceling overlap with the ground state.
As mentioned at the beginning of this paragraph, this analysis
applies to values obtained in an optimized model space only.
\par
The conclusion drawn from the newly found correlation in fig.~\ref{fig_a0} is
that every potential with the correct NN low-energy phase shifts and
appropriately tuned three-body interaction, \textit{e.g.}, to give the correct
triton binding energy, predicts not only the correct $B(\alpha)$ but also
the experimental $a_0(^3\text{{\small He-n}})$ within a NLO error range.
In table~\ref{tab_a0}, the results for $a_0(^3\text{{\small He-n}})$ of RGM
calculations with the EFT($\slashed{\pi})$ NLO potential and the \texttt{AV18} NN force model with
and without the three-body interaction models Urbana9 (\texttt{UIX}) and Illinois2 (\texttt{IL2})$^\text{\cite{ill}}$
are given.
\begin{table}
  \caption{\label{tab_a0}\small RGM predictions for the triton binding energy and the spin singlet, S-wave scattering length $a_0(^3\text{{\small He-n}})$
    for elastic $^3$He-n scattering of phenomenological nucleon potential models. The values for \texttt{AV18} and \texttt{AV18+UIX} are taken from~\cite{hmh-4he},
    while the \texttt{AV18+IL2} numbers are new results of this work. The imaginary part was investigated on a qualitative level, only.}
    \begin{tabular}{c|ccc}
\hline
      force&$B(t)~$[MeV]&$\text{Re}\lbrace a_0\rbrace~$[fm]&$\text{Im}\lbrace a_0\rbrace~$[fm]\\
      \hline\hline
      EFT($\slashed{\pi})$ NLO&$8.48~$(input)&$7.5(6)$&$-2.6(\text{?})$\\
      \hline
      \texttt{AV18}&$7.57$&$7.81$&$-4.96$\\
      \texttt{AV18+UIX}&$8.43$&$7.62$&$-4.07$\\
      \texttt{AV18+IL2}&$8.48$&$7.63$&$-4.28$\\
      \hline
      \multirow{2}{*}{exp}&\multirow{2}{*}{$8.48$}&$7.456(20)^\text{\cite{exp-a0-huffman}}$&\\
      &&$7.370(58)^\text{\cite{exp-a0-zimmer}}$&\\
\hline
    \end{tabular}
\end{table}
The values for the recent IL2 3NI have been calculated for this work, employing methods
described in~\cite{mythesis,hmh-4he}.  The IL2 prediction for
$\text{Re}\lbrace a_0(^3\text{{\small He-n}})\rbrace$ is almost identical
to the UIX value and not plotted separately in fig.~\ref{fig_a0}. UIX and IL2
have parameters fitted to, amongst others, the triton and \mbox{4-helium} binding
energies.  The observed deviation of the prediction of both models for
$a_0(^3\text{{\small He-n}})$ is therefore not a result of a deficiency of
the structure of the potential. The correlation between $B(t)$ and
$a_0(^3\text{{\small He-n}})$ supports the conjecture that this small deviation
can only be improved by the inclusion of higher order interactions.
\par
Analogously to UIX, the IL2 model easily satisfies the criteria mentioned in sect.~\ref{subsec_r3-ch},
namely to reproduce low-energy NN observables and the triton binding energy
at least with a $10$\% accuracy, and therefore its prediction
for $a_0(^3\text{{\small He-n}})$ should also lie within a $10$\% radius of the datum in fig.~\ref{fig_a0}.
This EFT prediction is confirmed here by explicit calculation.
\par
Without a calculation for the scattering length in the triplet channel, the results cannot be compared to the
most recent measurements$^\text{\cite{exp-a0-recent}}$. The older experimental values for $a_0(^3\text{{\small He-n}})$ (see tab.~\ref{tab_a0})
are included in the predicted universality band, and therefore we cannot resolve the discrepancy
between the two measurements.
\newpage
\subsection{Conclusions from A=4}\label{subsec_a4-summary}
The implications of these results for the scaling
of four-nucleon interactions (4NI) in EFT($\slashed{\pi}$) at LO and NLO
are summarized in this section.
\par
Simplistic dimensional analysis suggests that a momentum-independent 4NI
enters at N$^4$LO, but the unusual renormalization of the 3NI may also promote
the 4NI to contribute at lower orders to ensure renormalizability.
In the RGM approach with a Gau{\ss}ian regulator, the cutoff was varied for
the 4-helium binding energy and for the $^3$He-n scattering length from $400~$MeV to $700~$MeV.
This led to a correlation band which is nearly identical to the ones mapped out when
using different NN potentials (in part also at different cutoffs) or different
3NIs. All results were converged  numerically. Moreover, we demonstrated that
the theoretical uncertainty for $B(\alpha)$ decreases from LO to NLO by a factor
consistent with the \textit{a priori} expansion parameter estimate $Q\approx\frac{1}{3}$. In both
cases, the physical datum lies well inside the NLO correlation band.
\par
Assuming that a 4NI enters at NLO, not accounting for it would
then have two effects. First, the results could be unstable against cutoff
variations, indicating that a 4NI is
necessary at NLO to renormalize EFT($\slashed{\pi}$). No such effect is seen here, but a cutoff variation beyond the window chosen in
this exploratory study needs to be performed. Secondly,  the theoretical accuracy of the results
would be reduced to LO even if results independent of the cutoff can be
achieved, \textit{i.e.}, even if the cutoff can be removed to infinity without a
4NI. In other words, when one separately or collectively varies the cutoff
or the NN potential used or the 3NI strength, the different results should
spread in a corridor set by the size of LO corrections, \textit{i.e.}, $\lesssim 30$\%,
and not by the corridor of $\lesssim 10$\% expected in a NLO
calculation. Above, the residual
short-distance dependence of bound- and scattering-state observables is $\lesssim 10$\% of the central
value, taking as a conservative estimate the combination of errors which occur
when varying EFT($\slashed{\pi}$) at unphysically short distances: Different, phase-equivalent NN potentials; different 3NIs; different cutoffs. The correlation
bands and error estimates are thus quantitatively consistent with those
of a NLO calculation.
A residual 4NI which is not necessary for
renormalization but with unnaturally large coefficient can therefore be ruled out.
\par
My conclusion is that there is strong evidence that 4NIs do not enter at either LO or NLO in EFT($\slashed{\pi}$).
\section{A=5}\label{sec_a5}
In the preceding section, it was shown that EFT($\slashed{\pi}$) describes low-energy bound- and scattering- three- and four-nucleon systems systematically.
In this chapter, its application to 5-helium is considered. The five-nucleon system is found different from its t,$^3$He, and $^4$He predecessors. All its isotopes,
 \textit{i.e.}, $^5$n, $^5$H, $^5$He, $^5$Li, and $^5$Be are unstable.
In this respect, it has no analogy in gravitational or bosonic systems. Given reasonable initial conditions, a system of five planets will be bound by a classical
Newtonian interaction. For identical bosons interacting via short-range forces, bound states have been predicted for all systems of less than $15$ particles.
In $^5$He, both types of an interaction are present: a long-range Coulomb force and a short-range NN force with unnaturally large scattering lengths.
The short-range repulsion following from the Pauli exclusion principle is implicit in the antisymmetrized states.
It is the qualitative difference in the spectra of these systems - bound compared to unbound - emerging from interactions with very similar characteristics that
motivates this analysis. Besides that, it is of interest to test whether EFT($\slashed{\pi}$) can be used for a description of low-energy five-nucleon scattering
processes, a non-trivial generalization like the ones from two- to three-, and from three- to four-body systems. Adding a third nucleon made a modification of
the na\"ive power counting necessary, and there is no \textit{a priori} argument known why this cannot happen again when a fifth nucleon is added. Yet,
a five-nucleon force would have to be momentum dependent which suggests its promotion even more unlikely.
The spectrum of $^5$He in particular contains two resonant states$^\text{\cite{a5-spect}}$ in the
$J^\pi=\frac{3}{2}^-,\frac{1}{2}^-$ channels, at energies of $\approx 0.89~$MeV and $\gtrsim 1.5~$MeV above the $^4$He-n threshold. Both energies are small
compared to $B(\alpha)$, and the typical momenta of the neutron relative to the $\alpha$ core are then expected to be small as well.
A breakdown of EFT($\slashed{\pi}$) is therefore not expected for five nucleons.
\par
The universality of no shallow bound state is a different aspect.
Intuitively, this non-existence of a real bound state of five nucleons is not expected to be a prediction at LO or even NLO in the pionless EFT.
Considering the LO uncertainty in $B(\alpha)$ (see ch.~\ref{sec_a4}) of about $\pm 7~$MeV, a theoretical error in $B(^5\text{He})$ of similar magnitude
would na\"ively be expected to allow for a real instead of a virtual bound state.
I emphasize, this does not mean that 5-helium is outside the range of applicability of the pionless theory. The criterion of the validity of the theory
is rather the prediction of a state at LO whose energy is renormalization-scheme independent within reasonable limits, \textit{e.g.},
roughly the ones found in $B(\alpha)$. Furthermore, with NLO corrections the magnitude of this dependence must decrease for the theory to be
useful. Therefore, I search for a state of energy of about $\pm 7~$MeV relative to the $\alpha$-n threshold.
Different models of the short-distance structure of the NN interaction could then shift the physical two negative parity resonances
below threshold to become bound states. If such a state is shallow, the system would be similar to the deuteron but contrast the three- and four-nucleon bound systems
since its bound state is very shallow
relative to the lowest breakup threshold. The (un)bound $^5$He would thus require a particular short-distance structure and would not be a universal
consequence of an unnaturally large pair of NN scattering lengths and $B(t)$.
\par
The analysis is not intended to rival
the most recent, accurate descriptions of $\alpha$-n scattering, treated as a five-body problem. For the interaction models \texttt{AV18+UIX/IL2}, obtained
using a combined variational- and Green's function Monte Carlo technique, predictions are in good agreement with data for $\frac{1}{2}^+$,$\frac{3}{2}^-$,$\frac{1}{2}^-$
phase shifts and the resonance positions$^\text{\cite{an-gfmc}}$. A comparable agreement is achieved with the combined NCSM/RGM method for chiral N$^3$LO
interactions$^\text{\cite{an-ncsm}}$. The interaction in both analyses reproduced the binding energies of the deuteron-, triton-, and $^{3,4}$He subsystems precisely,
in contrast to our LO investigation. The essence of this section will be twofold: First, in contrast to the n-d system, where an accurately reproduced deuteron target
resulted in an improved accuracy for the phase shifts (see discussion of fig.~\ref{fig_d-n-ph}), a well described $^4$He at LO does not yield a similar accuracy
for the $\alpha$-n phase shifts. We also showed that the straightforward generalization of matching to a
Z-parameterized amplitude instead of the less appropriate ERE does not improve
convergence in $\alpha$-n phase shifts at LO in EFT($\slashed{\pi}$). Second, the results suggest that, counterintuitively, an unbound five-nucleon system
is a universal feature of the two- and three-nucleon interaction.
\par
The phenomena I will describe are similar to the findings for few-boson systems where two-boson interactions with a scattering length $|a|\to\infty$ and
repulsive three-boson interactions were considered$^\text{\cite{stecher}}$. If the triplet and singlet np scattering lengths $a_{s,t}$ are treated as infinite, the difference
to the nuclear systems, considered here, lies in the behavior of the interacting particles under permutations, namely identical bosons compared to two species of fermions.
In \cite{stecher} one five-body bound state with a binding energy of about ten times that of the associated trimer was found. In addition, either a second relatively weakly bound state,
or a state ``on the verge of being bound'', was discovered.
This finding is a guideline for our investigation. If a deeply bound state is found in the five-nucleon system, it would certainly not be describable with EFT($\slashed{\pi}$).
The sought-for useful description of the first excited, shallow state with EFT($\slashed{\pi}$) should not be affected by the possible existence of the deep ground state.
Again, gravitational systems provide a heuristic justification for this assertion. The possible formation of a black hole from five massive objects does not
render the Newtonian approach to the system under ``normal'' conditions useless.
\par
The analysis is comprised of two parts. First, EFT($\slashed{\pi}$) interactions are used at LO to predict low-energy $\alpha$-n phase shifts. In this course, both
of the aforementioned goals are addressed: Is the LO uncertainty reduced if the $\alpha$ target is reproduced accurately? In this respect, does the system behave
similar to d-n scattering (see ch.~\ref{subsec_nd-phase})? Furthermore, the behavior of the phase shifts should already indicate the existence of bound structures.
Second, a LO and NLO EFT($\slashed{\pi}$) analysis in the RGM bound-state formalism is conducted to solidify the implications of the P-wave phase shift results.
\par
Preceding the presentation of the results, a brief, qualitative comparison of the $\alpha$-n and the n-p system shall later allow for a better interpretation of them.
One can think of elastic $\alpha$-n, d-n, and n-p scattering as two-body problems with the distinguishing property for the expected behavior of the phase shifts in a given channel
being the relative position of the thresholds and bound states.
There is an important difference between the $\alpha$-n and the NN system. Treating the $\alpha$-n $l_\text{rel}=1$ amplitudes analogous to the np system would
suggest them as higher-order quantities not subject to constraints at LO. Only the $^2S_\frac{1}{2}$ amplitude would then be expected to be predicted within a defined uncertainty range.
The shallow virtual P-wave state, however, discriminates the $\alpha$-n from np. In \cite{bert-an} it was shown how the power counting in an $\alpha$-n halo EFT
for a shallow, real, or virtual bound P-wave state does not follow na\"ive dimensional analysis. The halo EFT considers the $\alpha$ as an inert DoF, and hence, $\alpha$-n as
a two-body problem. The enhancement of a certain class of diagrams then parallels that in the np system in EFT($\slashed{\pi}$), where it allows for the real
(virtual) state in the $^3S_1$ ($^1S_0$) channel (see ch.~\ref{sec_eft}) but not in the P-waves. This similarity of the $^2P_J$ channels in $\alpha$-n and $^{2S+1}S_J$ channels
in np motivates the attempt to describe the low-lying states with the LO EFT potentials derived in ch.~\ref{sec_pots}, for the reasons outlined below:
\par
First, possible bound states are expected to be formed by states with typical momenta between the nucleons in the core and the neutron that are small compared
to momenta conjugate to the range of the interaction between the $\alpha$ and the neutron. A halo structure similar to $^6$He is expected for a weakly bound $^5$He, which
implies a broad coordinate-space and correspondingly narrow momentum-space wave function.
\par
Second, $^2P_J$ bound states are the first candidates for universal properties associated with $a_{s,t}$ and $B(t)$ or $B(\alpha)$, analogous to the aforementioned bosonic five-body states which contain relative S-waves only.
Universal shallow $^5$He states cannot be pure S-waves, because such states are not accessible for three identical fermions.
As I will argue below, the $\alpha$ core is described in terms of an S-wave basis at LO and pNLO.
Therefore, one must allow at least one relative P-wave for a totally antisymmetric wave function.
Furthermore, the P-wave channels are considered here as a subsystem of the $^6$He
state with its two halo neutrons in relative P-wave states but a total angular momentum of zero (see ch.~\ref{sec_a6}).
\par
The construction of the RGM model space is presented in some detail here, because of its peculiar simplicity compared to the larger spaces employed in potential-model
calculations. Already in the deuteron, the deviation from most approximations to its wave function is apparent since it
is a pure S-wave state up to NLO in EFT($\slashed{\pi}$).
The construction of the $^5$He RGM model space starts with selecting an appropriate set of groupings of the two protons and three neutrons. A $\alpha$-n division only is used here.
As we consider the $\alpha$-n scattering system only for $E_\text{cm}<10~$MeV, \textit{i.e.}, below the lowest breakup threshold, this is a reasonable approximation. For the
hypothetical bound state, more care has to be taken. Assuming that such a state is similar in its shallowness, momentum distribution, charge- and matter radius to the
$^6$He halo structure, the approximation would certainly be justified, too. It is important that this assumption contradicts the possibility of a similarly large LO uncertainty
in the energy of the four- and five-nucleon system. According to this estimate, a bound state of about $7~$MeV below threshold would be admissible but certainly not exhibit a
pronounced halo structure. Nevertheless, we will justify \textit{a posteriori} that other groupings like t-d are negligible. The reasoning is the following:
if such a fragmentation is significant, so will states of the $\alpha$-n structures that resemble relatively small separations between the core and the neutron - both objects
are small. This is not found in our calculations, and the predictions are derived with the $\alpha$-n group structure only.
\par
Of the two fragments, only the $\alpha$ has a nucleon substructure. Its four constituents can be divided into two deuterons, or a triton and a proton, or a $^3$He and a neutron,
or unbound nn-pp and \mbox{d\hspace{-.55em}$^-$}-\mbox{d\hspace{-.55em}$^-$} spin singlets.
This leaves the two- and three-body matryoshkas to be opened: the triton reveals a triplet and singlet np system with a neutron, and
the 3-helium just replaces the single neutron with a proton.
Thereby we have defined all relevant groupings and the accompanying angular momentum coupling schemes and definitions of Jacobi coordinates. The specific values for the
orbital angular momenta and the Gau\ss ian width parameters are now to be chosen to appropriately represent the small dolls in the larger ones, \textit{i.e.} for example,
the triton within the $\alpha$.
\par
\begin{figure}
  \includegraphics[width=1.\textwidth]{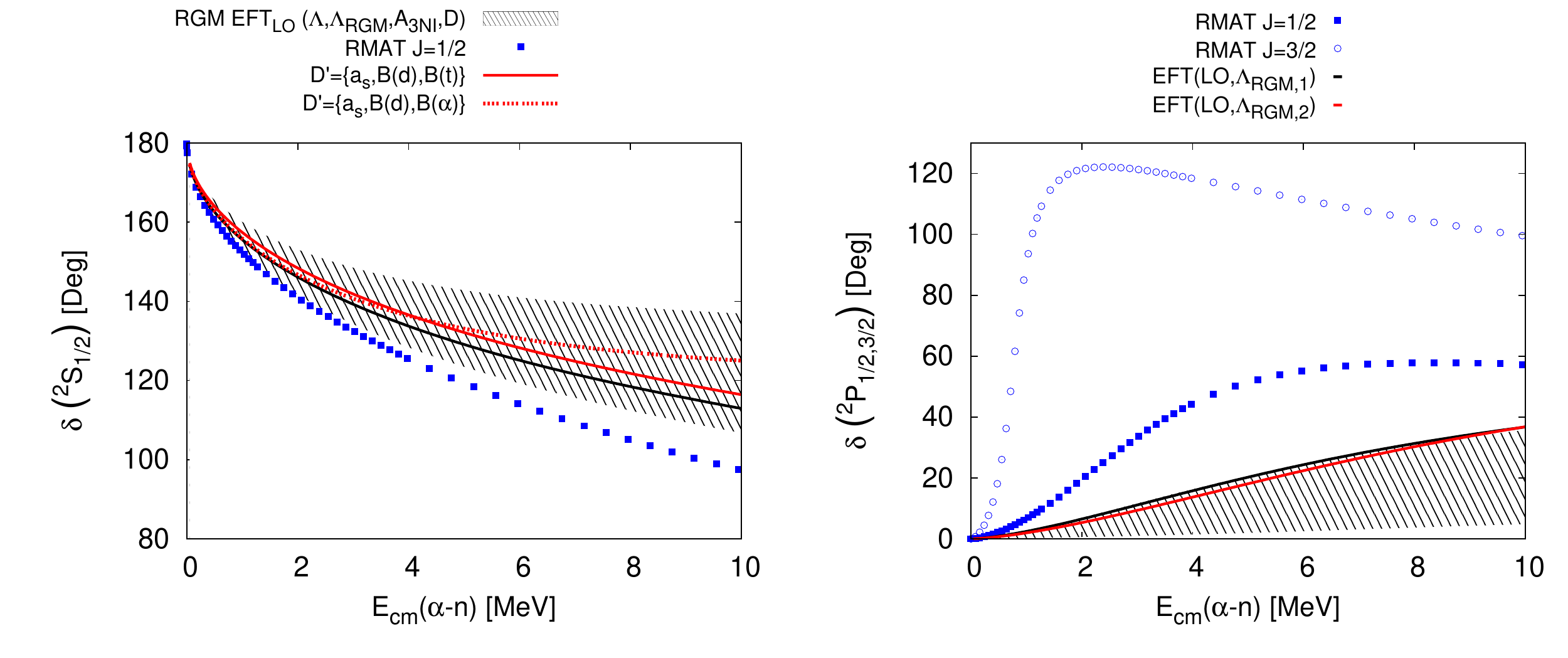}
  \caption{\label{fig_5-he-phases}\small Phase shifts for elastic $\alpha$-neutron scattering. Results of LO EFT($\slashed{\pi}$) potentials with cutoffs from
$300$~MeV - upper (lower) edge of the shaded area in the left (right) panel - to $1.6$~GeV - lower (upper) edge of the shaded area in the left (right) panel -
are compared to R-matrix$^\text{\cite{hale-priv-comm}}$ fits.
The shaded areas mark the assessed LO uncertainty by a variation of four renormalization parameters: the regulator parameters $\Lambda$ and
$\Lambda_\text{RGM}$ (corresponding to two model spaces for the $\alpha$ core, black and red lines),
the three-body interaction strength $A_{3\text{NI}}$, and two fit input sets (solid lines use $B(t)$, the dashed line uses $B(\alpha)$ with all other renormalization parameters fixed).
}
\end{figure}

The lightest bound structure is the deuteron, which is well described as an S-wave state up to NLO (ch.~\ref{sec_a2}). The bound $A=3$ systems are therefore in a first approximation
modeled by two S-waves. The angular momenta on the two Jacobi coordinates of the three-body system have to be the same, in any case, to couple to a total $l=0$.
Only states with total $l=0$ states are of interest, because only they have nonzero overlap with the four-nucleon ground state in this approach.
From a technical perspective, this is because the interaction
is even at pNLO a spherical tensor or rank zero, \textit{i.e}, a central operator that does not mix states with different total orbital angular momenta $l$. Contributions to the bound
three-nucleon binding energies at LO EFT($\slashed{\pi}$) from states with two P-waves were found to be less than $\sim 100~$keV. Even relative to NLO accuracy,
this is a small amount, and the omission of such configurations is justified. Now, with d, t, and $^3$He as S-wave states, an S-wave for the third coordinate of the
enclosing $^4$He doll is implied. Using a short-hand notation, the ansatz for a $^5$He wave function used in this calculation is:
\begin{equation}\label{eq_5he-wfkt}
\vert ^5\text{He}\rangle=\vert\text{d-d}\rangle_{118}\vert \text{n}\rangle_{14}+\vert\text{t-p}\rangle_{85}\vert \text{n}\rangle_{14}+\vert ^3\text{He-n}\rangle_{87}\vert \text{n}\rangle_{14}\;\;\;.
\end{equation}
The subscripts denote the number of basis vectors. For instance, there are $85(\text{triton-p})\times 14(\text{relative }\alpha\text{-p widths})$ t-p configurations in a
five-nucleon basis basis of dimension $4060$. Fourteen different width parameters,
$$\mathcal{W}_{\alpha-\text{n}}=\left\lbrace\begin{array}{c}1.6902,1.3423,0.8214,0.4447,0.2574,0.1690,0.1185\\
0.0843,0.0500,0.0257,0.0186,0.0097,0.0056,0.0028
\end{array}\right\rbrace$$ expand the relative motion between the neutron and the 290-dimensional $\alpha$. With the 20-dimensional deuteron w120 as base, and
the same $20$ width parameters to expand the d-d, d-n, d-p, d-p, and $^3$He-n functions, we selected two sets of 290 basis vectors from the $8000$ for $^4$He.
One set was optimized to yield maximal $B(\alpha)$ for $\Lambda=200~$MeV, the other for $\Lambda =1600~$MeV. The two model spaces of equal size define the
values $\Lambda_{\text{RGM},(1,2)}$ used in fig.~\ref{fig_5-he-phases}. For all $\Lambda$, the calculations are conducted in both model spaces as an additional
assessment of the renormalization-scheme dependence of the phase shifts parameterized by these two $\Lambda_\text{RGM}$ values.
\par
\newpage
With the variational basis defined, I will discuss the LO treatment of elastic $\alpha$-n scattering in the $^2S_\frac{1}{2}$ and $^2P_J$ channels.
The leading-order potentials specified in table~\ref{tab_lo-pots} are employed, thereby covering a cutoff range from $300$~MeV to $1.6$~GeV.
In order to search for bound states, I am looking for a rapid decrease of the phase shift in the respective channel through $\frac{\pi}{2}$ (see, \textit{e.g.},
\cite{newton-bound}) and a corresponding eigenvalue of the Hamiltonian below the lowest, \textit{i.e.}, $\alpha$-neutron, threshold.
This threshold is identical to $B(\alpha)$, and varies with $\Lambda_\text{RGM}$, \textit{i.e.}, the model space used for the $\alpha$ fragment.
\par
The results for the phase shifts are shown in fig.~\ref{fig_5-he-phases}.
Within LO accuracy, the phases do not indicate a bound state in either channel. No crossing through $\frac{\pi}{2}$ is observed. The shaded areas are mapped out by
all our LO predictions and assess the uncertainty by their maximal deviation from the R-matrix data (blue). The spread was induced by the different short-distance
structure of the interaction. The various models are parameterized here analogous to how we probed for renormalization-scheme dependence in $A\leq 4$ systems:
\newpage
\begin{enumerate}
\item Variation of the regulator parameter $\Lambda$ from $300~$MeV (upper ($^2S_\frac{1}{2}$)/lower ($^2P_J$) edge of the shaded area) to $1.6~$GeV
(lower ($^2S_\frac{1}{2}$)/upper ($^2P_J$) edge of the shaded area);\vspace{-.2cm}
\item Change of the variational basis, \textit{i.e.}, $\Lambda_\text{RGM}$, using two $\alpha$ core basis sets (solid red/black) with which
$B(\alpha)$ is predicted consistently within LO
and pNLO uncertainty; This corresponds to the above mentioned reduction of the basis size from $8000$ to $290$ vectors, without significant effect on the binding energy.\vspace{-.2cm}
\item Two sets of input data to determine the LO LECs:\\ $\mathcal{D}'_1=\lbrace a_s,B(d),B(t)\rbrace$ (solid lines), $\mathcal{D}'_2=\lbrace a_s,B(d),B(\alpha)\rbrace$ (dashed lines);\vspace{-.2cm}
\item Various values of the three-nucleon interaction parameter $A_{3\text{NI}}$ for which the resultant $B(\alpha)$ and $B(t)$ differ by less than $30\%$ from their
experimental values.
\end{enumerate}
The third option is to be compared to the d-n system. Here, it is the $\alpha$ which is either predicted less strongly bound ($\mathcal{D}'_1$) or taken as input ($\mathcal{D}'_2$).
In contrast to fitting to $B(d)$ in the three-nucleon case, no improvement in the predicted phase shifts is found for an $A_{3\text{NI}}$ fitted to $B(\alpha,\text{exp})$.
On the contrary, the example displayed in the left panel of fig.~\ref{fig_5-he-phases} shows for $\Lambda =300~$MeV and within one model space that using $\mathcal{D}'_2$ yields
a larger deviation (red dashed) from data relative to $\mathcal{D}'_1$ (red solid). This confirms the difference to n-d scattering where an
accurately reproduced deuteron fragment improved the accuracy of phase shift predictions significantly.
Regarding the fourth handle defined in the list above, we found
the phases to converge to the upper edge of the shaded area when $A_{3\text{NI}}$ was varied such that $B(\alpha,\Lambda_\text{RGM})$
is increased even beyond $B(\alpha,\text{exp})$. At a certain $A_{3\text{NI}}$, the system develops a resonance and the graph of the phase shifts discontinuously
changes its shape, dropping rapidly through $\frac{\pi}{2}$ at about $1~$MeV. This signals the appearance of a bound state for a sufficiently weak 3NI. I assume this state to be the
analog of the deep ground state found for bosons, and therefore refrain from a further investigation here, because of the inappropriate
model space and the expected breakdown of the pionless theory for such a deep state.
\par
Using $B(d)$ instead of $a_t$ for the determination of the LECs did not change $\alpha$-n phase shifts
significantly, although the respective $B(\alpha)$ differ by up to $5~$MeV. The small difference in phase shift predictions for the two $\Lambda_\text{RGM}$ values
(solid black and red) could be
compensated by a variation of $A_{3\text{NI}}$, thus verifying once more the interpretation of an appropriate truncation of the variational basis with a regularization
parameter. Further dependence on $\Lambda_\text{RGM}$ was probed by changing the number of included distortion channels from $800$ up to $1280$, with
no significant effect. This justifies \textit{a posteriori} the omission of the other five-body groupings - \textit{e.g.}, the closed t-d channel  - in the basis which are not
expected to contribute as distortion channels when channels which are already open yield only small improvements.
\par
The predicted phase shifts in the $^2P_\frac{1}{2}$ and $^2P_\frac{3}{2}$ channels are identical because the LO interaction is independent of $\vec{J}$ and $\vec{J}^2$.
The power counting suggested by halo EFT for shallow P-wave states includes $\vec{L}\cdot\vec{S}$ operators at LO$^\text{\cite{bert-an}}$. Consequently, the
two P-wave channels are coupled; a feature that would express itself not before N$^3$LO in the NN EFT($\slashed{\pi}$) considered in this work. Therefore, I do not
attempt to refine predictions for scattering phases with a pNLO EFT($\slashed{\pi}$) calculation, as it does not yet include $\vec{J}$ dependend operators, either.
Compared to the np P-wave phase shifts (fig.~\ref{fig_nn-phases}), the $\alpha$-n analogs are also small but are predictions and not input.
\par
In fig.~\ref{fig_he5-spect}, the results of RGM bound-state calculations in the $^2S_\frac{1}{2}$ and $^2P_J$ channels at LO and pNLO in EFT($\slashed{\pi}$) for
three cutoff values, $\Lambda =300~$MeV, $400~$MeV, and $500~$MeV, are presented.
The discussion in app.~\ref{app_rgm-pnlo} explains this limitation to relatively small values.
\input{he5_spect.tex}
When the model space was defined, it was already mentioned that a $^2S_\frac{1}{2}$ state of five nucleons is Pauli forbidden at LO because, at this order,
all nucleons are in relative S-wave states, \textit{i.e.}, the coordinate-space
wave function is symmetric like the isospin function of three nucleons of the same species. As there is no totally antisymmetric spin state for three spin-$\frac{1}{2}$
functions, the $^2S_\frac{1}{2}$ five-nucleon state vanishes.
Although this is precisely what we confirm for $\Lambda =300~$MeV and $500~$MeV in fig.~\ref{fig_he5-spect}, a detour on how such states could emerge as artifacts of the
finite RGM basis is given.
An asymmetry is introduced in the coordinate wave function by using different width parameter sets
to expand the radial wave functions of the various Jacobi coordinates. This will allow a $^2S_\frac{1}{2}$ state, nevertheless, in the sense that the explicit antisymmetrization
of the RGM basis (eq.~(\ref{eq_rgm-int-wfkt})) results in  mixed symmetry states in coordinate- and spin space. Only if a sufficiently large basis is chosen for all
Jacobi coordinates, a then symmetric coordinate space wave function will ensure that antisymmetrization will eliminate this state. In practice, this limit can be
identified by numerical instabilities due to numerically zero eigenvalues of the norm matrix because of the almost zero states. This leads via eq.~(\ref{eq_eigensystem})
to large eigenvalues of the Hamiltonian which in turn are sensitive to modifications of the basis at the level of adding/removing single vectors. In effect, the asymmetry introduced by
the small basis models, \textit{e.g.}, states where the $\alpha$ fragment includes components in its basis with two $l=1$ angular momenta, still forming a state of total $l=0$.
This suggests an interpretation of a converged $B(\alpha)$ in a relatively small RGM basis. The explicit addition of structures with two relative P-waves could have been found
small, partly, because they are simulated by an appropriately distorted pure S-wave basis. The four-body calculation was carried out in a relatively large model space,
in which a totally symmetric coordinate space wave function can be assumed. Reducing the number of basis states by removing those which do not decrease $B(\alpha)$
significantly, introduces the mixed symmetry states in the $^5$He basis. The important conclusion from this is not to exclude basis states, which an EFT power counting
would demote to higher orders, \textit{a priori}, when operating in small RGM model spaces.
Therefore, I will include configurations in $^6$He with $l=1$ on the coordinates of the two (halo) neutrons coupled to a total $l=0$.
An incomplete basis with all five relative coordinates in S-waves introduces such configurations implicitly but will also yield a bound system.
Therefore, bases with both P- and S-wave vectors bear more potential for the formation of numerical linear dependencies if the S-wave configurations with
more distorted S-wave configurations.
The extreme scenario would
model the six-nucleon state with a distorted pure S-wave basis only. This allows a severe reduction in computing time while exploiting the insignificant
contribution at (N)LO for nucleons in relative P-waves utilizing a property of our EFT($\slashed{\pi}$) potentials.
\par
Returning to the discussion of fig.~\ref{fig_he5-spect}, the $^2S_\frac{1}{2}$-calculation yields for both cutoffs an eigenvalue (dashed (LO) and solid (NLO) black)
slightly above the $\alpha$-n threshold -
determined by $B(\alpha)$ at the respective order (gray (LO) and red (NLO)). The corresponding state is identified as an $\alpha$ core with an unbound neutron.
This interpretation is verified by the
disappearance of the state from the spectrum, once the broad basis states which expand the continuum relative wave function between the $\alpha$ and
the neutron are omitted. The corresponding eigenvalue is not to be confused with the position of a potential five-nucleon $\alpha$-n resonant state.
Information about the latter mandates boundary conditions on the relative wave function between the $^4$He and the neutron appropriate for
a scattering state. Such a condition (eq.~\ref{eq_rgm-rel-wfkt}) is not enforced here, and although the state might have an overlap with a resonant state, no information about
the latter can be deduced. For that purpose the RGM scattering formalism will be used in a future analysis.
The bound state pNLO calculation in this channel is then just another, yet powerful demonstration of the $\Lambda_\text{RGM}$ independence of the pNLO potentials as
$B(\alpha,\text{exp})$ is accurately reproduced here within a different model space than the one used to fit the LECs.
\par
While this is the result expected from the Pauli principle, the qualitatively identical results in the $^2P_J$ channel are non-trivial with a profound implication.
For all three cutoff values the spectrum does not contain a bound state even in this channel.
This finding suggests that the unbound nature of a five-nucleon state does follow universally from unnatural $a_{s,t}$ and $B(t)$.
Explicitly, I conclude from the results, that a system of five fermions does not sustain a shallow bound state
when all particles interact via short-range forces characterized by $a_{s,t}$ and $B(t)$ and, additionally,
two particles repel each other with a Coulomb force. This does not stand in contrast to the bosonic case. A deeply bound state,
as it was found there, might also be an element of the $^5$He spectrum but have been cut off by the employed RGM regulator. In more detail, a state with a large binding
energy relative to the $\alpha$ is also expected to be accordingly more localized in coordinate space. Therefore, its expansion in terms of Gau\ss ian functions
requires narrow width parameters. As those parameters are absent in the employed basis, the small structure of the hypothetical deep ground state cannot be resolved.
\par
Finally, a limitation of the analysis presented here shall be mentioned. The universality, \textit{i.e.}, the independence with respect to details of the microscopic
structure of the interaction, of the unbound five-nucleon system was inferred from three models of this short-distance structure only, corresponding to the three cutoff values.
Yet, in combination with the phase shift results, even this
limited probe provides strong support for our conclusion. The dependence of the phase shifts on the renormalization-scheme
was analyzed much more thouroughly and results were found consistent with the nonexistence of a bound state. For instance, the phases predicted with $\Lambda =1.6~$GeV
do not differ to an extent that would suggest a profound difference of the spectrum relative to that of, \textit{e.g.}, $\Lambda =800~$MeV.
\input{he5_spect-nocoul.tex}
\par
This would contrast the appearance of a bound state for cutoffs or other renormalization parameters for which no explicit RGM bound-state calculation was
conducted. Instead of extending the analysis in this direction, we therefore deem the assessment of the role of the Coulomb repulsion - whether
it is decisive for rendering the system unbound - more interesting. The contribution from the
Coulomb interaction to the energy eigenvalue was throughout of the order of $+1~$MeV, \textit{i.e.}, the Coulombless system is more deeply bound.
Intuitively, one could imagine that it is this small amount which makes the
difference between the bound- and unbound system, especially because such an interaction was not considered in the calculations for bosons.
The predictions calculated with the proton charge set to zero (left spectrum) are compared in fig.~\ref{fig_he5-spect-nocoul} to the previously discussed
results (right spectrum) with charged protons. Protons and neutron are still distinguishable but the interaction is now a scalar in isospin space, \textit{i.e.},
nn-, np-, and pp interactions are identical. We used the same model space and one fixed regulator with $\Lambda =400~$MeV.
At LO, this results in a four-body bound state (gray line in fig.~\ref{fig_he5-spect-nocoul})
$\sim 0.91~$MeV below the $B(\alpha,\text{LO})$ result of fig.~\ref{fig_he5-spect}. This energy sets the lowest breakup threshold for the five-particle system, whose lowest LO
eigenvalue was calculated as $B(5,\text{LO})=23.27~$MeV (dashed line). Using the same potential as before with nonzero Coulomb interaction, the pNLO calculation
no longer yields $B(\alpha,\text{exp})$ but a deeper bound four-body state at about $28.88~$MeV (red solid, left column). In effect, not the binding energy of the $\alpha$
was used then to fit the LEC but this shifted value. This is admissible because there is no intentional experimental measurement for an uncharged four-nucleon system. Any value
of about the same distance from threshold is appropriate. Nevertheless, the right column in fig.~\ref{fig_he5-spect-nocoul} shows the results for a 3NI whose parameter
was adjusted at NLO to yield a four-body binding energy of $B(\alpha,\text{exp})$. In both approaches, the predicted lowest energy eigenvalue of the five-body
system (solid black in fig.~\ref{fig_he5-spect-nocoul}) was about $0.56~$MeV above this threshold at pNLO. This result suggests the generalization of our previous conjecture:
A system of two fermions of type one plus three fermions of type two which interacts via short-range forces characterized by two unnatural scattering lengths and a
three body binding energy, does sustain a four-body bound state but no shallow five-body bound state in its vicinity. The Coulomb repulsion between two of the
five particles is not responsible for this unexpected phenomenon. This finding is to be confirmed for a probe of the sensitivity of the difference $\vert B(4)-B(5)\vert$
towards a change in the short-distance behavior of the interaction. Here, the Coulombless case was only considered for one cutoff value and two renormalization
conditions. However, compared to the ``charged'' case, the threshold separation did not change significantly and na\"ively it would come as a surprise if
this pattern would change with $\Lambda$ or $\Lambda_\text{RGM}$.
\section{A=6}\label{sec_a6}
Within the limits regarding cutoff variation and model-space dependency, it was shown in the previous chapter that the physical unbound $^5$He system is a
prediction of the pionless EFT. In this section the aim is to predict not only qualitative features like this but also to demonstrate the convergence of the EFT
for the next heavier nucleus, 6-helium.
The study is exploratory, and for definite conclusions about the usefulness of the pionless theory for the description of the 6-helium ground state
we rely on computational resources unavailable at present. Nevertheless, the results demonstrate the insight the EFT approach is able to provide.
Furthermore, the variational RGM method is established as another numerical tool for the investigation of extended halo structures (see \textit{e.g.} \cite{papen-hagen-bacc}
for an investigation using the hyperspherical-harmonics and the coupled-cluster methods).
\par
In contrast to 5-helium, $^6$He sustains a real shallow bound state and not just shallow resonant states.
Besides lying on opposite sides of the breakup threshold, the two systems are similar with respect to their pole position relative to the lowest breakup threshold.
Experimentally, the $J^\pi=0^+$ ground state of $^6$He
is only bound by about $0.975~$MeV below the three-fragment $\alpha$-nn breakup threshold, set here by $B(\alpha)$. This value is of the same order
as the pole positions of the lowest $^2P_J$ $^5$He resonant states above threshold. It is small relative to the binding energy of the $\alpha$ fragment in the lowest
breakup channel.
\par
We have shown in ch.~\ref{sec_a4} that the $\alpha$ system can be described with EFT($\slashed{\pi}$), and because the two additional neutrons increase the energy
in the system only marginally, typical momenta of nucleons in the six-body system are also expected to be below the breakdown scale of the pionless theory.
Therefore, it is a worthwhile endeavor to apply EFT($\slashed{\pi}$) to the six-nucleon system with the expectation that the physical state is within its range of applicability
and hence predicted with increasing accuracy order by order. This would be a remarkable consequence because of the peculiar halo structure of the $^6$He ground state.
\par
The attribute ``halo'' refers to the existence of sufficiently distinct momentum scales in the system$^\text{\cite{vkolck-halo-def}}$:
\vspace{-.3cm}
\begin{itemize}
\item In an $A$-nucleon system, one scale ($M_\text{hi}\sim\sqrt{m_\text{\tiny N}E_c}$) is associated with the excitation energy $E_c$ of a core structure.
This core is composed of a subgroup of $A'<A$ particles.\vspace{-.2cm}
\item A second scale ($M_\text{low}\sim\sqrt{m_\text{\tiny N}E_h}$) which is small relative to $M_\text{hi}$ and is set by the separation energy $E_h$
of nucleons (one or more) from the bound $A$-body system.\vspace{-.2cm}
\end{itemize}
As a consequence of the low separation energy, the probability to detect neutrons at distances large compared to the size of the core is unusually high.
The large extension of the halo nucleus has in turn important consequences for low-energy scattering reactions. The corresponding cross sections are
then found very different from those where the target's size is given by the interaction range of its constituents. Another peculiarity of the $^6$He halo in
particular is its borromean structure, \textit{i.e.}, its lowest breakup channel consists of three fragments. That corresponds to the three interlocked rings
which separate if only one of them is broken. Analogously, if one neutron is removed from $^6$He, the remaining $\alpha$ and neutron will not form a bound
system.
\par
According to the given definition of a halo, we already discussed two halos in this work: first, the deuteron as the lightest nuclear halo.
Consistent with the pionless theory, the proton core of this two-body state can be treated as a point particle
for probes which transfer only momenta less than $M_\text{hi}\sim m_\pi$. Compared to this $M_\text{hi}$, the separation energy of the halo neutron, namely the deuteron
binding energy $B(d)\sim E_h$ is certainly small. The triton also meets the criteria and is an example of a two-neutron halo state. Again, the proton-core scale is
large relative to the neutron separation energy of about $E_h\sim B(t)-B(d)\approx 6.3~$MeV. The $\alpha$ particle does not exhibit a halo structure. Its neutron separation energy
of about $21~$MeV exceeds the excitation energy of the remaining 3-helium. With all five-nucleon systems unbound, $^6$He is the next largest, stable halo nucleus.
Like the triton a two-neutron halo, its core structure is an $\alpha$ particle with an excitation energy of approximately $E_c\approx 10~$MeV set by the t-p threshold.
Compared to this $E_c$, the neutron separation energy of 6-helium $E_h\approx 1~$MeV is reasonably small.
\par
In contrast to the deuteron and the triton, the halo character of 6-helium would be a prediction of EFT($\slashed{\pi}$). In the two-nucleon system, a finely tuned
short-range interaction produces unnaturally large scattering lengths and by that the halo. Without this fine tuning, \textit{e.g.}, between strength and range of the
interaction, a generic potential is more likely to produce a scattering length of the same order as the interaction range, which would correspond to a deeper lying
ground state. The shallow triton constrains the nuclear interaction further. Not all interactions that produce a shallow deuteron predict a shallow triton. A particular three-nucleon
parameter, or a specific cutoff value, as two specific examples for renormalization parameters, have to be chosen for the reproduction of the physical, shallow triton state.
The two- and three-nucleon halos are therefore input for EFT($\slashed{\pi}$), while the six-nucleon halo would, without modification of the power counting, be a prediction.
The halo character might, however,
express itself only at higher orders, which would make it a consequence of details of the interaction, \textit{e.g.}, specific values of higher-order effective-range parameters
($r_l$ and $v_l$ in eq.~\ref{eq_ere}) and not a universal property of the shallow deuteron and triton ($a_{s,t},B(t)$). This is because of the LO uncertainty assessed
for $B(\alpha)$ in ch.~\ref{sec_a4} of about $7~$MeV. An uncertainty in $B(6)$ of similar magnitude would allow for neutron separation energies of comparable size to
the approximate $10~$MeV $M_\text{low}$ scale set by the $\alpha$ core. Therefore, the clear separation of scales would develop at higher orders, where the usefulness of
EFT($\slashed{\pi}$) is determined by the amount by which the uncertainties of predictions for $\vert B(6)-B(\alpha)\vert$ decrease from LO to NLO.
\par
We employ RGM bound-state calculations in the $J^\pi=0^+$ 6-helium channel at LO and pNLO within EFT($\slashed{\pi}$) in order to assess
whether a bound six-nucleon system is a universal consequence of shallow two- and three-nucleon states, and if the halo character of this six-body object is
a perturbative effect.
First, the RGM model space is defined. I follow the same procedure as in the previous chapter:\vspace{-.25cm}
\begin{itemize}
\item selection of relevant groupings/fragmentations $\mathcal{G}'$ which determine the\\ Jacobi coordinates;\vspace{-.25cm}
\item selection of maximal angular momenta for the Jacobi coordinates and accompanying orbital- and spin-coupling schemes;\vspace{-.25cm}
\item selection of sets of Gau\ss ian width parameters appropriate for the expected spatial extension of the six-nucleon state in the range of applicability of EFT($\slashed{\pi}$).\vspace{-.2cm}
\end{itemize}
Consistent with the physical two-neutron halo structure around an $\alpha$ core, previous RGM analyses$^\text{\cite{wurzer-halo}}$ with model interactions identified the
$\alpha$n-n fragmentation as the dominant structure. This $^5$He grouping alone was found to bind $^6$He. Other group structures, \textit{e.g.}, triton-triton, typically contributed
less than $1~$MeV to the ground-state binding energy. An expected LO uncertainty of about $10~$MeV justifies not to include these configurations in the variational basis.
For this work, we therefore approximate the $^6$He ground state by the single \texttt{Y}-shaped $^5$He-n group (left sketch in fig.~\ref{fig_6he-fragm}).
The other $\alpha$-2n configuration (\texttt{T}-shape, right sketch in fig.~\ref{fig_6he-fragm}) differs from the \texttt{Y}-configuration only in the definition of the Jacobi coordinates.
\input{diss_6he-fragm.tex}
The truncation of the partial-wave expansion of the two neutron coordinates $\vec{n}_{1,2}$ discriminates the two structures. They are equivalent if an infinite (complete)
set of width parameters and angular momenta is used to expand the functions dependent on $\vec{n}_{1,2}$. The contribution of \texttt{T}-groups is
estimated to be of the same order of that which is due to an increase of the number of widths in the \texttt{Y}-groups. I choose the sets from which the
$\gamma_{i,(1,2)}$ widths (defined in fig.~\ref{fig_6he-fragm}) were taken such that adding new widths to those sets did
affect the eigenvalues of the LO Hamiltonian by less than $1~$MeV. These are the widths used for approximating the neutron halo around the core.
An explicit inclusion of \texttt{T}-groups is hence unnecessary considering the expected accuracy of
a (N)LO calculation.
\par
The angular momentum structure is defined next. For the $\alpha$ core, the same arguments as given in the $^5$He chapter justify a pure S-wave ansatz of
t-p, $^3$He-n, and d-d groups. For the two neutron coordinates, the respective angular momenta have to be equal because the total angular momentum has to be zero.
Again, this is because the central (pN)LO interaction cannot induce transitions between states differing in their total angular momentum. Hence, two sets
($l_{1,2}=0,1$ in fig.~\ref{fig_6he-fragm}) of angular momenta are considered: one with two S-waves and the other with two P-waves coupled to a total $L=0$.
\par
\begin{figure}
  \includegraphics[width=1.\textwidth]{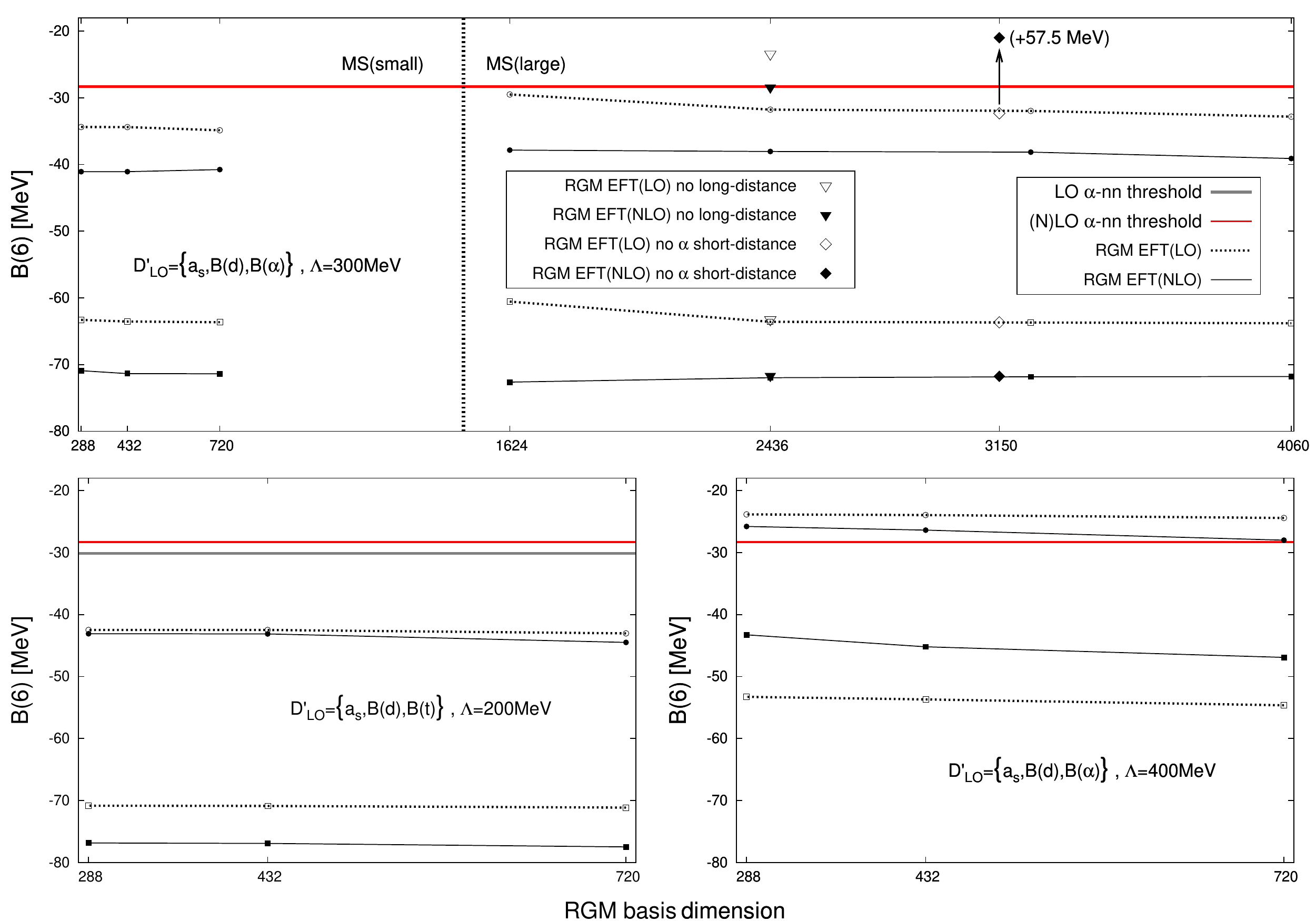}
  \caption{\label{fig_6he-spect}\small
EFT($\slashed{\pi}$) results for the (second) lowest energy eigenvalues ((circles) squares) in the 6-helium channel as a function of the basis size.
(pN)LO predictions ((full) empty) are shown for three cutoff values with either $B(t)$ ($\Lambda=200~$MeV) or $B(\alpha)$ ($\Lambda=300,400~$MeV)
as experimental input to fit LO LEC. For $\Lambda=300~$MeV, a small (left upper graph) and large (right upper graph) model space was analyzed including
values for a particularly (broad) localized basis ((diamonds) triangles). The red and gray horizontal lines mark the $\alpha$-nn breakup thresholds.
}
\end{figure}

An appropriate choice of width parameters for $\gamma_{i,(1,2)}$, \textit{i.e.}, the halo neutrons, and the $\alpha$ core completes the definition of the variational basis.
Initial calculations were carried out in a relatively large model space (MS(large) in top graph of fig.~\ref{fig_6he-spect}). This basis could be reduced to span a smaller
model space (MS(small) of dimension $\leq 720$; results in all graphs of fig.~\ref{fig_6he-spect}). First, I define the large space before its reduced derivative is introduced.
Heuristically, such a basis should be able to expand an $\alpha$ structure whose binding energy, and thereby approximate spatial extension, is known at each order of the calculation.
From the large four-body model space used in ch.~\ref{sec_a4}, a subset of $85$ t-p, $^3$He-n, and $117$ d-d components was selected such that $B(\alpha)$ calculated
in this $203$ dimensional space was only about $1~$MeV lower than in the large space. The parameter sets of the radial part of that component of the wave function which
depends on $\vec{n}_{1,2}$ were chosen under the following considerations:
Basis vectors which correspond to small separations between the core and a neutron are expected to have little overlap with a state within EFT($\slashed{\pi}$)'s range of
applicability. Such highly localized states are associated na\"ively with large typical momenta. We therefore allow only width parameters up to a certain $\gamma_\text{max}$
as larger widths correspond to narrower Gau\ss ians. I choose  the cutoff width for each regulator cutoff as: $\gamma_\text{max}\sim 1~\text{fm}^{-2}+\frac{\Lambda^2}{4}$.
Starting from this $\gamma_\text{max}$, three values were defined as means of normal distributions. The three distributions were used to generate $3\times 203$ width
parameters, where each set can be imagined as a compact $\alpha$ core with an attached neutron in a shell whose distance from the core is set by the respective
mean value. For the widths of the position of the second neutron, the same procedure was implemented but with five distributions generating $5\times 108$ different
parameters. The smallest of those five central values was chosen as $\gamma_\text{min}=0.0001~\text{fm}^{-2}$. Vectors including widths from this distribution correspond to
the broadest Gau\ss ians in the basis which are necessary to allow for the experimentally$^\text{\cite{he6-angle-exp-1,he6-angle-exp-2,exp-6he_rch}}$
and theoretically$^\text{\cite{he6-angle-theo}}$ established correlation between the neutron coordinates, which places them at almost opposite sides of the core
structure (fig.~\ref{fig_6he-fragm}, right \texttt{T}-configuration).
\par
I summarize the model space definition by giving the total number of basis vectors: A $203$-dimensional $\alpha$ core for each of the two sets of angular momentum
coupling schemes for the remaining two neutron coordinates; three sets of width parameters expand the $\alpha$-n coordinate, each covering a different range between the two
objects; five width sets expand the $\alpha$n-n Jacobi coordinate, allowing for both, an extended halo structure as well as a more deeply bound, localized state. Hence, the basis
is of dimension $2\times 203\times 3\times 5=6060~$\footnote{The computation was performed
on IA32 Xeon $2.66~$GHz nodes of the RRZE in Erlangen. For this basis, a total of $\approx 650~$CPUh were invested to calculate the LO NN+3NI and NLO+3NI
matrix elements up to the \texttt{QUAFL} stage of the program chain.}.
\par
A smaller basis was sought to facilitate the analysis of numerical instabilities associated with the RGM method, \textit{e.g.}, sensitivity of predictions to specific basis states
corresponding to either very broad or localized wave functions, the influence of the regularization of irregular Coulomb/Bessel functions, or the dependence on the energy used
to extract the scattering lengths for the LEC fit. For instance, we could sort out pNLO LEC sets which were numerically harmless up to the four-body problem but are unstable
in the six-body system due to the increase of the expectation values of single operators.
It was also possible to assess much quicker the impact of a shift of the interval from which the width parameters for the halo neutrons were selected. This enabled us to
optimize this range such that the second lowest eigenstate was reproduced more accurately in the small basis. Compare values marked by the full/empty circles in the upper
graph in fig.~\ref{fig_6he-spect} for a dimension $288$ and $4060$. The value obtained in the much smaller basis is converged to a lower value due to a wiser selection
of the interval for $\gamma_{i,(1,2)}$.
\par
The smaller model space was derived from the large, defined above, by
using the same \texttt{Y}-group structure with its two angular momentum coupling schemes but reducing the number of vectors responsible for the expansion of the $\alpha$
core and the first attached neutron. This basis consisted only of $144\times n$ vectors, where $n$ is the number of width parameters used to expand the dependency on
the coordinate of the second neutron. The results are presented for $n=2,3,5$ in fig.~\ref{fig_6he-spect} with each increment adding more localized basis states, only.
For $\Lambda=300~$MeV and $B(\alpha)\in\mathcal{D}'_\text{LO}$, it was found that this reduction from $6060$ to $720$ basis states affected the lowest eigenvalues
insignificantly or even predicted lower, better converged values.
\par
The EFT interactions we utilized in this section are specified next. To assess the sensitivity of predictions on particular models of the supposedly unobservable
short-distance structure of the theory, three regulator values were chosen, $\Lambda=200,300,400~$MeV. The reason for using relatively small values for the cutoff,
were the relatively accurate predictions for $B(\alpha)$ if $B(t)$ was LO input, \textit{e.g.}, for $\Lambda=200~$MeV and
$B(t,\text{LO})=B(t,\text{exp})\;\;\Rightarrow\;\;B(\alpha,\text{LO})\approx 30~$MeV. For increasing $\Lambda$, a smaller value for $B(\alpha,\text{LO})$ was predicted,
deviating by as much as $7~$MeV from experiment (fig.~\ref{fig_tjon}). As a consequence, the necessary pNLO corrections to $B(\alpha)$ increase with $\Lambda$, while corrections
to $a_t$, one of the other constraints used to determine the pNLO LECs, remain relatively small. Specifically, a $7~$MeV pNLO contribution constitutes about $25~$\% of
$B(\alpha)$, while typical corrections to $a_t$ (condition (ii) in ch.~\ref{sec_pots}) do not exceed $5~$\% of the physical $a_t$. The magnitude of the resulting
LECs was found to increase rapidly with $\Lambda$ in consequence. Correspondingly, the contributions from individual operators increase, too, only to cancel
in yielding a small number. Here, I avoided those numerical difficulties by using small cutoff values but stress that one is not bound to them, neither by the RGM, and certainly not
by the EFT framework.
\par
NLO corrections were included perturbatively with LECs determined as described in ch.~\ref{sec_pots}.
The specification of the LO fit input $\mathcal{D}'$ concludes the discussion of the interactions. We used
$\mathcal{D}'_\text{LO}=\left\lbrace a_s,B(d),B(\alpha)\right\rbrace$ to determine the LEC for $\Lambda=300,400~$MeV, and
$\mathcal{D}'_\text{LO}=\left\lbrace a_s,B(d),B(t)\right\rbrace$ for $\Lambda=200~$MeV. We used $B(\alpha)$ as input for the cutoff values
which would have predicted a value for it deviating more than $4~$MeV from experiment if $B(t)$ is used instead. This was done to reduce
the pNLO correction to $A>2$ binding energies in order to yield LECs which kept individual operator matrix elements of about the same order as
the binding energies. Thereby, the numerical calculations became feasible. We could use $B(t)$ for $\Lambda=200~$MeV because its
prediction for $B(\alpha)$ was off by less than $2~$MeV. Chosen for practical reasons, this input is just another way of renormalizing the theory and
does not violate established EFT regulations.
\par
Figure~\ref{fig_6he-spect} displays the predictions for the lowest eigenvalues of the LO Hamiltonians in the two-proton four-neutron $0^+$ channel for
various basis dimensions. For $\Lambda=300~$MeV (upper graph), the values were calculated in both the small and large model spaces defined above.
The larger basis does not place significantly lower bounds on the binding energy. We even find the smaller basis more appropriate to expand the states closer to
threshold because of the refined choice of halo-width parameters (see above). As the cutoff is only varied by $\pm 100~$MeV and does not exceed the constant RGM
regulator for the asymptotic functions (eq.~\ref{eq_reg-irreg-fkt}, $\beta=1.1~\text{fm}^{-1}$) significantly, the other calculations were carried out in the smaller model space.
Increasing the dimension means adding basis vectors with larger and larger width parameters on the $\vec{n}_2$ coordinate (fig.~\ref{fig_6he-fragm}).
By that, more and more localized structures are included. The stability of our results confirms that even the smallest bases are apt to expand the states to small
enough distances. The newly added vectors are thence identified with such localized
states, which represent short-distance structure that should according to EFT($\slashed{\pi}$) tenet not contribute significantly to observables.
As it cannot be assumed that a halo structure emerges in EFT($\slashed{\pi}$) at all, it is imperative to increase the basis further to allow for more
localized states, even if a certain eigenvalue is found stable.
However, these eigenstates will eventually assume energies too large for a description with EFT($\slashed{\pi}$).
The states close to threshold - above or below - are na\"ively thought to be in the range of applicability.
\par
To further the understanding of the interplay between EFT and RGM regulators at LO and NLO, I selectively removed basis vectors that correspond to relatively short distances
between nucleons in the $\alpha$ core. As expected, the LO predictions (upper graph, empty diamonds for dimension $3150$) are almost unaffected due to the relatively
long-range nature of the interaction at LO. At NLO, the short-distance structure becomes more relevant. From a technical perspective, this increasing importance resembles
matrix elements of respective operators which receive most of their contribution from small distances. Consistently, the NLO correction for the shallower of the two states
is found considerably different, albeit the dimension is practically the same compared to the prediction of a basis without this deficiency (compare filled diamond and circle
at dimensions $3150$ and $3248$, respectively). Without the short-distance components the basis cannot expand states irrelevant at LO but of apparent significance
at NLO. This concludes the discussion of the convergence in the variational basis.
\par
For $\Lambda=200,300~$MeV, two bound states (empty circles and squares) below the $\alpha$-nn breakup threshold (red/gray lines) are found at LO.
Referring to the case of identical bosons interacting via comparable,
short-range forces$^\text{\cite{stecher}}$, multiple bound states with binding energies differing by orders of magnitude do not come as a surprise.
However, such states are expected to be outside of the range of applicability of EFT($\slashed{\pi}$), \textit{i.e.}, exhibit a strong $\Lambda$ and $\mathcal{D}'$
dependence, while those states which are in the range are excited states in the spectrum.
\par
\begin{figure}
  \includegraphics[width=.5\textwidth]{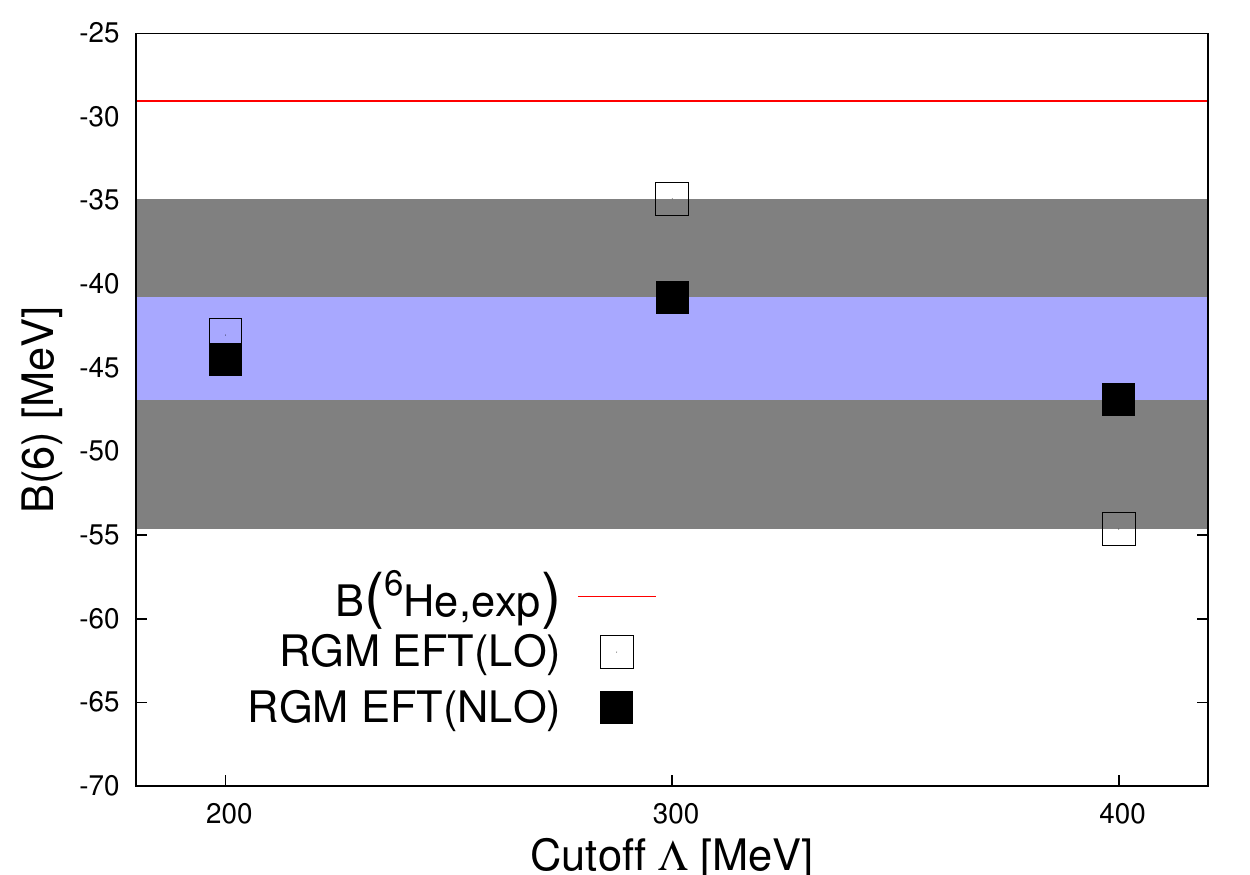}
  \caption{\label{fig_6he-b-of-lambda}\small
Predictions for the shallowest six-nucleon energy eigenvalue in the six-nucleon system with EFT($\slashed{\pi}$) as a function of $\Lambda$. The (blue) gray band
shows (N)LO uncertainties assessed via cutoff and input variation.
}
\end{figure}
Na\"ively, the deeply bound states should not be in the spectrum at all if the regularization truly eliminates modes of high momentum, as suggested.
It is the variational basis through which such high-momentum modes are reintroduced, because I am at present not able to derive bounds for the Gau{\ss}ian
width parameters which define the basis from $\Lambda$. The practical approach is the one described above, namely, to make the basis successively narrower.
In that process, the number of bound states will increase, and the states apt for an EFT($\slashed{\pi}$) description are those which are, one, relatively close
to threshold, and two, stable under the addition of narrower basis states, \textit{i.e.}, independent of $\Lambda_\text{RGM}$.
All graphs show the lowest breakup threshold set by $B(\alpha)$. At NLO, this threshold is at the experimental $\alpha$ binding
energy independent of $\Lambda$ as it is included in $\mathcal{D}'$.
For $\Lambda=300,400~$MeV, predictions were made with interactions whose three-body parameter was fitted to $B(\alpha,\text{exp})$ at LO. Hence, the threshold is at the
same position for (N)LO and for all three $\Lambda$ values. As the potential with $\Lambda=200~$MeV uses $B(t,\text{exp})$ to fix $A_{3\text{NI}}$ at LO,
$B(\alpha)$ (gray line) is a prediction at this order.
\par
For each interaction, we tentatively identify the bound eigenstate closest to the threshold with the EFT($\slashed{\pi}$) prediction for the $^6$He system.
At LO, bound states (empty squares in fig.~\ref{fig_6he-b-of-lambda}) with the following energies are predicted with the 720-dimensional basis:
$B_\text{\tiny LO}(6,\Lambda=400~\text{MeV})\approx 55~$MeV,
$B_\text{\tiny LO}(6,\Lambda=300~\text{MeV})\approx 35~$MeV, and
$B_\text{\tiny LO}(6,\Lambda=200~\text{MeV})\approx 43~$MeV. This implies a LO uncertainty of about $20~$MeV (gray band in fig.~\ref{fig_6he-b-of-lambda}).
At NLO, the corrections place the states (filled squares in fig.~\ref{fig_6he-b-of-lambda}) at:
$B_\text{\tiny NLO}(6,\Lambda=400~\text{MeV})\approx 47~$MeV,
$B_\text{\tiny NLO}(6,\Lambda=300~\text{MeV})\approx 41~$MeV, and
$B_\text{\tiny NLO}(6,\Lambda=200~\text{MeV})\approx 44~$MeV.
The pNLO contribution to the binding energy is attractive for the smaller two cutoff values and
repulsive for $\Lambda=400~$MeV. EFT($\slashed{\pi}$) predicts thus at NLO a binding energy for the state between $47~$MeV and $41~$MeV
(blue band in fig.~\ref{fig_6he-b-of-lambda}). Therefore,
for this state, the theory does not converge to the physical, shallow $^6$He state. Yet, this result does not necessarily mandate a promotion of a four-,
five-, or six-nucleon interaction for an EFT($\slashed{\pi}$) description of the six-nucleon system which converges to experiment. If such an unusual
promotion of a momentum-dependent six-body vertex would be necessary, a behavior of the six-nucleon spectrum similar to that of the three-nucleon spectrum
is expected following a change in the renormalization scheme without the additional constraint that fixes, \textit{e.g.}, $B(t)$. In the six-nucleon sector, the
missing constraint should result analogously in a shift of the spectrum (limit cycle behavior). Instead, the considered eigenvalues are found within defined
uncertainty ranges which decrease from LO to NLO. This observation suggests that the RGM basis was chosen wisely for the states investigated here, but
poorly for the physical states.
\par
First, it could be insufficient for an expansion of a
shallow state bound by a relatively small amount below threshold. Specifically, a much broader range of width parameters for the expansion of the two
radial functions of the halo-neutron coordinates $\vec{n}_{1,2}$ has to be considered in order to discriminate such states from $\alpha$-nn continuum states.
Misinterpreting such a free-$\alpha$ state as a bound six-nucleon state amounts to a four-body calculation without predictive power.
To illustrate this point, we included the predictions for such a state in fig.~\ref{fig_6he-spect} for $\Lambda=400~$MeV (filled and empty circles, bottom right graph).
An unbound $\alpha$-nn state of about $24~$MeV is there not to be confused with a bound $^6$He at NLO, although the pNLO correction places the state very
close to threshold.
If the width parameters for $\vec{n}_{1,2}$ cover a range beyond the expected halo extension, the numerical ambiguity between $\alpha$-nn continuum and $^6$He bound
states can be resolved as follows. The continuum states are naturally much more sensitive to changes of the basis regarding vectors that correspond to
very large separations of a neutron from the core. The problem is to induce such changes without affecting the expansion of an also widely extended shallow bound state.
\par
An example of such a modification in the large model space is shown in the upper graph for $\Lambda=300~$MeV in fig.~\ref{fig_6he-spect}. Two bases, each of dimension
$2436$, were chosen which differ only in the width set used to expand the dependency on the coordinate of the second halo neutron. In one basis, the smallest width of this set
was replaced by an order-of-magnitude larger value. In effect, this more localized basis expands the ground state (squares) with $B(6)>60~$MeV as well as the
broader basis but deviates significantly in its prediction of the shallower state (compare triangles (``more localized'' basis) to circles (``broader'' basis)).
A similar sensitivity to the removal of the broadest basis vectors will eventually only be observed for the $\alpha$-nn continuum states if those vectors are sufficiently broad
to have negligible overlap with the halo states. With the resources available, this investigation cannot be included in this work. Restating the problem, in general one deals with the
slower convergence regarding the basis size of a variational calculation for excited states. We propose here that the $^6$He halo state is an
excited state in the spectrum of the employed potentials, in contrast to the deuteron and triton. For the latter two systems, the regularization through the
EFT cutoff $\Lambda$ or the model space $\Lambda_\text{RGM}$ did not allow for (admissible) deeper lying states. In the six-nucleon sector, the chosen
regularization parameters admit such states and complicate the numerical analysis. The three-body analog would be to identify the triton as an excited Efimov state
for an interaction which allows for more than one bound state. The other states in this spectrum are unphysical. Yet, the formal approach is the same as the one
where those deeper states are observable, for instance in the $^{20}$C isotope. For this nucleus, the possibility of an excited $^{18}$C-2n halo state was
investigated$^\text{\cite{ham-can-2n}}$ in an effective three-body framework.
\par
In contrast to this hypothetical carbon halo, the P-wave resonances in the $\alpha$-n subsystem are supposed to dominate the 6-helium ground state.
In the previous chapter, no such resonances were found at LO (see fig.~\ref{fig_5-he-phases}). A future scattering calculation will determine whether only
interactions which do predict an $\alpha$-n P-wave resonance also result in a shallowly bound 6-helium. The results presented here would be consistent with this finding.
Here, we employed only potentials without such a five-body resonance. As a consequence, the only bound six-body structures found are unphysical, \textit{i.e.},
the system is unbound relative to the $\alpha$-nn threshold. The EFT does not claim to predict whether the state is bound or unbound. Assuming uncertainties
for the shallow state of the same order as assessed here for the deeper states, even at NLO, higher-order modifications (of order $7~$MeV, see above)
could push the state above threshold.
\par
Before I conclude, the limitation to small cutoff values relative to the ones used in $A<6$ calculations, and how it can be overcome in future analyses is explained.
For a simplification of the numerical treatment of larger cutoff values, I propose to increase $\Lambda$ to values for which the $\alpha$ is no longer the ground state
of the resultant potential but rather the shallowest state of a tower of four-body bound states. This would utilize a phenomenon observed in the three-body sector
where ``new'' bound states enter the spectrum for discrete values of $\Lambda$ at threshold. Increasing $\Lambda$ from such a value also renders the initially shallow
state steadily deeper and deeper, until the next bound structure enters, again at threshold (see \textit{e.g.} ch.~3.4 in \cite{bra-ham}).
Analogously, I expect additional four-body bound states entering the spectrum - either from above or below - at the $^3$He-n threshold at certain cutoff values $\Lambda_t$.
Increasing $\Lambda$ from such
a $\Lambda_t$ until the lowest state is bound by approximately $B(\alpha,\text{exp})$ and identifying this instead of the, by then, much more deeply bound state(s) with the
$\alpha$ would also mandate only smaller pNLO corrections.
One would monitor $B(t)$ and $B(\alpha)$ simultaneously for $A_{3\text{NI}}=0$ and select cutoff values for which the 3NI has to contribute relatively little in the course of
renormalization. This avoids
cutoff ranges where the LO renormalization requires large modifications from the 3NI.
The complications associated with large 3NI LO modifications can be understood by considering the demands on the variational basis. Two states of very different spatial extent
have to be expanded: the unrenormalized deeply bound and thereby localized state, and the more extended states renormalized to the physical $B(t)$ and $B(\alpha)$.
Initially, I saw no need to draw this option, and used $\Lambda$ values as given above.
\par
In conclusion, the convergence of relatively deep states in the six-nucleon spectrum from LO to NLO strongly supports the assertion that  a shallow state would be
predicted by EFT($\slashed{\pi}$) without a promotion of a higher-order vertex. The finding is in contrast to the limit-cycle behavior in the triton channel of three-nucleon system.
The identification of the discovered states with $^6$He would be false but their existence and convergence are hints of the existence of a more shallow state in the spectrum.
Whether this state was not resolved due to the small RGM basis, or the fact that there is no bound 6-helium for the employed interactions, but rather a shallow resonant
state, needs to be addressed in a future larger-scale bound and scattering computation.
\section{Summary}\label{sec_summ}
\textit{Nihil est dictum, quod non sit dictum prius} - Advances in the understanding of particular areas in the parameter space of the physical world - to recapitulate this
object from ch.~\ref{sec_intro} where the effective field theory formalism was introduced - are guided and inspired by experience in other areas.
This process of generalizing theories - developed to describe a particular system - to others, refining particular aspects while discarding other, irrelevant
components of the theory, is more efficient if the principal path between the areas is known.
Technical advancement makes larger areas of the parameter space available for scientific investigation, and by that discovers paths connecting one area of
nature to another where the understanding might be either more mature or still in relative infancy. In nuclear physics, this general perspective focuses on three
current areas of interest.
\par
The first is the relation of the nuclear interaction to its underlying theory of QCD. Recent developments in calculating nuclear properties from this high-energy
theory include the prediction of the neutron-proton scattering length$^\text{\cite{bedaque-anp}}$, the nucleon masses$^\text{\cite{np-masses}}$, electric and
magnetic moments of nucleons$^\text{\cite{nucl-mom-lee,nucl-mom-trib}}$, and even tri-nucleon binding energies$^\text{\cite{qcd-triton}}$.
\par
Similarly,
taking the NN interaction as fundamental, its systematic extension to describe many-body systems in terms of a density functional theory$^\text{\cite{scidac}}$
falls into the category of relating different parameter subspaces of the physical world. Both developments depend on, but also drive,
the advance in EFT methodology ($\chi$PT, $\chi$EFT$^\text{\cite{epel-chieft}}$, EFT($\slashed{\pi}$)) and numerical methods (lattice techniques, quantum Monte Carlo).
\par
While in these areas increasing computational power makes certain systems accessible for theoretical investigation, experimental skills in
controlling atomic systems at very low temperatures allow the investigation of universal properties of bosonic$^\text{\cite{cold-bosons}}$ and
fermionic$^\text{\cite{cold-fermions}}$ systems. The modification of Feshbach resonances$^\text{\cite{newton-feshbach}}$ of trapped atoms via magnetic fields
results in a change in the atom-atom scattering length $a_{AA}$ and measurements of loss rates out of the trap yield information about
dimer, trimer, quadrumer, \textit{etc.}, formation as a function of $a_{AA}$. By that, aspects of atomic physics are related to nuclear physics, where
similar experimental techniques are nonexistent.
\par
This sets the context for this work, where I have presented a study of an effective field theory without pions in few-nucleon ($A\leq 6$) systems. The EFT formalism
was used to construct the interaction (ch.~\ref{sec_pots}) and to refine it model-independently. The EFT($\slashed{\pi}$) is the low-energy derivative of a pionful theory and
thus linked naturally to QCD.
\par
We demonstrated explicitly the usefulness of the theory by presenting results for binding energies of the triton and $\alpha$ particle, the
charge radius of the triton, and the S-wave neutron-deuteron scattering length in the triton channel
of increasing accuracy from leading to next-to-leading order. The first-time (except for the nd scattering length and $B(t)$) application of NLO EFT($\slashed{\pi}$)
to those observables confirmed assertions based previously for $A\geq 4$ on LO calculations only, which I summarize:
the 4-helium bound state is a universal consequence of $a_{s,t}$ and $B(t)$; neither for the three- nor for the four-nucleon system, the power counting
had to be modified by, \textit{e.g.}, the promotion of a four-nucleon interaction to NLO, in order to be consistent with \textit{a priori} accuracy predictions of EFT($\slashed{\pi}$).
More specifically, the effective field theory formalism can be used to explain empirically found
correlations amongst few-nucleon observables like the Phillips and Tjon line.
In these two cases, it relates the deviations from data to, first, an incomplete renormalization in the
three-nucleon sector, and second, to higher-order interactions omitted in the course of
the EFT expansion. The theoretical uncertainty at every order of the calculation can be quantified
in this approach.
Numerical inaccuracies were demonstrated to be negligible (see discussion of figs.~\ref{fig_sgl-kett} \& \ref{fig_a0ofe}).
Different results for different short-distance parameterizations therefore have
other origins.
The size of the expansion parameter $p_\text{\tiny typ}/\Lambda_\text{\tiny breakdown}$
determines how fast an EFT expansion converges and decides its usefulness for
the calculation of an observable in a given system.
In heavier systems like \mbox{4-helium}, the expansion parameter can \textit{a priori} be as large as $1$,
so they are border line. However, we find that next-to-leading order corrections to leading-order
results are still parametrically small in the four-nucleon system. This confirms a pattern
already seen in two- and three-nucleon systems at momenta which approach the \textit{a priori}
breakdown scale but where convergence is still found, see \textit{e.g.}~\cite{christl-ddis,tni-gang-3,hgrie-zpara,rupak-npdgamma}.
\par
The feasibility of calculating few-body observables with the numerical Resonating Group Method and EFT($\slashed{\pi}$) potentials was thus established.
The RGM has so far been employed in conjunction with potential models of the nuclear (\textit{e.g.} the \texttt{AV18+UIX/IL2} predictions in this work for $a_0(^3\text{{\small He-n}})$)
and quark$^\text{\cite{quark-rgm}}$ interaction only. Its practicality to obtain predictions
with the very different radial structure of the EFT($\slashed{\pi}$) interaction - single Gau{\ss}ian functions (eq.~(\ref{eq_nlo-pot-coord-space})) - was not guaranteed.
The reliability of the method was assessed in a neutron-deuteron scattering calculation, where the predictions of the RGM agreed with the Faddeev results and also
showed the impact of different input data on the estimate of LO uncertainty. The conclusion from this analysis, in combination with the correlations of bound-state
properties for $A=3,4$ (fig.~\ref{fig_rch-t} \& \ref{fig_trinucl-splitt} \& \ref{fig_tjon}) is that for an estimate of the renormalization-scheme dependence of observables
the data input and the regulator should be varied.
\par
For two correlations for which leading-order calculations exist, namely between the binding energy and charge radius
of the triton (fig.~\ref{fig_rch-t}) and the binding energies of \mbox{4-helium} and the triton (fig.~\ref{fig_tjon}),
our coordinate space EFT($\slashed{\pi}$) calculations at NLO report the expected improvement from LO to NLO
consistent with an expansion parameter $p_\text{\tiny typ}/\Lambda_\text{\tiny breakdown}\approx\frac{1}{3}$. By that we demonstrated that
a consistent description of the $\alpha$ particle is possible at NLO in EFT($\slashed{\pi}$).
In fig.~\ref{fig_trinucl-splitt}, we also report a correlation between the triton
binding energy and its difference to the $^3$He binding energy. As the
EFT($\slashed{\pi}$) potential is at NLO isospin-symmetric, this
model-independent difference is attributed entirely to Coulomb interactions, which
are included in the RGM. At the physical triton binding energy, this value
agrees well both in magnitude and uncertainty with estimates of
charge-symmetry breaking and Coulomb contributions to $^3$He binding.
In fig.~\ref{fig_a0}, a new correlation between the triton binding energy and the real part of the singlet S-wave scattering length
of \mbox{3-helium-neutron} scattering similar to the Tjon line is presented. This, and the three aforementioned correlation bands, let us also conclude that no
four-body contact interaction is required to renormalize the system at next-to-leading order.
The position of all four bands, which represent universal properties of the two-nucleon system,
was determined by fitting nine NN potentials differing at short distances but with identical
long-distance behavior, by variations of the 3NI strengths, and by changing the cutoff.
Consistent with a basic tenet of EFT, namely model independence,
we also showed that the results of the phenomenological models \texttt{AV18(+UIX/IL2)}, which share the
input of our EFT($\slashed{\pi}$) NLO potentials, agree with their results within NLO accuracy.
\par
The first-time exploratory application of EFT($\slashed{\pi}$) at LO and NLO (perturbative) to the 5- and 6-helium systems considers low-energy $\alpha$-n scattering (LO only)
for a large range of cutoffs, and bound-state calculations for both systems at LO and NLO. The renormalization-scheme dependence for LO EFT($\slashed{\pi}$) for the
$\alpha$-n S-wave scattering phase shift was assessed thoroughly and was found not to exceed the na\"ive 30\% deviation from data even for $\Lambda\to\infty$.
The accurate prediction of $B(\alpha,\text{exp})$, either by an appropriate choice for the cutoff or by enforcing it as a condition when fitting the LECs, did not reduce
the uncertainty.
\par
The five-nucleon system was found unbound in the $^5$He channel at LO and pNLO. A calculation without the Coulomb force led to the same conclusion that
an unbound 5-helium system is a universal property of the nuclear force.
\par
The RGM bound-state LO and pNLO EFT($\slashed{\pi}$) calculations in the six-nucleon system identified deeply bound structures. The uncertainty in the
binding energy was found to decrease from LO to NLO, but no agreement with experiment within the assessed NLO uncertainty was observed.
The absence of a shallow bound state in the spectrum
and the convergence of deeper lying states was interpreted as either a shortcoming of the numerical method or an unbound shallow resonant state. The latter
is deemed to be the more likely scenario, because we could not reproduce P-wave resonances in the $^5$He subsystem. The results do not contradict EFT tenets.
The observed decrease of the uncertainty in the six-nucleon binding energies rather supports the assertion that no higher-order operator is to be promoted
for an EFT($\slashed{\pi}$) description of a shallow six-nucleon state.
\par
We argued that a larger RGM basis is imperative to test the applicability of EFT($\slashed{\pi}$) to the shallow states in both, 5- and 6-helium.
By considering NLO shifts which leave the state bound, this can be done in the computationally less expensive bound-state framework, while the general case
of a transition from a shallow real- to a virtual bound state and \textit{vice versa}, can only be analyzed in a larger-scale scattering calculation. This work also
provides the feasibility study for this forthcoming investigation.
\newpage
\appendix
\section{Next-to-leading Order Potential in Coordinate Space}\label{app_FT}
The transformation from a potential in momentum representation $\langle p'|\hat{V}|p\rangle$ to coordinate representation, which can be used in our
RGM implementation, is given in this section. Without specifying the form of the potential operator, the form of interest is given by:
\begin{equation}\label{eq_pot-coord-sgl}
\langle\vec{r}'\vert\hat{V}\vert\Psi\rangle=\frac{1}{(2\pi)^3}\int d^3r\;\langle r'\vert \hat{V}\vert r\rangle\langle r\vert \Psi\rangle\;\;\;.
\end{equation}
Changing the basis yields:
\begin{eqnarray}\label{eq_mom-rep-coord-rep}
\langle\vec{r}'\vert\hat{V}\vert\vec{r}\rangle&=&\frac{1}{(2\pi)^6}\int d^3p'd^3p\;\langle r'\vert p'\rangle\langle p'\vert \hat{V}\vert p\rangle\langle p\vert r\rangle=
\frac{1}{(2\pi)^6}\int d^3p'd^3p\;e^{i\vec{r}'\cdot\vec{p}'}V(\vec{p}',\vec{p})e^{-i\vec{r}\cdot\vec{p}}\nonumber\\
&=&\frac{1}{(2\pi)^6}\int d^3kd^3q\;e^{i\vec{r}'\cdot(\vec{k}+\frac{\vec{q}}{2})}V(\vec{k},\vec{q})e^{-i\vec{r}\cdot(\vec{k}-\frac{\vec{q}}{2})}\nonumber\\
&=&
\frac{1}{(2\pi)^6}\int d^3kd^3q\;e^{i\vec{k}\cdot(\vec{r}'-\vec{r})}V(\vec{k},\vec{q})e^{i\vec{q}\cdot(\frac{\vec{r}+\vec{r}'}{2})}\;\;\;.\nonumber\\
\end{eqnarray}
The operators in $V^\text{(NLO)}(\vec{q},\vec{k})$ (eq.~(\ref{eq_nlo-pot-mom-space})) are polynomials in $\vec{k}$.
Replacing $\vec{k}=i\vec{\nabla}_r$ in the potential operator with the derivative acting on the exponential only, yields an equivalent equation.
This replacement must not be confused with replacing $\vec{k}$ by its coordinate space representation, in which case, $\vec{r}$ would be the conjugate
variable to $\vec{k}$.
\par
Consider, for example, the operator multiplying $C_7^\text{NLO}$ in eq.~(\ref{eq_nlo-pot-mom-space}),
$V(\vec{q},\vec{k})=e^{-\frac{\vec{q}^2}{\Lambda^2}}\,\vec{\sigma}_1\cdot\vec{k}\;\vec{\sigma}_2\cdot\vec{k}$.
\begin{eqnarray}
\langle\vec{r}'\vert\hat{V}\vert\vec{r}\rangle&=&
\frac{1}{(2\pi)^6}\int d^3q\;e^{i\vec{q}\cdot(\frac{\vec{r}+\vec{r}'}{2})}V(\vec{q},i\vec{\nabla}_r)\int d^3ke^{i\vec{k}\cdot(\vec{r}'-\vec{r})}\nonumber\\&=&
\frac{1}{(2\pi)^3}\int d^3q\;e^{i\vec{q}\cdot(\frac{\vec{r}+\vec{r}'}{2})}V(\vec{q},i\vec{\nabla}_r)\delta(\vec{r}'-\vec{r})\nonumber\\
&=&-\frac{1}{(2\pi)^3}\int d^3qe^{i\vec{q}\cdot(\frac{\vec{r}+\vec{r}'}{2})}e^{-\frac{\vec{q}^2}{\Lambda^2}}\sigma_{1,i}\sigma_{2,j}\nabla_{r,i}\nabla_{r,j}\delta(\vec{r}'-\vec{r})\nonumber\\
&=&-\underbrace{\frac{\Lambda^3}{8\pi^{\frac{3}{2}}}e^{-\frac{\Lambda^2}{4}(\frac{\vec{r}+
\vec{r}'}{2})^2}}_{\equiv I_0\left(\frac{\vec{r}+\vec{r}'}{2}\right)}\sigma_{1,i}\sigma_{2,j}\nabla_{r,i}\nabla_{r,j}\delta(\vec{r}'-\vec{r})\;\;\;.
\end{eqnarray}
This operator acts on a wave function, $\Psi(\vec{r})\equiv\langle\vec{r}\vert\Psi\rangle$:
{\small
\begin{eqnarray}
\langle\vec{r}'\vert\hat{V}\vert\Psi\rangle&=&
-\sigma_{1,i}\sigma_{2,j}\int d^3rI_0\left(\frac{\vec{r}+\vec{r}'}{2}\right)\nabla_{r,i}\nabla_{r,j}\left[\delta(\vec{r}'-\vec{r})\right]\Psi(\vec{r})\label{eq_v7-coord-1}\nonumber\\
&=&-\sigma_{1,i}\sigma_{2,j}\int d^3r\Big[I_0\left(\frac{\vec{r}+\vec{r}'}{2}\right)\Psi(\vec{r})\Big]\nabla_{r,i}\nabla_{r,j}\Big[\delta(\vec{r}'-\vec{r})\Big]\nonumber\\
&=&\sigma_{1,i}\sigma_{2,j}\int d^3r\nabla_{r,i}\Big[I_0\left(\frac{\vec{r}+\vec{r}'}{2}\right)\Psi(\vec{r})\Big]\nabla_{r,j}\Big[\delta(\vec{r}'-\vec{r})\Big]\nonumber\\
&=&-\sigma_{1,i}\sigma_{2,j}\int d^3r\nabla_{r,i}\nabla_{r,j}\Big[I_0\left(\frac{\vec{r}+\vec{r}'}{2}\right)\Psi(\vec{r})\Big]\delta(\vec{r}'-\vec{r})\\
&=&-\sigma_{1,i}\sigma_{2,j}\Bigg( \frac{1}{4}\nabla_{r',i}\nabla_{r',j}\Big[I_0(r')\Big]\Psi(\vec{r}')+\frac{1}{2}\nabla_{r',j}\Big[I_0(r')\Big]\nabla_{r',i}\Big[\Psi(\vec{r}')\Big]\nonumber\\ &&\;\;\;\;\;\;\;\;\;\;\;\;\;\;\;\;\;\;\;+\frac{1}{2}\nabla_{r',i}\Big[I_0(r')\Big]\nabla_{r',j}\Big[\Psi(\vec{r}')\Big]+I_0(r')\nabla_{r',i}\nabla_{r',j}\Big[\Psi(\vec{r}')\Big]\Bigg)\nonumber\\
&=&-I_0(r)\vec{\sigma}_1\cdot\vec{\nabla}\,\vec{\sigma}_2\cdot\vec{\nabla}\,\Psi(\vec{r})+
\frac{\Lambda^2I_0(r)}{4}\Big(\vec{\sigma}_2\cdot\vec{r}\,\vec{\sigma}_1\cdot\vec{\nabla}\nonumber\\
&&+
\vec{\sigma}_1\cdot\vec{r}\,\vec{\sigma}_2\cdot\vec{\nabla}+\frac{1}{2}\vec{\sigma}_1\cdot\vec{\sigma}_2-
\frac{\Lambda^2}{4}\vec{\sigma}_1\cdot\vec{r}\,\vec{\sigma}_2\cdot\vec{r}\Big)\,\Psi(\vec{r})\nonumber
\label{eq_v7-coord-2}
\end{eqnarray}}
Derivatives are understood to act on the expression in the first square bracket on their right, \textit{i.e.},
on the delta function in eq.~(\ref{eq_v7-coord-1}) and not on the wave function, because they were introduced to 'bring down' the $\vec{k}$'s from the exponential.
Analogous transformations of all operators in eq.~(\ref{eq_nlo-pot-mom-space}) allow for an expression of the potential in coordinate space as given in
table~\ref{tab_pot-coord-lec-spect}. Hermiticity is explicit in this form, and the anticommutator structures allow an efficient analytic evaluation of the
corresponding matrix elements relative to a form like na\"ively derived in eq.~(\ref{eq_v7-coord-2}).\\
\input{tab_pot-coord-space.tex}
\newpage
\section{Genetic Search Algorithm}\label{app_gen-alg}
With the function chosen to regulate the contact interactions (eq.~(\ref{eq_reg-fkt-mom})), an analytic calculation of the LECs, even at leading order, is not possible.
Furthermore, a certain $\Lambda$ only suggest how select and does not provide a rigorous way to derive values for the boundaries of an interval from
which the Gau{\ss}ian width parameters $\gamma_n$, which define the RGM basis, are to be taken.
Optimized values for both the LECs and the $\gamma_n$'s are determined here with a genetic algorithm$^\text{\cite{gen-alg}}$.
This stochastic method was chosen because:
\begin{itemize}
\item The entire parameter space is probed. This made the determination of nNLO LECs feasible even without an estimate of their magnitude.
Such an estimate, available for the LO LECs through eq.~(\ref{eq_lo-lec}), would be based on the naturalness of the $C_i^\text{(N)LO}$, only, and therefore be
unreliable because the LECs are redefined. The $A_i$'s in eq.~(\ref{eq_nlo-pot-coord-space}) are linear combinations, $A_i=\sum_{j=1}^9a_{ij}\Lambda^{n(j)}C_j$,
and hence a change in the original $C_i^\text{(N)LO}$'s at the percent level could induce relatively large changes in the $A_i$'s because of the $\Lambda$
dependence.
\item The search parameters can be adjusted to reduce the probability of identifying a local instead of a global minimum. Local minima could,
for instance, correspond to an interaction with a deep two-body bound state in addition to the relatively shallow deuteron. This ground state, however, is
beyond the range of applicability of the EFT, \textit{i.e.}, $\Lambda_\text{RGM}$ might not even allow for its expansion. In general, its energy will
depend nonperturbatively on all renormalization parameters.
\item The computation factorizes into independent blocks and scales almost linearly with the number of processors used.
\end{itemize}
The task can be abstracted to the problem of finding a set of parameters $\lbrace a_i\rbrace$ that minimizes the function $W\left(\lbrace a_i\rbrace\right)$. Specifically,
\begin{description}
\item[class (i)] In a given complete model space, the LECs are to be fitted to data, \textit{i.e.},
$W\left(\lbrace a_i\rbrace\right)=\sum\limits_{o\in\mathcal{D}'}\frac{|o(\text{exp})-o(\text{trial})|}{|o(\text{exp})|}$ with $\lbrace a_i\rbrace=\lbrace A_{1,\ldots,9}\rbrace$
and $\mathcal{D}'$ as defined in eq.~(\ref{eq_nlo-fit-para}) (implemented for this EFT($\slashed{\pi}$) application).
\item[class (ii)] For a given potential, dimension, and angular momentum structure, the $\gamma_n$ are optmized to maximize the binding energy, \textit{i.e.},
$W\left(\lbrace a_i\rbrace\right)=-B(d,t,\alpha)$ and $\lbrace a_i\rbrace=\lbrace \gamma_n\rbrace$ (implementation from \cite{gen-winkl}).
\end{description}
The incarnation of the algorithm used here proceeds along 8 basic steps:
\input{enum_genalg.tex}
Steps $(3)$ to $(8)$ are repeated $k$ times. This iteration together with the preparation phase $(1),(2)$ is called run.
At the end of a run the individuum with the lowest $W$ value is returned. Consecutive runs were used, each time only the intervals - set in step $(2)$ - were
adjusted. At the end of a run, they are centered around the values of the best individuum of that run. Additionally, the size of the intervals is reduced by a factor
of $\frac{3}{4}$. After that, the next run is started with the new bounding intervals. Typical populations consisted of $Z\simeq\mathcal{O}\left(10^3\right)$ individuums,
which created their respective \textit{Vitruvian Man} within less than $\mathcal{O}\left(10^3\right)$ generations.
\section{How an Integral is done}\label{app_rgm-coord-me}
The analytic expressions for a coordinate space matrix element of the operator $\propto A_9$ in $V^\text{(NLO)}$ (eq.~(\ref{eq_nlo-pot-coord-space})), in the
bound-state basis (eq.~(\ref{eq_frag-int-eigenfkts})), \textit{i.e.}, set parameters $\lbrace f,j,\mathbf{l}_j,\mathbf{s}_j,\mathbf{\gamma}_j\rbrace$ in \textit{bra} and \textit{ket},
are derived in this section. The operator structure is given by:\\
\begin{equation}\label{eq_tensor-p-structure}
\mathcal{P}\left[\nabla_{ij,\nu}\otimes\nabla_{ij,\nu'}\right]^{2q}=\sum_{\tau,\tau'}\gamma_{\tau\tau'}^{\lambda\lambda'}\sum_{\nu\nu'}(1\nu 1\nu'\vert 2q)\nabla_{s_\tau,\nu}\nabla_{s_{\tau'},\nu'}\;\;\;;
\end{equation}
with
\begin{eqnarray}
\nabla_{ij(s),\nu}&:&\text{\footnotesize spherical component}\;\nu\;\text{\footnotesize of the derivative}\nonumber\\
&&\text{\footnotesize acting on the relative coordinate between}\nonumber\\
&&\text{\footnotesize particles}\;i\;\text{\footnotesize and}\;j\;\text{\footnotesize Jacobi coordinate }\tau^{(')}\;;\\
\mathcal{P}\nabla_{ij}&=&\gamma^{\lambda\lambda'}_l\nabla_{\vec{s}_l}\;\;,\;\;\lambda^{(')}=\mathcal{P}(i(j))\;\;\;.\nonumber
\end{eqnarray}
\newpage
The integral $J(\mathcal{P};L_1,L_{1_z},\ldots,L_z,L_{z_z})$ (eq.~(\ref{eq_bv-me})) is obtained from a generating integral which is defined in terms of the generating function of the
spherical harmonics (eq.~(\ref{eq_gen-fkt})) by:
\begin{eqnarray}\label{eq_gen-int}
I^{ij,q}(a_1,b_1,\ldots,a_z,b_z)&=&\int d^3s_1\ldots d^3s_{A-1}e^{-\sum_{\mu,\mu'}^{A-1}\rho'_{\mu\mu'}\vec{s}_\mu\cdot\vec{s}_{\mu'=1}+\sum_{n=n_{c_r}}^za_n\vec{b}_n\cdot\vec{Q}_n}\cdot\nonumber\\
&&\cdot\sum_{\tau,\tau'}\gamma_{\tau\tau'}^{\lambda\lambda'}\sum_{\nu\nu'}(1\nu 1\nu'\vert 2q)\nabla_{s_\tau,\nu}\nabla_{s_{\tau'},\nu'}
e^{-\sum_{\mu,\mu'}^{A-1}\rho''_{\mu\mu'}\vec{s}_\mu\cdot\vec{s}_{\mu'}+\sum_{n=1}^{n_{c_r}-1}a_n\vec{b}_n\cdot\vec{Q}_n}\nonumber\\
&=&\sum_{l_1,m_1,\ldots,l+z,m_z}\left(\prod_{n=1}^z\frac{C_{l_nm_n}}{l_n!}a_n^{l_n}b_n^{l_n-m_n}\right)J^q(\mathcal{P};l_1,m_1,\ldots,l_z,m_z)\;\;\;.
\end{eqnarray}
The number of particles in the \textit{ket} state is $n_{c_r}$. For all states used in this work, this number is identical to the corresponding number $n_{c_l}$ in the \textit{bra}, and
defined here only as a reminder that a grouping of nucleons in S-wave clusters within a fragment, leading to possibly different coordinate structures in \textit{bra} and \textit{ket},
is an unused option. The exponential of the interaction operator, $\exp(-\frac{\Lambda^2}{4}\vec{r}^2)$ is implicit in the transformation matrix $\rho'_{\mu\mu'}(\rho''_{\mu\mu'})$
for each summand of the anticommutator. A principal axis transformation,
\begin{equation}\label{eq_princ-axis-trafo}
\vec{s}_\mu=\sum_{\lambda=\mu}^{A-1}T_{\mu\lambda}\vec{t}_\lambda\;\;\;\text{and}\;\;\;\vec{Q}_n=\sum_{\mu'=1}^{A-1}P_{n\mu'}\vec{t}_{\mu'}\;\;\;\text{such that}\;\;\;
\sum_{\mu,\mu'}^{A-1}(\rho'_{\mu\mu'}+\rho''_{\mu\mu'})\vec{s}_\mu\cdot\vec{s}_{\mu'}\stackrel{!}{=}\sum_\lambda^{A-1}\beta_\lambda\vec{t}_\lambda^2\;\;\;,
\end{equation}
followed by a translation to complete the square in the exponent,
\begin{equation}\label{eq_compl-squeres}
\vec{t}_\lambda=\vec{u}_\lambda+\frac{\sum_n^za_nP_{n\lambda}\vec{b}_n}{2\beta_\lambda}\;\;\;,
\end{equation}
results in an expression of the integral in terms of the norm integral, defined analogously to the operator generating integral in eq.~(\ref{eq_gen-fkt}):
\begin{equation}\label{eq_j-von-inorm}
I^{ij,q}(a_1,b_1,\ldots,a_z,b_z)=I^{ij}_{\text{\tiny Norm}}(a_1,b_1,\ldots,a_z,b_z)\sum_{n,n'}^zZ_{nn'}\cdot a_na_{n'}F_q
\end{equation}
with
\begin{eqnarray}\label{eq_z-nn}
Z_{nn'}=&\sum_{\tau,\tau'}\gamma_{\tau\tau'}^{\lambda\lambda'}&\Bigg[\sum_{\mu,\mu'}^{A-1}\left(\rho''_{\tau'\mu}+
\rho''_{\mu \tau'}\right)\left(\rho''_{\tau\mu'}+\rho''_{\mu'\tau}\right)\cdot\nonumber\\
&&\sum_{\alpha,\alpha'}^{A-1}\frac{T_{\mu\alpha}T_{\mu'\alpha'}}{4\beta_\alpha\beta_{\alpha'}}P_{n\alpha}P_{n'\alpha'}+
\theta(n_{c_r}-1-n)\theta(n_{c_r}-1-n')\xi_{n'\tau}\xi_{n\tau'}\nonumber\\
&&-\sum_{\mu=1}^{A-1}\left(\rho''_{\tau\mu}+\rho''_{\mu\tau}\right)\theta(n_{c_r}-1-n)\xi_{n\tau'}\sum_{\alpha=\mu}^{A-1}\frac{T_{\mu\alpha}P_{n'\alpha}}{\beta_\alpha}\Bigg]\nonumber
\end{eqnarray}
and
$$
F_q=\left\{
\begin{array}{rcl}
2b_{n}^2b_{n'}^2 && q=2\\
-2(b_{n}^2b_{n'}+b_{n'}^2b_{n}) && q=1\\
\sqrt{\frac{2}{3}}(b_{n}^2+4b_{n}b_{n'}+b_{n'}^2) &\;\;\;\;\;\;\;\text{for}\;\;\;\;\;\;\;& q=0\\
-2(b_{n}+b_{n'}) && q=-1\\
2 && q=-2\\
\end{array}\right.
$$
Relating the coefficients of $a_n^\mu b_n^l$ and $a_n^\mu b_n^la_{n'}^{\mu'} b_{n'}^{l'}$ in eq.~(\ref{eq_gen-int}) and eq.~(\ref{eq_j-von-inorm}) gives the matrix
elements for the five spherical components of the operator:
\begin{eqnarray*}
J_{\lbrace l,m\rbrace}^{ij,q=2}&=&
\sum_{n\neq n'}^z2 Z_{nn'}\frac{l_nl_{n'}}{C_{l_n,m_n}C_{l_{n'},m_{n'}}}C_{l_n-1,m_n+1}C_{l_{n'}-1,m_{n'}+1}
J^{\text{\tiny Norm}}_{(\mathcal{P};\ldots ,l_n-1,m_n+1,\ldots,l_{n'}-1,m_{n'}+1,\ldots)}\\
&&+\sum_n^z2Z_{nn}\frac{(l_n-1)l_nC_{l_n-2,m_n+2}}{C_{l_n,m_n}}J^{\text{\tiny Norm}}_{(\mathcal{P};\ldots ,l_n-2,m_n+2,\ldots)}\;\;\;,
\end{eqnarray*}
\begin{eqnarray*}
J_{\lbrace l,m\rbrace}^{ij,q=1}&=&
\sum_{n\neq n'}^z(-2) Z_{nn'}\frac{l_nl_{n'}}{C_{l_n,m_n}C_{l_{n'},m_{n'}}}\Big[C_{l_n-1,m_n+1}C_{l_{n'}-1,m_{n'}}
J^{\text{\tiny Norm}}_{(\mathcal{P};\ldots ,l_n-1,m_n+1,\ldots,l_{n'}-1,m_{n'},\ldots)}\\
&&+C_{l_n-1,m_n}C_{l_{n'}-1,m_{n'}+1}J^{\text{\tiny Norm}}_{(\mathcal{P};\ldots ,l_n-1,m_n,\ldots,l_{n'}-1,m_{n'}+1,\ldots)}\Big]\\
&&+\sum_n^z(-4)Z_{nn}\frac{(l_n-1)l_nC_{l_n-2,m_n+1}}{C_{l_n,m_n}}J^{\text{\tiny Norm}}_{(\mathcal{P};\ldots ,l_n-2,m_n+1,\ldots)}\;\;\;,
\end{eqnarray*}
{\small
\begin{eqnarray}
J_{\lbrace l,m\rbrace}^{ij,q=0}=&
\sum\limits_{n\neq n'}^z\sqrt{\frac{2}{3}} Z_{nn'}\frac{l_nl_{n'}}{C_{l_n,m_n}C_{l_{n'},m_{n'}}}&
\Big[C_{l_n-1,m_n+1}C_{l_{n'}-1,m_{n'}-1}J^{\text{\tiny Norm}}_{(\mathcal{P};\ldots ,l_n-1,m_n+1,\ldots,l_{n'}-1,m_{n'}-1,\ldots)}\nonumber\\
&&+C_{l_n-1,m_n-1}C_{l_{n'}-1,m_{n'}+1}J^{\text{\tiny Norm}}_{(\mathcal{P};\ldots ,l_n-1,m_n-1,\ldots,l_{n'}-1,m_{n'}+1,\ldots)}\nonumber\\
&&+4C_{l_n-1,m_n}C_{l_{n'}-1,m_{n'}}J^{\text{\tiny Norm}}_{(\mathcal{P};\ldots ,l_n-1,m_n,\ldots,l_{n'}-1,m_{n'},\ldots)}\Big]\nonumber\\
&&+\sum_n^z\left(6\sqrt{\frac{2}{3}}\right)Z_{nn}\frac{(l_n-1)l_nC_{l_n-2,m_n}}{C_{l_n,m_n}}J^{\text{\tiny Norm}}_{(\mathcal{P};\ldots ,l_n-2,m_n,\ldots)}\;\;\;,\nonumber
\end{eqnarray}}
\begin{eqnarray*}
J_{\lbrace l,m\rbrace}^{ij,q=-1}&=&
\sum_{n\neq n'}^z(-2) Z_{nn'}\frac{l_nl_{n'}}{C_{l_n,m_n}C_{l_{n'},m_{n'}}}\Big[C_{l_n-1,m_n}C_{l_{n'}-1,m_{n'}-1}
J^{\text{\tiny Norm}}_{(\mathcal{P};\ldots ,l_n-1,m_n,\ldots,l_{n'}-1,m_{n'}-1,\ldots)}\\
&&+C_{l_n-1,m_n-1}C_{l_{n'}-1,m_{n'}}J^{\text{\tiny Norm}}_{(\mathcal{P};\ldots ,l_n-1,m_n-1,\ldots,l_{n'}-1,m_{n'},\ldots)}\Big]\\
&&+\sum_n^z(-4)Z_{nn}\frac{(l_n-1)l_nC_{l_n-2,m_n-1}}{C_{l_n,m_n}}J^{\text{\tiny Norm}}_{(\mathcal{P};\ldots ,l_n-2,m_n-1,\ldots)}\;\;\;,
\end{eqnarray*}
\begin{eqnarray*}
J_{\lbrace l,m\rbrace}^{ij,q=-2}&=&
\sum_{n\neq n'}^z2 Z_{nn'}\frac{l_nl_{n'}}{C_{l_n,m_n}C_{l_{n'},m_{n'}}}C_{l_n-1,m_n-1}C_{l_{n'}-1,m_{n'}-1}
J^{\text{\tiny Norm}}_{(\mathcal{P};\ldots ,l_n-1,m_n-1,\ldots,l_{n'}-1,m_{n'}-1,\ldots)}\\
&&+\sum_n^z2Z_{nn}\frac{(l_n-1)l_nC_{l_n-2,m_n-2}}{C_{l_n,m_n}}J^{\text{\tiny Norm}}_{(\mathcal{P};\ldots ,l_n-2,m_n-2,\ldots)}\;\;\;,
\end{eqnarray*}
with $C_{LM}$ defined in eq.~(\ref{eq_gen-fkt}) and $\lbrace l,m\rbrace=(l_1,m_1,\ldots,l_z,m_z)$. Given the values for the norm matrix elements$^\text{\cite{winkler-rgm-me}}$,
the operator elements are thus obtained for permutations $\mathcal{P}$ representing double cosets$^\text{\cite{selig-cosets}}$, only.
\section{The pNLO calculation with the RGM}\label{app_rgm-pnlo}
Thissection documents the modifications to the adopted RGM implementation in order to carry out perturbative NLO calculations as explained in ch.~\ref{sec_pots}.
The stage of the RGM implementation at which the extensions were made and the technical notation used here is reported in \cite{mythesis} for bound-state
calculations and in \cite{thielmann} for the scattering problem.
\par
Obtaining perturbative corrections to bound state energies was already implemented in the program chain, while the calculation of the operators with
coordinate space structure $\vec{r}^2$ and $\lbrace e^{-\frac{\Lambda^2}{4}\vec{r}^2},\vec{\nabla}^2\rbrace$ is new. The $\vec{r}^2$ operator was
implemented as the fourth operator in the \texttt{LUDW} part of the program chain. For this operator, \texttt{LUDW} was modified to prepare the calculation of
reduced spacial matrix elements (see eq.~(\ref{eq_op-me-prod-red-me})) for
\begin{equation}\label{eq_r-squ}
\vec{r}^2=\underbrace{-\sqrt{3}\cdot\frac{4\pi}{3}}_{\equiv F}\cdot\frac{3}{4\pi}\left[r^p\otimes r^q\right]^{00}\;\;\;,
\end{equation}
with spherical components of the relative coordinate between interacting particles equal to solid spherical harmonics$^\text{\cite{edmonds}}$,
$r^q=\sqrt{\frac{4\pi}{3}}\cdot\mathcal{Y}_{1q}(r)$, and a prefactor $F$ implemented in the \texttt{KOBER} part of the program chain.
\par
The peculiar anticommutator structure chosen to represent the other momentum-dependent NLO contribution (compare to the form in \cite{vkolck-thesis-pub})
made it possible to calculate its matrix elements by a linear combination of modified matrix elementsof the kinetic energy operator. The changes were made in the
\texttt{QUAF} code, in which the width parameters of the basis and the interaction are specified. The derivation of an analytic formula for a generic matrix element
parallels the one given in app.~\ref{app_rgm-coord-me}. In comparison to the more complicated case of an operator of rank 2 as considered in app.~\ref{app_rgm-coord-me},
the structure for pNLO is a scalar:
\begin{equation}\label{eq_p2-tensor}
\vec{\nabla}^2=\left[\nabla^\nu\otimes\nabla^{\nu'}\right]^{00}\;\;\;.
\end{equation}
The two summands of the anticommutator yield expressions for their matrix elements in the RGM basis which differ only in the definition of the transformation
matrices: $\rho'_{\mu\mu'}$ is modified for the $e^{-\frac{\Lambda^2}{4}\vec{r}^2}\vec{\nabla}^2$ element and $\rho''_{\mu\mu'}$ for
$\vec{\nabla}^2e^{-\frac{\Lambda^2}{4}\vec{r}^2}$ (see eq.~(\ref{eq_gen-int})). This ingenious treatment\footnote{HMH, private communication.}
makes the explicit calculation of operators of the form $\vec{r}\cdot\vec{\nabla}$ and $\left(\vec{\sigma}\cdot\vec{r}\,\vec{\sigma}\cdot\vec{\nabla}\right)$ unnecessary.
\par
With those extensions, all numerical tools are available to realize the NLO constraints (iii)-(v) as defined in ch.~\ref{sec_pots}. In distinction to the ground state wave
functions for the deuteron, triton, and $^4$He and the asymptotic fragment relative functions, the scattering states $\langle r\vert\Psi_l^\pm(k)\rangle$ are complex
valued. The reality of a bound-state wave function $\langle r\vert d,t,\alpha\rangle$ is a consequence of the real RGM basis and the real potential, while a boundary condition enforces
real valued relative motion radial wave functions between the fragments. In the $h_l^\pm$ basis (eq.~(\ref{eq_asymp-wfkt})), the NLO amplitude - given diagrammatically in
fig.~\ref{fig_eft-exp} - is
{\small
\begin{equation}\label{eq_nlo-ampl}
\langle h_l^-\vert\hat{T}\vert h_l^+\rangle=\langle h_l^-\vert\hat{T}_\text{LO}\vert h_l^+\rangle+\langle\Psi_0^-\vert\hat{V}_\text{NLO}\vert\Psi_0^+\rangle
\end{equation}
}
with
$$\left(\hat{T}+\hat{V}_\text{LO}\right)\vert\Psi_0^+\rangle=\frac{k^2}{2m_\text{\tiny red}}\vert\Psi_0^+\rangle\;\;\;.$$
The wave functions $\langle r\vert\Psi_0^+\rangle$ are obtained by calculating the S-matrix in LO $S_\text{LO}$ (see steps leading to eq.~(\ref{eq_s-from-a})).
Substituting the Gau{\ss}ian
expansions for $h_l^\pm$ in eq.~(\ref{eq_asymp-wfkt}) and the LO S-matrix then yields the expansion coefficients for $\langle r\vert\Psi_0^+\rangle$. The coefficients for
$\langle r\vert\Psi_0^-\rangle$ are calculated by $\langle r\vert\Psi_0^+\rangle=S_\text{LO}\langle r\vert\Psi_0^+\rangle$. A modified version of the \texttt{SPOLE}
component of the program chain writes an input file in the standard shown in table~\ref{tab_inent}.
\begin{table}
\begin{tabular}{|l|l|l|l|}
\hline
$N_e$&&&\\
\hline
$\text{Re}\lbrace k_1=\sqrt{2m_\text{\tiny red}E_1}\rbrace$&$\text{Im}\lbrace k_1\rbrace$&&\\
\hline
$\text{Re}\lbrace f_1(k_1)\rbrace$&$\text{Im}\lbrace f_1(k_1)\rbrace$&$\text{Re}\lbrace g_1(k_1)\rbrace$&$\text{Im}\lbrace g_1(k_1)\rbrace$\\
\hline
\multicolumn{1}{|c|}{$\vdots$}&\multicolumn{1}{|c|}{$\vdots$}&\multicolumn{1}{|c|}{$\vdots$}&\multicolumn{1}{|c|}{$\vdots$}\\
\hline
$\text{Re}\lbrace f_N(k_1)\rbrace$&$\text{Im}\lbrace f_N(k_1)\rbrace$&$\text{Re}\lbrace g_N(k_1)\rbrace$&$\text{Im}\lbrace g_N(k_1)\rbrace$\\
\hline
$\text{Re}\lbrace S_\text{\tiny LO}(E_1)\rbrace$&$\text{Im}\lbrace S_\text{\tiny LO}(E_1)\rbrace$&&\\
\hline
$\text{Re}\lbrace p_1^+(k_1)\rbrace$&$\text{Im}\lbrace p_1^+(k_1)\rbrace$&&\\
\hline
\multicolumn{1}{|c|}{$\vdots$}&\multicolumn{1}{|c|}{$\vdots$}&&\\
\hline
$\text{Re}\lbrace p_N^+(k_1)\rbrace$&$\text{Im}\lbrace p_N^+(k_1)\rbrace$&&\\
\hline
$\text{Re}\lbrace p_1^-(k_1)\rbrace$&$\text{Im}\lbrace p_1^-(k_1)\rbrace$&&\\
\hline
\multicolumn{1}{|c|}{$\vdots$}&\multicolumn{1}{|c|}{$\vdots$}&&\\
\hline
$\text{Re}\lbrace p_N^-(k_1)\rbrace$&$\text{Im}\lbrace p_N^-(k_1)\rbrace$&&\\\hline
\end{tabular}
\caption{\label{tab_inent}\small
Format of the modified $\texttt{INEN}$ file generated by $\texttt{S-POLE\_DWBA.f}$ and processed by $\texttt{DR2END\_DWBA.f}$. The head of the file remains unchanged
with $\texttt{NBAND1=10(13)}$ for a NN(NN+3NI) calculation and $\texttt{NZZ=-2}$ to read the LO data. The additions are attached to the end of the file. The coefficients
expand the functions: $G_l(kr)=\sqrt{\frac{k}{m_\text{\tiny red}}}r^{l+1}\sum_i^Ng_i(k) e^{-\gamma_ir^2}$,
$F_l(kr)=\sqrt{\frac{k}{m_\text{\tiny red}}}r^{l+1}\sum_i^Nf_i(k) e^{-\gamma_ir^2}$ and
$\langle r\vert\Psi_l^\pm(k)\rangle=\sqrt{\frac{k}{m_\text{\tiny red}}}r^{l+1}\sum_i^Np_i^\pm(k) e^{-\gamma_ir^2}$with width parameters $\gamma_i$ set in \texttt{INQUA\_N}.}
\end{table}
\begin{table}
\begin{tabular}{|r|r|r|r|r|r|r|}
\hline
{\scriptsize $\text{Re}\lbrace k_1=\sqrt{2m_\text{\tiny red}E_1}\rbrace$}&
{\scriptsize $\text{Re}\lbrace S(k_1)\rbrace$}&{\scriptsize $\text{Im}\lbrace S(k_1)\rbrace$}&
{\scriptsize $\text{Re}\lbrace S_\text{LO}(k_1)\rbrace$}&{\scriptsize $\text{Im}\lbrace S_\text{LO}(k_1)\rbrace$}&
{\scriptsize $\text{Re}\lbrace S_\text{NLO}(k_1)\rbrace$}&{\scriptsize $\text{Im}\lbrace S_\text{NLO}(k_1)\rbrace$}\\
\hline
\multicolumn{1}{|c|}{$\vdots$}&\multicolumn{1}{|c|}{$\vdots$}&\multicolumn{1}{|c|}{$\vdots$}&\multicolumn{1}{|c|}{$\vdots$}&
\multicolumn{1}{|c|}{$\vdots$}&\multicolumn{1}{|c|}{$\vdots$}&\multicolumn{1}{|c|}{$\vdots$}\\
\hline
{\scriptsize $\text{Re}\lbrace k_{N_e}\rbrace$}&
{\scriptsize $\text{Re}\lbrace S(k_{N_e})\rbrace$}&{\scriptsize $\text{Im}\lbrace S(k_{N_e})\rbrace$}&
{\scriptsize $\text{Re}\lbrace S_\text{LO}(k_{N_e})\rbrace$}&{\scriptsize $\text{Im}\lbrace S_\text{LO}(k_{N_e})\rbrace$}&
{\scriptsize $\text{Re}\lbrace S_\text{NLO}(k_{N_e})\rbrace$}&{\scriptsize $\text{Im}\lbrace S_\text{NLO}(k_{N_e})\rbrace$}\\
\hline
\end{tabular}
\caption{\label{tab_dwbasout}
\small Format of the output file generated by $\texttt{DR2END\_DWBA.f}$.}
\end{table}
This file specifies: the functions $\langle r\vert\Psi_0^\pm\rangle$
and $G_l,F_l$ in terms of complex expansion coefficients; $S_\text{LO}$; the energies for which the states were calculated. The information is processed by
an updated \texttt{DR2END} code which writes the pNLO S-matrix to an output file with standard as in table~\ref{tab_dwbasout}. The LO and NLO part of the S-matrix,
{\small\begin{eqnarray}
S_\text{NLO}=S_\text{LO}+2i&\Big[&-\frac{2}{\hbar c}\int drr^2\left(\sum c_i(\Psi^-)e^{-\frac{\gamma_i}{2}r^2}\right)
\left(c_ce^{-\frac{\Lambda^2}{4}r^2}+c_rr^2e^{-\frac{\Lambda^2}{4}r^2}+c_p\lbrace e^{-\frac{\Lambda^2}{4}\vec{r}^2},\vec{\nabla}^2\rbrace\right)\cdot\nonumber\\
&&\left(\sum c_j(\Psi^+)e^{-\frac{\gamma_j}{2}r^2}\right)\Big]\;\;\;,\label{eq_nlo-smatrix-rgm}
\end{eqnarray}}
are printed separately.
\par
For LECs $c_{c,r,p}$ of order $1$, the $S_\text{NLO}$ matrix elements between the scattering states, with expansion coefficients $c_i(\Psi^\pm)$ determined in \texttt{SPOLE},
typically assume values between $10^{-3}$ and $10^{-9}$. Comparing magnitudes of real- and imaginary parts of the $S_\text{NLO}$ matrix elements
(eq.~(\ref{eq_nlo-smatrix-rgm})) for: the three NLO two-body operators, for a single operator, and for individual operators the matrix elements in the $^1S_0$- and $^3S_1$
channel, differences up to $10^4$ are common. The discrepancies are increased by the regularization of $G_l$ (eq.~(\ref{eq_reg-irreg-fkt}))
which reduces the size of the matrix elements. Increasing the parameter $\beta$ - defined in eq.~(\ref{eq_reg-irreg-fkt}) - to allow for a non-zero $\tilde{G}_l$ at smaller
distances, and therefore a larger overlap of the suppots of the wave function and the Gau{\ss}ian-shape potentials, comes at the expense of a less accurate expansion of
$\tilde{G}_l$ in the Gau{\ss}ian basis. Instead of increasing the basis size, I chose smaller cutoff values $\Lambda$ for which the
$S_\text{NLO}$ integrals (eq.~(\ref{eq_nlo-smatrix-rgm})) with expanded functions $\tilde{G}_l$ were practically identical to the one calculated with the
unregularized $G_l$. The parameter $\beta$ constitutes another part of $\Lambda_\text{RGM}$ and was set to $\beta=1.1~\text{fm}^{-1}$. Although, the
insensitivity of NN phase shifts to this parameter has been shown$^\text{\cite{winkler-rgm-me}}$, the results found here for five- and six-body systems
inferred from interactions based on the newly implemented pNLO calculation, require for a new investigation. In addition to the suggested refinement of the
RGM bases in ch.~\ref{sec_a5} \& \ref{sec_a6}, the $\beta$ dependence of $B(5,6)$ has to be assessed. In other words, insensitivity towards high-energy physics
parameterized by $\beta$ remains to be proved. \textit{A priori}, there is no reason why $\beta$ should be less significant in this respect than the truncation of the
basis by a finite set of width parameters.
\section{Ansatz for the triton wave function}\label{app_rgm-wfkt}
In this section, we use the triton system as an explicit example for the construction of the RGM trial wave function. The chapter is thus an
amendment to the general discussion in ch.~\ref{sec_rrgm} following eq.~(\ref{eq_rgm-int-wfkt}).\\
The proton and the two neutrons can assemble into three different two-fragment groupings: deuteron-proton ($\alpha=1$),
singlet-deuteron(\mbox{d\hspace{-.55em}$^-$})-neutron ($\alpha=2$), and (di-neutron)-proton ($\alpha=3$). For $\alpha=1,2$,
the definition of the Jacobi coordinates $\vec{\rho}_{1,2}$
is identical but the structures differ in the intermediate coupling $S_{12}$ of the spins in the two-particle fragment (see fig.~\ref{fig_t-fragm}).
Using the notation introduced in eq.~(\ref{eq_rgm-int-wfkt}), the explicit ansatz for the antisymmetrized ($\mathcal{A}$) variational wave function
with total angular momentum $J=\frac{1}{2}$ reads:
\begin{equation}\label{eq_t-trial}
\phi^{\frac{1}{2}}=\mathcal{A}\left\lbrace\sum_{\alpha=1}^3\phi_\alpha^{\frac{1}{2}}\right\rbrace=
\mathcal{A}\left\lbrace\sum_{\alpha=1}^3\left[\Phi^l_\alpha\otimes\Xi^S_\alpha\right]^{\frac{1}{2}}\right\rbrace\;\;\;.
\end{equation}
The square brackets indicate an angular momentum coupling as defined in eq.~(\ref{eq_angl-mom-coupl}).
The spin functions $\Xi$ corresponding to the respective groupings are given as:
\begin{eqnarray}
\Xi^\frac{1}{2}_1&=&\left[\left[\text{np}\right]^1\text{n}\right]^{\frac{1}{2}m}=\sum_{m_{1,2,3,4}}\left(s_nm_1s_pm_2\vert 1m_3\right)\left(1m_3s_nm_4\vert 1/2m\right)
\vert s_n,m_1\rangle\vert s_p,m_2\rangle\vert s_n,m_4\rangle\;\;\;,\nonumber\\
\Xi^\frac{1}{2}_2&=&\left[\left[\text{np}\right]^0\text{n}\right]^{\frac{1}{2}m}\;\;\;,\\
\Xi^\frac{1}{2}_3&=&\left[\left[\text{nn}\right]^0\text{n}\right]^{\frac{1}{2}m}\;\;\;,\nonumber
\end{eqnarray}
where the individual nucleon spins $\vert s_{n,p}=\frac{1}{2},m\rangle$ are coupled to a total spin $S=\frac{1}{2}$ equal to $J$ because the orbital angular momentum is zero.
The product states are superimposed with standard Clebsch-Gordan coefficients.
The ansatz for the coordinate-space component $\Phi$ of the wave function, assuming S-waves only, is:
\begin{equation}\label{eq_t-ort}
\Phi^0_\alpha=\sum_{\alpha\alpha'}c_{\alpha\alpha'}\Phi_{\alpha\alpha'}=
\sum_{\alpha\alpha'}c_{\alpha\alpha'}\exp{\left\lbrace -\gamma_1(\alpha\alpha')\vec{\rho}_1^2-\gamma_2(\alpha\alpha')\vec{\rho}_2^2\right\rbrace}\cdot
\left[\mathcal{Y}_0\left(\vec{\rho_1}\right)\otimes\mathcal{Y}_0\left(\vec{\rho_2}\right)\right]^{l=0}\;\;\;,
\end{equation}
with a total orbital angular momentum $l=0$ and variational parameters $c_{\alpha\alpha'}$.
A basis vector is then labeled by its grouping $\alpha$ and a specific pair of width parameters $\gamma_{1,2}$ indexed by $\alpha'$.
The convention of \cite{edmonds} is used for the solid harmonics $\mathcal{Y}$. With the trial wave function in this form, as a linear combination of nonorthogonal
basis functions,
\begin{equation}
\phi_{\alpha\alpha'}^\frac{1}{2}=\left[\Phi_{\alpha\alpha'}\otimes\Xi_\alpha\right]^\frac{1}{2}\;\;\;,
\end{equation}
the (Ritz) variational method reduces to an eigenvalue problem of the Hamiltonian inside the space spanned by the $\phi_{\alpha\alpha'}$.
\input{diss_t-fragm.tex}
In table~\ref{tab_t-toy}, a toy basis is defined with four $\alpha=1$, three $\alpha=2$, and two $\alpha=3$ vectors. The widths were chosen of similar size to represent either
compact or extended distributions. For example, the $\alpha=1$ set $\gamma_1=11.1~\text{fm}^{-2}$, $\gamma_2=8.2~\text{fm}^{-2}$ has a very small support
relative to the vector with $\gamma_1=0.9~\text{fm}^{-2}$, $\gamma_2=0.1~\text{fm}^{-2}$. The former is needed to refine the approximation of the wave function
at short distances, while the latter would be more important to expand extended structures or scattering functions.
\par
With a LO interaction of the form
\begin{equation}\label{eq_toy-pot}
\hat{V}=
\sum\limits_{i<j}^3e^{-0.26\cdot\vec{r_{ij}}^2}\left(-40.6-4.8\,\vec{\sigma}_i\cdot\vec{\sigma}_j\right)\;\;\;.
\end{equation}
and a variational basis of dimension $d=1$ to $d=9$, the generalized eigenvalue problem,
\begin{equation}\label{eq_gen-ev-app}
\sum_{i=1}^d\left(\langle\phi_{j}^\frac{1}{2}\vert\hat{H}\vert\phi_{i}^\frac{1}{2}\rangle-\text{EV}(3)\langle\phi_{j}^\frac{1}{2}\vert\hat{N}\vert\phi_{i}^\frac{1}{2}\rangle\right)c_i=0\;\;\;,
\end{equation}
has to be solved. To simplify the notation the pair $\alpha\alpha'$ is replaced by a single index.
From this equation, the lowest eigenvalues as listed in the fourth column
of table~\ref{tab_t-toy} are calculated. The basis is defined by eq.~(\ref{eq_t-ort}), and widths are given in table~\ref{tab_t-toy} where $d=1$ uses the width pair
of the first line, $d=2$ the first and second line, \textit{etc}..
\begin{table*}
\renewcommand{\arraystretch}{1}
  \caption{\label{tab_t-toy}{\small Width parameters defining the variational space for a toy triton when used in eq.~(\ref{eq_t-ort}). In the fourth column,
the displayed smallest eigenvalue of the Hamiltonian was obtained with a potential of the form eq.~(\ref{eq_toy-pot}) and a basis which includes all vectors above
the respective line.}}
\footnotesize
\begin{tabular}{lcc|cc|c}
\hline
&$\alpha$&$\alpha'$&$\gamma_1\;\;[\text{fm}^{-2}]$&$\gamma_2\;\;[\text{fm}^{-2}]$&EV$(3)\;\;[\text{MeV}]$\\
\hline\hline
d-n&1&1&$5.9$&$3.5$&$480$\\
&1&2&$0.9$&$0.1$&$10.5$\\
&1&3&$11.1$&$8.2$&$10.5$\\
&1&4&$3.3$&$0.5$&$9.48$\\\hline
\mbox{d\hspace{-.55em}$^-$}-n&2&1&$0.7$&$0.4$&$4.48$\\
&2&2&$1.1$&$0.4$&$3.34$\\
&2&3&$16.1$&$11.1$&$3.33$\\\hline
nn-p&3&1&$3.3$&$4.2$&$3.33$\\
&3&2&$0.9$&$0.28$&$2.48$\\
\hline
    \end{tabular}
\end{table*}
Adding vectors to the basis which correspond to highly localized structures relative to the range of the potential - line 3 and 7 in the table - have no significant
effect on the eigenvalue even spaces of such a low dimension.
In $d=9$, the eigenstate with the lowest eigenvalue EV$(3)$, as given in table~\ref{tab_t-toy}, is, for example,
expanded with coefficients $c_i$ (eq.~(\ref{eq_gen-ev-app})) as given in table~\ref{tab_t-ev}.
\begin{table*}
\renewcommand{\arraystretch}{1}
  \caption{\label{tab_t-ev}{\small Expansion coefficients $c_{\alpha\alpha'}$ for the three-nucleon ground state for a basis of dimension 9 as defined in table~\ref{tab_t-toy}.
Row number labels $\alpha$ and the column $\alpha'$.}}
\footnotesize
\begin{tabular}{c|cccc}
\hline
&1&2&3&4\\
\hline\hline
1&0.65223&0.51435&-1.7785&-0.54223\\
2&-1.7899&1.6454&-1.0757&-\\
3&0.63488&0.47931&-&-\\
\hline
    \end{tabular}
\end{table*}
As fragment $f1$ in a triton-proton scattering wave function, this state would be $\phi^{\frac{1}{2}_{f1}}$ in the channel wave function as given in eq.~(\ref{eq_rgm-ch-wfkt}).
\newpage
\bibliographystyle{unsrt}
\bibliography{diss_revtex}
\end{document}

%% file: diss_fig_bubble.tex
\begin{figure}
\begin{tikzpicture}
\coordinate (c1) at (0.5,.75);
\coordinate (c2) at (2,.75);
\coordinate (c3) at (3,.75);
\coordinate (c4) at (4,.75);
\coordinate (c5) at (5,.75);
\coordinate (c6) at (6,.75);
\coordinate (c7) at (7,.75);
\coordinate (c8) at (8,.75);
\coordinate (c9) at (9,.75);
\draw (12,.75) node(eq) [] {\large$\mathcal{O}\left(\frac{1}{M\aleph}\right)$};
\draw (0,0) node[below] {{\tiny$p=(\frac{E}{2},\vec{p})$}} --(0,1.5);
\draw[->] (0,0)--(0,.375);
\draw[->] (0,0)--(0,1.125);
\draw (1,0)--(1,1.5) node[above] {{\tiny$-p'=(\frac{E}{2},-\vec{p}')$}};
\draw[->] (1,0)--(1,.375);
\draw[->] (1,0)--(1,1.125);
\draw[fill=gray!50,draw] (c1) ellipse (0.6cm and 0.3cm);
\draw (c1) node(TM) [] {T};
\draw (c2) node(eq) [] {=};
\draw (2.5,0)--(3.5,1.5);
\draw (3.5,0)--(2.5,1.5);
\draw (c3) node(eq) [circle,fill=white,draw,inner sep=0.03cm] {\footnotesize 6};
\draw (c4) node(eq) [] {+};
\draw (5,1.5)--(4.5,1.75);
\draw (5,1.5)--(5.5,1.75);
\draw (4.5,.75) node(eq) [] {$\wedge$};
\draw (5.5,.75) node(eq) [] {$\wedge$};
\draw (c5) ellipse (0.5cm and 0.75cm);
\draw (5,0)--(4.5,-.25);
\draw (5,0)--(5.5,-.25);
\draw (5,1.5) node(eq) [circle,fill=white,draw,inner sep=0.03cm] {\footnotesize 6};
\draw (5,0) node(eq) [circle,fill=white,draw,inner sep=0.03cm] {\footnotesize 6};
\draw (c5) node(ll) [] {\tiny$(l_0,\vec{l})$};
\draw (c6) node(eq) [] {+};
\draw (7,2.25)--(6.5,2.5);
\draw (7,2.25)--(7.5,2.5);
\draw (7,1.5) ellipse (0.5cm and 0.75cm);
\draw (7,0) ellipse (0.5cm and 0.75cm);
\draw (7,-.75)--(6.5,-1);
\draw (7,-.75)--(7.5,-1);
\draw (c7) node(eq) [circle,fill=white,draw,inner sep=0.03cm] {\footnotesize 6};
\draw (7,2.25) node(eq) [circle,fill=white,draw,inner sep=0.03cm] {\footnotesize 6};
\draw (7,-.75) node(eq) [circle,fill=white,draw,inner sep=0.03cm] {\footnotesize 6};
\draw (c8) node(eq) [] {+};
\draw (c9) node(eq) [] {$\vdots$};
\draw[->] (-.8,2.5) -- (-.8,.5) node[midway,sloped,below] {vertex};
\draw[->] (-.7,2.6) -- (1.3,2.6) node[midway,sloped,above] {loop};
\coordinate (c21) at (0.5,-3);
\coordinate (c22) at (2,-3);
\coordinate (c23) at (3,-3);
\coordinate (c24) at (4,-3);
\coordinate (c25) at (5,-3);
\coordinate (c26) at (6,-3);
\coordinate (c27) at (7,-3);
\coordinate (c28) at (8,-3);
\coordinate (c29) at (9,-3);
\draw (12,-3) node(eq) [] {\large$\mathcal{O}\left(\frac{1}{M^2}\right)$};

\draw (c22) node(eq) [] {+};
\draw (2.5,-3.75)--(3.5,-2.25);
\draw (3.5,-3.75)--(2.5,-2.25);
\draw (c23) node(eq) [circle,double,fill=white,draw,inner sep=0.03cm] {\footnotesize 8};
\draw (c24) node(eq) [] {+};
\draw (4.5,-2.5)--(4.5,-2);
\draw (5.5,-2.5)--(5.5,-2);
\draw (4.5,-2.65)--(5.5,-4.15);
\draw (5.5,-2.65)--(4.5,-4.15);
\draw[fill=gray!20,draw] (5,-2.5) ellipse (0.6cm and 0.3cm);
\draw (5,-2.5) node(TM) [] {T$^\text{(LO)}$};
\draw (5,-3.4) node(eq) [circle,double,fill=white,draw,inner sep=0.03cm] {\footnotesize 8};
\draw (c26) node(eq) [] {+};
\draw (6.5,-4.15)--(6.5,-3.65);
\draw (7.5,-4.15)--(7.5,-3.65);
\draw (6.5,-3.5)--(7.5,-2);
\draw (7.5,-3.5)--(6.5,-2);
\draw[fill=gray!20,draw] (7,-3.65) ellipse (0.6cm and 0.3cm);
\draw (7,-3.65) node(TM) [] {T$^\text{(LO)}$};
\draw (7,-2.75) node(eq) [circle,double,fill=white,draw,inner sep=0.03cm] {\footnotesize 8};
\draw (c28) node(eq) [] {+};
\draw (8.5,-3.9)--(8.5,-4.4);
\draw (9.5,-3.9)--(9.5,-4.4);
\draw (8.5,-3.75)--(9.5,-2.25);
\draw (9.5,-3.75)--(8.5,-2.25);
\draw (8.5,-2.1)--(8.5,-1.6);
\draw (9.5,-2.1)--(9.5,-1.6);
\draw[fill=gray!20,draw] (9,-3.9) ellipse (0.6cm and 0.3cm);
\draw[fill=gray!20,draw] (9,-2.1) ellipse (0.6cm and 0.3cm);
\draw (9,-3.9) node(TM) [] {T$^\text{(LO)}$};
\draw (9,-2.1) node(TM) [] {T$^\text{(LO)}$};
\draw (c29) node(eq) [circle,double,fill=white,draw,inner sep=0.03cm] {\footnotesize 8};
\end{tikzpicture}
\caption{\label{fig_eft-exp}{\small Lowest order contributions to the amplitude T. The vertices are labeled by the mass dimension of the operator. The first line of diagrams is T$^\text{(LO)}$.}}
\end{figure}

%% file: diss_rgm-cluster.tex
\usetikzlibrary{calc}
\begin{figure}
\begin{tikzpicture}
\begin{scope}[xshift=0.0cm,yshift=0.0cm,scale=1.0]
\draw[draw=gray] (5.4,-2.9) rectangle +(6.9,6);
\draw[draw=gray] (-2.4,-2.9) rectangle +(7.2,6);
\coordinate (rho1) at (1.5,2);
\coordinate (rho2) at (1.3333,1.6666);
\coordinate (rho3) at (1.5,1.5);
\path (1,1) node(a) [circle,draw,fill=gray,inner sep=1pt] {\tiny n$_3$}
          (1,2) node(b) [circle,draw,fill=gray!10!white,inner sep=1pt] {\tiny p$_1$}
          (2,2) node(c) [circle,draw,fill=gray!10!white,inner sep=1pt] {\tiny p$_2$}
          (2,1) node(d) [circle,draw,fill=gray,inner sep=1pt] {\tiny n$_4$};
\draw (node cs:name=b) -- (node cs:name=c);
\node [] (asfd) at ($ (a)!.5!(b) $) {$\vec{\rho}_2$};
\node [above] (asfd) at ($ (b)!.5!(c) $) {$\vec{\rho}_1$};
\node [below] (asfd) at (rho3) {$\vec{\rho}_3$};
\draw (node cs:name=a) -- (node cs:name=rho1);
\draw (node cs:name=d) -- (node cs:name=rho2);
\node [above] (asfd) at ($ (rho3)!.518!(8.5,1) $) {$\vec{\rho}_\text{\tiny rel}$};
\path (8,0.5) node(e) [circle,draw,fill=gray,inner sep=1pt] {\tiny n$_2$}
          (8,1.5) node(f) [circle,draw,fill=gray!10!white,inner sep=1pt] {\tiny p$_1$}
          (9,1.5) node(g) [circle,draw,fill=gray!10!white,inner sep=1pt] {\tiny p$_3$}
          (9,0.5) node(h) [circle,draw,fill=gray,inner sep=1pt] {\tiny n$_4$};
\draw (node cs:name=e) -- (node cs:name=f);
\draw (node cs:name=g) -- (node cs:name=h);
\node [right] (asfd) at ($ (g)!.5!(h) $) {$\vec{\rho}_2$};
\node [label=235:\tiny $\vec{\rho}_1$] (asfd) at ($ (e)!.5!(f) $) {};
\node [below] (asfd) at (8.5,1) {$\vec{\rho}_3$};
\draw (8,1) -- (9,1);
\draw (1.5,1.5) -- (8.5,1);
\draw[rotate=0]  (1.5,1.5) ellipse  (1 and 1);
\draw[rotate=0] (8.5,1) ellipse  (1 and 1);
\node [below] (asfd) at (8.8,-1) {Fragment R (f2): deuteron-deuteron};
\node [below] (asfd) at (8.8,-1.5) {$\left[\left[\left[s_1\otimes s_2\right]^1\otimes\left[s_3\otimes s_4\right]^1\right]^{S=0}\otimes L=0\right]^{J_2=0}$};
\node [below] (asfd) at (1.2,-1) {Fragment L (f1): 3-helium-neutron};
\node [below] (asfd) at (1.2,-1.5) {$\left[\left[\left[\left[s_1\otimes s_2\right]^{0,1}\otimes s_3\right]^{\frac{1}{2}}\otimes s_4\right]^{S=0}\otimes L=0\right]^{J_1=0}$};
\node [below] (asfd) at (5,-3) {$\Rightarrow\;\;S_c=0\;\;\wedge\;\;J=L_\text{\tiny rel}$};
\end{scope}
\end{tikzpicture}
\caption{\label{fig_rgm-cluster}{\small Example of a decomposition of berillium-8 in two helium-4 cluster. Subscripts indicate the number of the proton or neutron
used to label spin- and orbital coordinates. Within a fragment, standard Jacobi coordinates are used:
$\vec{\rho}_i=\vec{r}_i-\frac{\sum_{n=1}^{i-1}m_\text{\tiny N}\vec{r}_n}{\sum_{n=1}^{i-1}m_\text{\tiny N}}$.}}
\end{figure}

%% file: he5_spect.tex
\usetikzlibrary{calc}
\begin{figure}
\begin{minipage}[l]{.9\textwidth}
\begin{tikzpicture}
\begin{scope}[xshift=-2.0cm,yshift=0.0cm,scale=1.0]
\begin{scope}[xshift=-1.5cm,yshift=0.0cm,scale=1.0]
\draw [lightgray,line width=3pt](1,3.311) -- (3,3.311);
\draw [red](1,0.935) -- (3,0.935);
\node [right,gray] (asfd) at (2.9,3.45) {\footnotesize\renewcommand{\arraystretch}{0.1} $\begin{array}{l}
23.98~\text{MeV}\\(\alpha-\text{n LO})\end{array}$};
\node [right,red] (asfd) at (2.9,0.8) {\footnotesize\renewcommand{\arraystretch}{0.1} $\begin{array}{l}28.30~\text{MeV}\\
(\alpha-\text{n NLO})\end{array}$};
\node [gray] (asfd) at (1.5,2.25) {$^2S_\frac{1}{2}$};
\node [gray] (asfd) at (2.5,2.25) {$^2P_J$};
\node [left] (lo-s) at (1,3.355) {\footnotesize $23.90~$MeV};
\node [above] (lo-s) at (1.5,3.355) {$\text{\tiny LO}$};
\draw [dashed](1,3.355) -- (2,3.355);
\node [left] (lo-s) at (1,0.979) {\footnotesize $28.22~$MeV};
\node [above] (lo-s) at (1.5,0.979) {$\text{\tiny NLO}$};
\draw [](1,0.979) -- (2,0.979);
\draw [->](1.15,3.355) -- (1.15,0.979);
\node [right] (lo-s) at (3,3) {\footnotesize $23.99~$MeV};
\node [above] (lo-s) at (2.5,3.31) {$\text{\tiny LO}$};
\draw [dashed](2,3.31) -- (3,3.31);
\node [right] (lo-s) at (3,1.298) {\footnotesize $27.64~$MeV};
\node [above] (lo-s) at (2.5,1.298) {$\text{\tiny NLO}$};
\draw [](2,1.298) -- (3,1.298);
\draw [->](2.15,3.31) -- (2.15,1.298);
\node [below,draw,rectangle,inner sep=4pt] (lo-s) at (2,-0.25) {\small $\Lambda=300~$MeV};
\draw (1,0) -- (1,5.5);
\draw [dotted](2,0) -- (2,5.5);
\draw (3,0) -- (3,5.5);
\end{scope}
\begin{scope}[xshift=3.2cm,yshift=0.0cm,scale=1.0]
\draw [lightgray,line width=3pt](1,4.21) -- (3,4.21);
\draw [red](1,0.935) -- (3,0.935);
\node [right,gray] (asfd) at (3,4.21) {\footnotesize $22.35~$MeV};
\node [gray] (asfd) at (2,2.25) {$^2P_J$};
\node [below] (lo-s) at (2,4.21) {\footnotesize $22.35~$MeV};
\node [above] (lo-s) at (1.5,4.21) {$\text{\tiny LO}$};
\draw [dashed](1,4.21) -- (3,4.21);
\node [below] (lo-s) at (2,1.254) {\scriptsize $27.72~$MeV};
\node [above] (lo-s) at (1.5,1.254) {$\text{\tiny NLO}$};
\draw [](1,1.254) -- (3,1.254);
\draw [->](1.15,4.21) -- (1.15,1.254);
\node [below,draw,rectangle,inner sep=4pt] (lo-s) at (2,-0.25) {\small $\Lambda=400~$MeV};
\draw (1,0) -- (1,5.5);
\draw (3,0) -- (3,5.5);
\end{scope}
\begin{scope}[xshift=8.0cm,yshift=0.0cm,scale=1.0]
\draw [lightgray,line width=3pt](1,4.345) -- (3,4.345);
\draw [red](1,0.935) -- (3,0.935);
\node [right,gray] (asfd) at (3,4.345) {\tiny $22.10~$MeV};
\node [gray] (asfd) at (1.5,2.25) {$^2S_\frac{1}{2}$};
\node [gray] (asfd) at (2.5,2.25) {$^2P_J$};
\node [left] (lo-s) at (1,4.5) {\footnotesize $22.13~$MeV};
\node [above] (lo-s) at (1.5,4.33) {$\text{\tiny LO}$};
\draw [dashed](1,4.33) -- (2,4.33);
\node [left] (lo-s) at (1,0.979) {\footnotesize $28.22~$MeV};
\node [above] (lo-s) at (1.5,0.979) {$\text{\tiny NLO}$};
\draw [](1,0.979) -- (2,0.979);
\draw [->](1.15,4.33) -- (1.15,0.979);
\node [right] (lo-s) at (3,4.0) {\tiny $22.15~$MeV};
\node [above] (lo-s) at (2.5,4.32) {$\text{\tiny LO}$};
\draw [dashed](2,4.32) -- (3,4.32);
\node [right] (lo-s) at (3,1.3) {\tiny $27.63~$MeV};
\node [above] (lo-s) at (2.5,1.3) {$\text{\tiny NLO}$};
\draw [](2,1.3) -- (3,1.3);
\draw [->](2.15,4.32) -- (2.15,1.3);
\node [below,draw,rectangle,inner sep=4pt] (lo-s) at (2,-0.25) {\small $\Lambda=500~$MeV};
\draw (1,0) -- (1,5.5);
\draw [dotted](2,0) -- (2,5.5);
\draw (3,0) -- (3,5.5);
\end{scope}
\end{scope}
\end{tikzpicture}
\par\vspace{0pt}
\end{minipage}
\caption{\label{fig_he5-spect}{\small Energy levels of the $^5$He system for $J^\pi=\frac{1}{2}^+,J^-$ calculated with the RGM for EFT($\slashed{\pi}$) interactions.
For three values of the regulator parameter $\Lambda$ (eq.~(\ref{eq_reg-fkt-mom})), the LO (dashed) and pNLO (solid) predictions are displayed. The $\alpha$-n thresholds are set by
$B(\alpha)$ and are predictions at LO (gray) and input at pNLO (red).}}
\end{figure}

%% file: he5_spect-nocoul.tex
\usetikzlibrary{calc}
\begin{figure}
\renewcommand{\arraystretch}{0.35}
\centering
\begin{minipage}[b]{.95\textwidth}
\centering
\begin{tikzpicture}
\begin{scope}[xshift=-0.5cm,yshift=0.0cm,scale=1.0]
\draw [lightgray,line width=3pt](1,4.49) -- (4,4.49);
\draw [red](1,0.75) -- (2.5,0.75);
\draw [red](2.5,1.13) -- (4,1.13);
\draw [dashed](1,4.49) -- (4,4.49);
\draw [](1,1.12) -- (2.5,1.12);
\draw [->](1.15,4.49) -- (1.15,1.12);
\draw [](2.5,1.51) -- (4,1.51);
\draw [->](2.65,4.49) -- (2.65,1.51);

\node [left,gray] (asfd) at (1,4.49) {\scriptsize $\begin{array}{c}23.26~\text{MeV}\\ B(\alpha,\text{LO})\end{array}$};
\node [left,red] (asfd) at (1,0.75) {\scriptsize $\begin{array}{c}28.88~\text{MeV}\\ B(\alpha,\text{NLO})\end{array}$};
\node [right,red] (asfd) at (4,1.13) {\scriptsize $28.30~\text{MeV}$};
\node [above] (lo-s) at (1.8,4.49) {\tiny $23.27$MeV};
\node [above] (lo-s) at (1.8,1.12) {\tiny $28.32$MeV};
\node [above] (lo-s) at (3.3,1.51) {\tiny $27.73$MeV};
\node [below,draw,rectangle,inner sep=4pt] (lo-s) at (5.5,-0.25) {\small $\Lambda=400~$MeV};
\draw (1,0) -- (1,6);
\draw [dotted] (2.5,0) -- (2.5,6);
\draw (4,0) -- (4,6);
\node [gray] (lo-s) at (5.5,3) {\huge $^2P_J$};
\node [fill=white,draw,rectangle,inner sep=4pt] (lo-s) at (2.5,3) {\small Coulombless};
\draw [lightgray,line width=3pt](7,4.59) -- (8.5,4.59);
\draw [red](7,1.13) -- (8.5,1.13);
\node [right,gray] (asfd) at (8.5,4.59) {\footnotesize $22.35~$MeV};
\node [right,red] (asfd) at (8.5,1.13) {\scriptsize $28.30~\text{MeV}$};
\node [above] (lo-s) at (7.8,4.59) {\tiny $22.35~$MeV};
\draw [dashed](7,4.59) -- (8.5,4.59);
\node [above] (lo-s) at (7.8,1.368) {\tiny $27.72~$MeV};
\draw [](7,1.368) -- (8.5,1.368);
\draw [->](7.15,4.59) -- (7.15,1.368);
\draw (7,0) -- (7,6);
\draw (8.5,0) -- (8.5,6);
\end{scope}
\end{tikzpicture}
\par\vspace{0pt}
\end{minipage}
\caption{\label{fig_he5-spect-nocoul}{\small Energy levels of the $^5$He system in a $^2P_J$ state calculated with the RGM for EFT($\slashed{\pi}$) interactions
without (left) and with (right) Coulomb interaction.
For a regulator parameter of $\Lambda =400~$MeV (eq.~(\ref{eq_reg-fkt-mom})), the LO (dashed) and pNLO (solid) predictions are displayed. The $\alpha$-n thresholds are set by
$B(\alpha)$ and are predictions at LO (gray) and input at pNLO (red). The left column uses the same potential as employed for fig.~\ref{fig_he5-spect} while the right one
refits the 3NI parameter $A_{3\text{NI}}$ to yield $B(\alpha,\text{exp})$ even without Coulomb repulsion.}}
\end{figure}

%% file: diss_6he-fragm.tex
\usetikzlibrary{calc}
\begin{figure}
\begin{minipage}[b]{.45\textwidth}
\centering
\begin{tikzpicture}
\begin{scope}[xshift=1.2cm,yshift=0.0cm,scale=1.5]
\coordinate (rho1) at (1.5,2);
\coordinate (rho2) at (1.4,0.4);
\coordinate (alpha) at (1.5,0);
\coordinate (n1) at (1,2);
\coordinate (n2) at (2,2);
\path (alpha) node(a) [circle,draw,fill=gray,inner sep=5pt] {\small $\alpha$}
          (n1)       node(b) [circle,draw,fill=gray!10!white,inner sep=1pt] {\small n}
          (n2)       node(c) [circle,draw,fill=gray!10!white,inner sep=1pt] {\small n};
\draw (node cs:name=a) -- (node cs:name=b);
\node [left] (asfd) at ($ (a)!.5!(b) $) {$\sum_ic_ie^{-\gamma_{i,1}\vec{n}_1^2}\mathcal{Y}_{l_1}$};
\node [right] (asfd) at ($ (rho2)!.5!(c) $) {$\sum_jc_je^{-\gamma_{j,2}\vec{n}_2^2}\mathcal{Y}_{l_2}$};
\draw (node cs:name=c) -- (node cs:name=rho2);
\end{scope}
\end{tikzpicture}
\par\vspace{0pt}
\end{minipage}
\hfill
\begin{minipage}[b]{.45\textwidth}
\centering
\begin{tikzpicture}
\begin{scope}[xshift=0.0cm,yshift=0.0cm,scale=1.2]
\coordinate (rho1) at (1,2);
\coordinate (alpha) at (1,0);
\coordinate (n1) at (-1,2);
\coordinate (n2) at (3,2);
\path (alpha) node(a) [circle,draw,fill=gray,inner sep=5pt] {\small $\alpha$}
          (n1)       node(b) [circle,draw,fill=gray!10!white,inner sep=1pt] {\small n}
          (n2)       node(c) [circle,draw,fill=gray!10!white,inner sep=1pt] {\small n};
\draw (node cs:name=b) -- (node cs:name=c);
\draw (node cs:name=a) -- (node cs:name=rho1);
\draw (1,1.1) arc (90:45.4:1.1);
\node [] (asfd) at (1.29,0.7) {$\theta$};
\draw[dashed] (node cs:name=a) -- (node cs:name=c);
\end{scope}
\end{tikzpicture}
\par\vspace{0pt}
\end{minipage}
\caption{\label{fig_6he-fragm}{\small The $^4$He-neutron-neutron group structure of the RGM basis for 6-helium.
The theoretically predicted$^\text{\cite{bert-op-angle}}$ opening angle
$\theta=(83^\circ)^\text{\tiny +20}_\text{\tiny -10}$ motivates the choice of width parameters $\gamma_{i,(1,2)}$.}}
\end{figure}

%% file: tab_pot-coord-space.tex
\begin{sidewaystable}
 \caption{\label{tab_pot-coord-lec-spect}\small Coordinate space representation of the potential $V^\text{(NLO)}$ expressed in LECs as used in eq.~\ref{eq_nlo-pot-mom-space}
(third column) and in the spectroscopic LECs (fourth column). The first column specifies the number of the structure in the modified \texttt{QUAFL\_EFT\_new} part of the code.
The Coulomb prefactor contains the electron mass $m_e$ and charge $e$, and $I_0=e^{-\frac{\Lambda^2}{4}\vec{r}^2}$ as in eq.~\ref{eq_nlo-pot-coord-space}.}
\begin{tabular}{c|cll}
\hline
\texttt{No.}&$V_{0}$&&Spectroscopic\\
\hline\hline
$(1)$&$\frac{2}{A(A-1)}$&$\RGB{1 0 0}{\mathbb{1}}\otimes\RGB{0 0 1}{\mathbb{1}}$&$\RGB{1 0 0}{\mathbb{1}}\otimes\RGB{0 0 1}{\mathbb{1}}$\\
$(2)$&$
-\frac{\hbar^2}{2m_\text{\tiny N}}=-\frac{\hbar^2c^2}{m_nc^2+m_pc^2}$&$\RGB{1 0 0}{\vec{\nabla}^2}\otimes\RGB{0 0 1}{\mathbb{1}}$&
$\RGB{1 0 0}{\vec{\nabla}^2}\otimes\RGB{0 0 1}{\mathbb{1}}$\\
$(3)$&$\frac{2}{\sqrt{\pi}}\,m_ec^2\cdot\frac{\vert e\vert}{m_ec^2}$&$\RGB{1 0 0}{\frac{1}{r}}\otimes\RGB{0 0 1}{\mathbb{1}}$&
$\RGB{1 0 0}{\frac{1}{r}}\otimes\RGB{0 0 1}{\mathbb{1}}$\\
$(4)$&$\frac{1}{2}I_0$&
$\RGB{1 0 0}{\mathbb{1}}\otimes\Bigg[\left[\Lambda^2\left(3C^\text{\tiny NLO}_1-\frac{3}{4}C^\text{\tiny NLO}_2\right)+2C^\text{\tiny NLO}_1\right]\RGB{0 0 1}{\mathbb{1}}$&
$\RGB{1 0 0}{\mathbb{1}}\otimes\frac{1}{32\pi}\Bigg[\Big[4C^\text{\tiny NLO}_{_{\left(^1S_0\right)}}+12C^\text{\tiny NLO}_{\left(^3S_1\right)}
-3\Lambda^2\left(3C^\text{\tiny NLO}_{\left(^1P_1\right)}+C^\text{\tiny NLO}_{\left(^3P_0\right)}+3C^\text{\tiny NLO}_{\left(^3P_1\right)}+5C^\text{\tiny NLO}_{\left(^3P_2\right)}\right)\Big]\RGB{0 0 1}{\mathbb{1}}+$\\
&&$\;\;\;\;\;\;\;+\big[\Lambda^2\left(3C^\text{\tiny NLO}_3-\frac{3}{4}C^\text{\tiny NLO}_4+C^\text{\tiny NLO}_6-\frac{1}{4}C^\text{\tiny NLO}_7\right)$&$
\;\;\;\;\;\;\;\;\;\;\left[-4C^\text{\tiny NLO}_{\left(^1S_0\right)}+4C^\text{\tiny NLO}_{\left(^3S_1\right)}-
\Lambda^2\left(-9C^\text{\tiny NLO}_{\left(^1P_1\right)}+C^\text{\tiny NLO}_{\left(^3P_0\right)}+3C^\text{\tiny NLO}_{\left(^3P_1\right)}+5C^\text{\tiny NLO}_{\left(^3P_2\right)}\right)\right]\RGB{0 0 1}{\vec{\sigma}_1\cdot\vec{\sigma}_2}\Bigg]$\\
&&$\;\;\;\;\;\;\;+2C^\text{\tiny NLO}_2\big]\RGB{0 0 1}{\vec{\sigma}_1\cdot\vec{\sigma}_2}\Bigg]$&\\
$(5)$&$\frac{1}{2}\cdot\frac{1}{16\pi^{\frac{3}{2}}}\Lambda^3$&$
\RGB{1 0 0}{\Big\lbrace e^{-\frac{\Lambda^2}{4}\vec{r}^2},\vec{\nabla}^2\Big\rbrace}\otimes
\Big[C^\text{\tiny NLO}_2\;\RGB{0 0 1}{\mathbb{1}}$&$\RGB{1 0 0}{\mathbb{1}}\otimes\frac{1}{48\pi}\Bigg[\left(9C^\text{\tiny NLO}_{\left(^1P_1\right)}+
6C^\text{\tiny NLO}_{\left(^1S_0\right)}+3C^\text{\tiny NLO}_{\left(^3P_0\right)}+9C^\text{\tiny NLO}_{\left(^3P_1\right)}+15C^\text{\tiny NLO}_{\left(^3P_2\right)}+18C^\text{\tiny NLO}_{\left(^3S_1\right)}\right)\RGB{0 0 1}{\mathbb{1}}+$\\
&&$\;\;\;\;\;\;\;+\left(-C^\text{\tiny NLO}_4-\frac{1}{3}C^\text{\tiny NLO}_7\right)\RGB{0 0 1}{\vec{\sigma}_1\cdot\vec{\sigma}_2}\Big]$&
$\;\;\;\;\;\;\;\;\;\;\left(9C^\text{\tiny NLO}_{\left(^1P_1\right)}+6C^\text{\tiny NLO}_{\left(^1S_0\right)}-
C^\text{\tiny NLO}_{\left(^3P_0\right)}-3C^\text{\tiny NLO}_{\left(^3P_1\right)}-5C^\text{\tiny NLO}_{\left(^3P_2\right)}-
6C^\text{\tiny NLO}_{\left(^3S_1\right)}\right)\RGB{0 0 1}{\vec{\sigma}_1\cdot\vec{\sigma}_2}\Bigg]$\\
$(6)$&$\frac{1}{4}\Lambda^4I_0$&$\RGB{1 0 0}{\vec{r}^2}\otimes\Big[\left(-C^\text{\tiny NLO}_1+C^\text{\tiny NLO}_2\right)\RGB{0 0 1}{\mathbb{1}}+$&$\RGB{1 0 0}{\mathbb{1}}\otimes
\frac{1}{64\pi}\Bigg[\left(15C^\text{\tiny NLO}_{\left(^1P_1\right)}+6C^\text{\tiny NLO}_{\left(^1S_0\right)}+5C^\text{\tiny NLO}_{\left(^3P_0\right)}+
15C^\text{\tiny NLO}_{\left(^3P_1\right)}+25C^\text{\tiny NLO}_{\left(^3P_2\right)}+18C^\text{\tiny NLO}_{\left(^3S_1\right)}\right)\RGB{0 0 1}{\mathbb{1}}+$\\
&&$\;\;\;\;\;\;\;\;\left(-C^\text{\tiny NLO}_3+C^\text{\tiny NLO}_4-C^\text{\tiny NLO}_6+\frac{1}{4}C^\text{\tiny NLO}_7\right)\RGB{0 0 1}{\vec{\sigma}_1\cdot\vec{\sigma}_2}\Big]$&$\left(-15C^\text{\tiny NLO}_{\left(^1P_1\right)}-6C^\text{\tiny NLO}_{\left(^1S_0\right)}+
C^\text{\tiny NLO}_{\left(^3P_0\right)}+6C^\text{\tiny NLO}_{\left(^3P_1\right)}+8C^\text{\tiny NLO}_{\left(^3P_2\right)}+6C^\text{\tiny NLO}_{\left(^3S_1\right)}-
6\sqrt{2}C^\text{\tiny NLO}_{\left(\text{\tiny SD}\right)}\right)\RGB{0 0 1}{\vec{\sigma}_1\cdot\vec{\sigma}_2}\Bigg]$\\
$(7)$&$\frac{1}{2}\Lambda^2I_0$&$C^\text{\tiny NLO}_5\,\RGB{1 0 0}{\vec{L}}\cdot\RGB{0 0 1}{\vec{S}}$&$\frac{1}{16\pi}\left(2C^\text{\tiny NLO}_{\left(^3P_0\right)}+3C^\text{\tiny NLO}_{\left(^3P_1\right)}-5C^\text{\tiny NLO}_{\left(^3P_2\right)}\right)$\\
$(8)$&$\frac{1}{4}\Lambda^4I_0$&$
\left(-C^\text{\tiny NLO}_6+\frac{1}{4}C^\text{\tiny NLO}_7\right)\cdot\left[\vec{\sigma}_1\cdot\vec{r}\vec{\sigma}_2\cdot\vec{r}-\frac{1}{3}\vec{r}^2\vec{\sigma}_1\cdot\vec{\sigma}_2\right]$&
$-\frac{1}{32\pi}\left(2C^\text{\tiny NLO}_{\left(^3P_0\right)}-3C^\text{\tiny NLO}_{\left(^3P_1\right)}+C^\text{\tiny NLO}_{\left(^3P_2\right)}\right)$\\
$(9)$&$-\frac{1}{2}\cdot\frac{\sqrt{5}}{16\pi^{\frac{3}{2}}}\Lambda^3$&$C^\text{\tiny NLO}_7\cdot\Bigg\lbrace e^{-\frac{\Lambda^2}{4}\vec{r}^2},\Big[\big[\partial^r\otimes
\partial^s\big]^{2}\otimes\big[\sigma_1^p\otimes\sigma_2^q\big]^{2}\Big]^{00}\Bigg\rbrace\;$&$
-\frac{1}{16\pi}\left(-6\sqrt{2}C^\text{\tiny NLO}_{\left(\text{\tiny SD}\right)}+2C^\text{\tiny NLO}_{\left(^3P_0\right)}-3C^\text{\tiny NLO}_{\left(^3P_1\right)}+C^\text{\tiny NLO}_{\left(^3P_2\right)}\right)$\\
\hline
\end{tabular}
\end{sidewaystable}

%% file: enum_genalg.tex
\begin{enumerate}
\item A set $\left\lbrace a_i\right\rbrace_{i\in\lbrace 1..\text{N}\rbrace}$ is mapped onto a bit representation which is called individuum.\vspace{.5cm}\\
\begin{tikzpicture}
\begin{scope}[xshift=0.0cm,yshift=-0.0cm,scale=1.0]
\foreach \x in {1,...,9}
{
\draw (\x,0) +(-0.5,-0.25) rectangle ++(0.5,0.25);
}
\foreach \x in {2,4,5,...,8}
{
\draw (\x,0) node{$\ldots$};
}
\draw (1,0) node{$a_1$};
\draw (3,0) node{$a_i$};
\draw (9,0) node{$a_\text{\tiny N}$};
\foreach \x in {1,1.4,...,5}
{
\draw (\x,-1) +(-0.2,-0.2) rectangle ++(0.2,0.2);
}
\foreach \x in {1,1.4,1.8,2.6,5}
{
\draw (\x,-1) node{1};
}
\foreach \x in {2.2,3,3.4,3.8,4.2}
{
\draw (\x,-1) node{0};
}
\draw (4.6,-1) node{\tiny $\ldots$};
\draw[draw,dotted,thick] (2.5,-.25) -- (.8,-.8);
\draw[draw,dotted,thick] (3.5,-.25) -- (5.2,-.8);
\end{scope}
\end{tikzpicture}\\
In this work in particular, $a_i$'s were either LECs $A_i$ or RGM basis width parameters $\gamma$.
\item A random set (initial population) of $Z$ individuums is generated from a set of boundary conditions of allowed values. Every
individuum is assigned a fitness value through the function $W$ (\textit{e.g.}, eq.~\ref{eq_nlo-fit-para}).
The individuums are ordered according to this fitness value $w_i$\vspace{.5cm}\\
\begin{tikzpicture}
\begin{scope}[xshift=0.0cm,yshift=-0.0cm,scale=1.0]
\foreach \x in {1,1.6,...,4.6}
\foreach \y in {1,2.2,3.4}
{
\draw (\x,\y) +(-0.3,-0.2) rectangle ++(0.3,0.2);
}
\draw (1,1) node{$a_1$};
\draw (1,2.2) node{$b_1$};
\draw (1,3.4) node{$c_1$};
\draw (4,1) node{$a_\text{\tiny N}$};
\draw (4,2.2) node{$b_\text{\tiny N}$};
\draw (4,3.4) node{$c_\text{\tiny N}$};
\draw (5.,3.4) node[right]{$w_1=w_\text{max}\;\;\;$
{\scriptsize\renewcommand{\arraystretch}{.6}
$\begin{array}{l}
\text{strongest individuum,}\\
\text{highest probability of becoming a parent;}
\end{array}$
}};
\draw (5.,2.2) node[right]{$w_j$};
\draw (5.,1) node[right]{$w_\text{Z}=w_\text{min}\;\;\;$      {\scriptsize weakest individuum, smallest probability to procreate;}};
\draw[draw,dotted,thick] (1.6,2.6) -- (1.6,3.0);
\draw[draw,dotted,thick] (5.3,2.6) -- (5.3,3.0);
\draw[draw,dotted,thick] (1.6,1.4) -- (1.6,1.8);
\draw[draw,dotted,thick] (5.3,1.4) -- (5.3,1.8);
\end{scope}
\end{tikzpicture}
\\
with $w_i<w_j\;\;\forall i<j\;\;\;.$
\item Two individuums (parents) are selected from the population. The rank ($b$) of the two parents in the hierarchy introduced in the previous step, is determined by
a random number $x\in[0,1]$ and $b=x\sqrt{\frac{2}{x}ln(x)}\cdot\frac{Z}{2}\;$. If $b<Z$, the element $b$ is one chosen parent. This procedure ensures a higher probability
of selecting parents with a small $W$ value but does not entirely neglect entries with high $W$.
\item The two parents are used to create two individuums (offspring) via uniform crossover (equations for bits $x_i$ and $z_i$)
with a random individuum $m$ (mask).\vspace{.5cm}\\
\begin{tikzpicture}
\begin{scope}[xshift=0.0cm,yshift=-0.0cm,scale=1.0]
\foreach \x in {1,1.6,...,4.6,7.9,8.5,...,11.5}
\foreach \y in {1,2.6}
{
\draw (\x,\y) +(-0.3,-0.2) rectangle ++(0.3,0.2);
}
\foreach \x in {4.7,5.2,...,7.2}
{
\draw (\x,1.55) +(-0.25,-0.2) rectangle ++(0.25,0.2);
}
\draw (2.8,2.6) node{$w_i$};
\draw (9,2.6) node{$x_i$};
\draw (2.2,1) node{$y_i$};
\draw (9.6,1) node{$z_i$};
\draw (5.7,1.55) node{\small $m_i$};
\draw (10.9,2.) node{$x_i=(w_i\wedge m_i)\vee (y_i\wedge \bar{m}_i)$};
\draw (11.5,1.55) node{$z_i=(w_i\wedge \bar{m}_i)\vee (y_i\wedge m_i)$};
\draw (2.0,0.3) node{parents};
\draw (8.9,0.3) node{offspring};
\draw (6,2.4) node{mask};
\draw[draw,->] (4.6,2) -- (7.3,2);
\end{scope}
\end{tikzpicture}\\
Uniform crossover is only one of infinitely many choices for a rule to create offspring.
\item Each offspring is mutated, i.e. every bit is inverted with a probability of $p$. Here, $p=0.01$ was used.
\item The $W$ values of the offspring are calculated.
\item The offspring is inserted in the population according to its $W$ value.
\item The two individuums with the highest $W$ value are removed (only the strong survive) from the population.
\end{enumerate}

%% file: diss_t-fragm.tex
\usetikzlibrary{calc}
\begin{figure}
\begin{minipage}[b]{.49\textwidth}
\centering
\begin{tikzpicture}
\begin{scope}[xshift=1.2cm,yshift=0.0cm,scale=1.]
\coordinate (rho1) at (1.5,1);
\coordinate (rho2) at (1.5,1);
\coordinate (n1) at (1.5,0);
\coordinate (p) at (1.5,2);
\coordinate (n2) at (4,0);
\path (n1)     node(a) [circle,draw,fill=gray!10!white,inner sep=5pt] {\small n}
          (p)       node(b) [circle,draw,fill=gray,inner sep=5pt] {\small p}
          (n2)     node(c) [circle,draw,fill=gray!10!white,inner sep=5pt] {\small n};
\draw (node cs:name=a) -- (node cs:name=b);
\node [left] (asfd) at ($ (a)!.5!(b) $) {\Large $\vec{\rho}_1$};
\node [above] (asfd) at ($ (rho2)!.5!(c) $) {\Large $\vec{\rho}_2$};
\draw (node cs:name=c) -- (node cs:name=rho2);
\draw[rotate=0]  (1.5,1) ellipse  (0.7 and 1.8);
\node [right] (asfd) at (1.8,2.7) {\Large $S_{12}$};
\node [right] (asfd) at (2.9,1.8) {\large $\left(\alpha=1,2\right)$};
\end{scope}
\end{tikzpicture}
\par\vspace{0pt}
\end{minipage}
\hfill
\begin{minipage}[b]{.49\textwidth}
\centering
\begin{tikzpicture}
\begin{scope}[xshift=1.2cm,yshift=0.0cm,scale=1.]
\coordinate (rho1) at (1.5,1);
\coordinate (rho2) at (1.5,1);
\coordinate (n1) at (1.5,0);
\coordinate (n2) at (1.5,2);
\coordinate (p) at (4,0);
\path (n1)     node(a) [circle,draw,fill=gray!10!white,inner sep=5pt] {\small n}
          (p)       node(b) [circle,draw,fill=gray,inner sep=5pt] {\small p}
          (n2)     node(c) [circle,draw,fill=gray!10!white,inner sep=5pt] {\small n};
\draw (node cs:name=a) -- (node cs:name=c);
\node [left] (asfd) at ($ (a)!.5!(c) $) {\Large $\vec{\rho}_1$};
\node [above] (asfd) at ($ (rho1)!.5!(b) $) {\Large $\vec{\rho}_2$};
\draw (node cs:name=b) -- (node cs:name=rho2);
\draw[rotate=0]  (1.5,1) ellipse  (0.7 and 1.8);
\node [right] (asfd) at (1.8,2.7) {\Large $S_{12}$};
\node [right] (asfd) at (2.9,1.8) {\large $\left(\alpha=3\right)$};
\end{scope}
\end{tikzpicture}
\par\vspace{0pt}
\end{minipage}
\caption{\label{fig_t-fragm}{\small Three-cluster fragmentations of the triton. The left panel shows the dominant grouping including
a bound deuteron fragment with intermediate spin $S_{12}=1$ and the singlet deuteron with $S_{12}=0$. The structure on the right shows
the nn grouping which allows for $S_{12}=0$ only.}}
\end{figure}
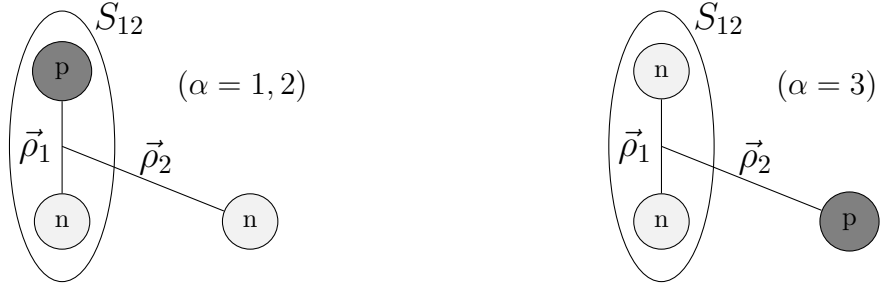